\documentclass[a4paper,11pt]{article}
\usepackage{jinstpub} 

\title{\boldmath Reconstructing High-Fidelity Light Yield Maps for Surface LArTPCs Using Crossing Cosmic Muons}

\author[a,1]{A. Heindel,\note{Corresponding authors.}}
\author[b,1]{W. Shi,}
\author[c]{A. Ralaikoto,}
\author[d]{S. Zhang,}
\author[e]{L. Paulucci,}
\author[e]{F. Marinho,}
\author[f]{M. P. Sampaio,}
\author[g]{S. Jain}

\affiliation[a]{Stony Brook University, SUNY, Stony Brook, New York 11794, USA}
\affiliation[b]{Brookhaven National Laboratory, Upton, NY, 11973, USA}
\affiliation[c]{University of Antananarivo, Antananarivo-101, Madagascar}
\affiliation[d]{Indiana University Bloomington, IU, Bloomington, Indiana 47405, USA}
\affiliation[e]{Instituto Tecnológico de Aeronáutica (ITA), São José dos Campos/SP, 12228, Brasil}
\affiliation[f]{Universidade Estadual de Campinas (Unicamp), Campinas/SP, 13083, Brasil}
\affiliation[g]{University of Texas at Austin, UT, Austin, Texas 78712, USA}
\emailAdd{alex.heindel@stonybrook.edu} 
\emailAdd{wshi2@bnl.gov}

\abstract{
Photon detection systems in liquid argon time projection chambers provide prompt scintillation light information that can improve triggering, timing, calorimetry, and interaction reconstruction. These applications require an understanding of the spatial dependence of the detected light yield (LY), which is not adequately described by a single detector-wide average. We present a method for reconstructing voxelized 3D light yield maps in surface LArTPCs using crossing cosmic muons. For each selected muon, the path length through every crossed voxel is converted to deposited energy using a characteristic energy deposition for cosmic muons at surface level, producing a linear system relating the unknown voxel light yields to the total detected photon signal. The resulting inverse problem is solved using nonnegative least squares, with an additional $L_2$ smoothness penalty used to stabilize weakly constrained voxel values. The method is studied in simulation using the ProtoDUNE-VD detector geometry and photon detection system. Reconstructed maps recover the dominant spatial structure of a visibility-based truth reference and respond as expected to changes in the photon detector configuration. Smoothness regularization reduces zero-valued voxels and localized fluctuations while leaving the detector-average light yield approximately unchanged at the selected regularization strengths. The voxel-wise RMSE relative to the truth reference is reduced by approximately $35\%$ for the regularized reconstruction.
}

\keywords{Noble liquid detectors, Time projection Chambers (TPC), Detector modeling and simulations I, Performance of High Energy Physics Detectors}

\arxivnumber{2608.21581} 

\begin{document}
\maketitle
\flushbottom

\section{Motivation}
\label{sec:motivation}
Since its conception, liquid argon time projection chamber (LArTPC) technology has seen rapid development and several large-scale implementations, including MicroBooNE \cite{microboone_2017}, ICARUS \cite{icarus_2004}, and the Short-Baseline Near Detector (SBND) of the Short Baseline Neutrino (SBN) program at Fermilab \cite{sbn_program1}. The upcoming Deep Underground Neutrino Experiment (DUNE) \cite{dune_tdr_v1} is currently being constructed and will be the largest and most advanced LArTPC experiment to date. Along with the improved resolution of the charge readout system, significant improvements to the photon detection system (PDS) used in these detectors have also been made, allowing for new opportunities to expand their use in physics analyses. 

Traditionally, the main purpose of PDS is for triggering, since photons traverse the detector on a nanosecond timescale, compared to the millisecond time scale that electrons drift toward the charge collection plane. Sorel \cite{Sorel_2014} provided an early simulation-based case that enhanced scintillation light collection in LArTPCs could also be used to improve neutrino energy reconstruction and other reconstruction capabilities at the GeV scale. Work has also been done to study how the PDS can be used to improve neutrino interaction reconstruction at different energy scales, such as MeV \cite{mev_light_2025}, sub-GeV \cite{subgev_light_2026}, and GeV energy scales \cite{gev_light_2025}. The first experimental demonstration of light-augmented calorimetry in a LArTPC came from the LArIAT (Liquid Argon In A Testbeam) experiment \cite{lariat_2020}, which evaluated Michel electrons from stopping muons at Fermilab. The LArIAT Collaboration demonstrated that scintillation light can be used to perform light-augmented calorimetry for 5-50 MeV electrons \cite{lariat_light_2020}. Using Michel electrons, the analysis combined the traditional charge-based energy measurement with information from its PDS in a maximum likelihood framework, showing that the inclusion of light signals improves the reconstructed energy resolution relative to charge only calorimetry.

The results from LArIAT have yet to be demonstrated by a large-scale LArTPC experiment, and SBND is an ideal testbed for this purpose. Located 110 m from the Booster Neutrino Beam (BNB) and with an active mass of 112 tons, SBND will collect an unprecedented level of data, with millions of neutrino interactions expected to be recorded every year. SBND also has the most advanced PDS ever installed in a LArTPC, utilizing 120 photomultiplier tubes (PMTs) and 192 of the silicon photomultiplier (SiPM) based X-ARAPUCA light traps \cite{Machado_2018}. The SBND Collaboration has presented a detailed study of the expected performance of its PDS \cite{sbnd_2024}, and they emphasize that a high-yield optical system is expected to improve timing resolution, allow for light-enhanced particle identification, and has the potential to enable coarse 3D position reconstruction using light alone.

In preparation for DUNE far detector (FD), which will be the largest and most advanced LArTPC ever constructed, several prototypes (called ProtoDUNE) have been operated at the CERN Neutrino Platform. These prototypes have been instrumental in developing and testing the various PDS technologies planned for the first two far detector modules. One major development tested in both the dual-phase (ProtoDUNE-DP) \cite{PDDP_2022} and single-phase (ProtoDUNE-SP) \cite{PDSP_2024} prototypes is the doping of the liquid argon with xenon, which has been shown to significantly improve the detected light yield over large distances~\cite{PDSP_2024}. Compared to the 127 nm scintillation light of argon, the 178 nm emission from xenon has less Rayleigh scattering. This light is more readily detected by photon detectors, making it advantageous for improving photon collection efficiency in large liquid argon volumes. Importantly, the introduction of xenon produced no observable deterioration in charge collection, indicating that PDS performance can be improved without degrading TPC performance.

In order to perform calorimetric analysis with PDS data, it is crucial to understand the light yield (LY), which is the amount of detected scintillation light produced per unit of deposited energy. This quantity is typically reported as a detector-wide average, such as the 18 PE/MeV value reported in the LArIAT study \cite{lariat_light_2020}. However, such a characterization implicitly assumes a spatially uniform light yield throughout the detector volume, which is only an approximation. The SBND Collaboration reported an expected light yield as a function of the mean drift distance \cite{sbnd_2024}, showing how the yield can vary significantly throughout the detector. 

In this study, we report a method for reconstructing high fidelity 3D light yield maps for a detector volume, utilizing crossing cosmic muons present in surface LArTPCs. This work is complementary to ongoing work by the DUNE collaboration to develop a technique of charge and light calibration using a pulsed neutron source \cite{pns_2026}. From neutron capture and de-excitation on argon, a gamma cascade with a total released energy of 6.1 MeV is produced, which can be used as a standard candle for calibration in DUNE's low-energy physics program. 

\section{Technique}
\label{sec:technique}

The technique described in this study utilizes high energy cosmic muon events\footnote{In this study, an event denotes one simulated crossing muon together with its associated energy depositions and photon detector response. The specifics of the simulation are described in section~\ref{sec:sim}}, which are an unavoidable facet of running detectors on the surface. These cosmic muons will cross through the detector at all points and leave tracks of both light and charge that are detected. The underlying principle for this technique is that muons crossing at different angles and through different parts of the detector will have a different number of photons detected by the PDS. With enough crossing muons, a linear system can be formed and solved to reveal a light yield map for the specific arrangement of the PDS in a detector.

For this technique, the trajectory of each muon through the detector must be known, and this information can be obtained in several ways. The simplest method is to use a cosmic ray tagger (CRT), which is made of at least two scintillator panels on opposite sides of the main detector volume. CRT systems are crucial for tagging and rejecting the muons that continually pass through a detector. Time-matched hits in two opposing CRT panels provide two spatial points that can be used to approximate the crossing muon trajectory as a straight line.

Alternatively, the muon trajectory can be reconstructed from the TPC charge readout, supplemented by PDS timing information. In its simplest implementation, the reconstructed entry and exit points can again be connected by a straight line, analogous to the CRT-based approach. More detailed charge reconstruction can provide additional information about the trajectory, including deviations from straight line motion caused by scattering or detector effects. Charge-based reconstruction is also more flexible because it does not depend on the placement or coverage of an external CRT system, but it is more complicated and must account for effects such as space-charge distortions and drift nonuniformities.

\begin{figure}[htbp]
\centering
\includegraphics[width=.6\textwidth]{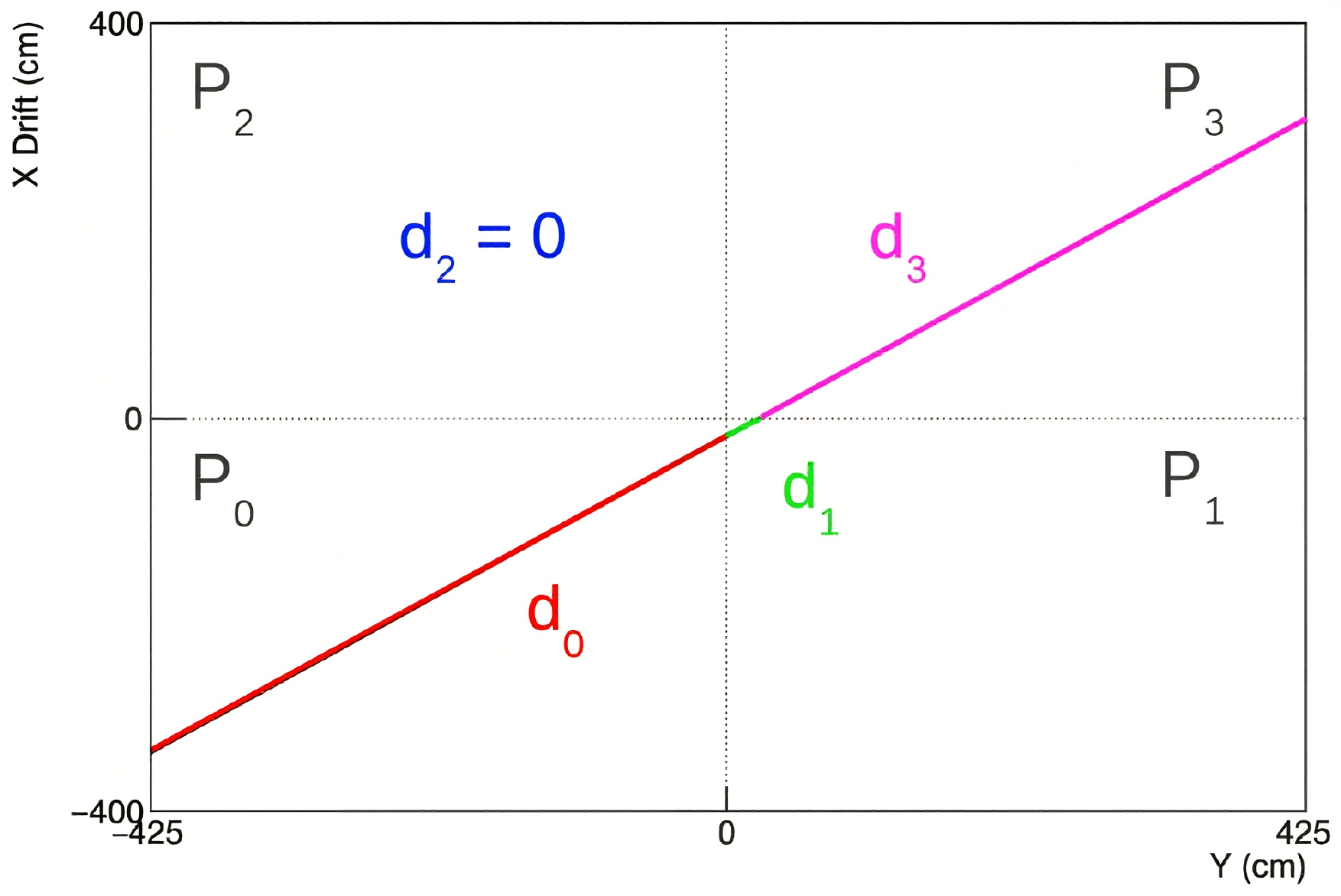}
\caption{Example in 2D showing track crossing a $2\times2$ pixel map. The distances the track traverses through each pixel is shown color coded, with the $P_2$ pixel having zero distance traversed. Each pixel will also have a light yield $P_i$ (i.e., detected photons or photoelectrons per unit energy deposit).\label{fig:demo}}
\end{figure}

For the work presented in this study, the simpler straight line crossing muon approach was used. The specific source of the trajectory information, whether obtained from a CRT system or from TPC charge reconstruction, is not essential for the purposes of this study. An example for two dimensions is shown in figure~\ref{fig:demo}, with the region of interest divided into a $2\times2$ arrangement of pixels (voxels for 3D). 

For each crossing muon event, we can assemble a linear equation consisting of the energy deposition multiplied by the light yield for that voxel ($P_i$ [PE/MeV]), which is summed over all voxels and ideally set equal to the total photons detected by all channels of the photon detection system. Since these crossing muons primarily lose energy through ionization with negligible radiative losses at the characteristic energies considered here, their deposited energy can be estimated from the mean muon stopping power in liquid argon. The cosmic-muon energy distribution used in this study is based on a realistic CORSIKA surface-muon flux and has a median kinetic energy of approximately $5.4~\mathrm{GeV}$. We therefore adopt a representative stopping power corresponding to this median energy rather than the absolute minimum-ionizing value.

The Particle Data Group tabulation of muon stopping power in liquid argon gives total mass stopping powers of $1.877~\mathrm{MeV}$ $\mathrm{cm^2/g}$ at $4~\mathrm{GeV}$ and $1.991~\mathrm{MeV}$ $\mathrm{cm^2/g}$ at $8~\mathrm{GeV}$ \cite{pdg_2024}. Using a liquid-argon density of $\rho = 1.396~\mathrm{g/cm^3}$ and interpolating to the median kinetic energy of $5.4~\mathrm{GeV}$, the corresponding linear stopping power is

\begin{equation}
\rho
\left(\frac{1}{\rho}\frac{dE}{dx}\right)
\simeq
(1.396)
(1.92)
\simeq
2.7~\frac{\mathrm{MeV}}{\mathrm{cm}}.
\label{eq:muon_dep}
\end{equation}

By finding the distance ($d_i$) the muon traverses through a voxel and multiplying by this $2.7~\mathrm{MeV/cm}$ representative energy deposition, the energy deposited in each voxel can be estimated. The equation for a single event of the example $2 \times 2$ configuration from figure~\ref{fig:demo} is shown in \eqref{eq:lin}.

\begin{equation}
[P_0 d_0+ P_1 d_1 + P_2 d_2 + P_3 d_3]\times 2.7 \frac{\text{MeV}}{\text{cm}} = \text{Total Photoelectrons Detected}
\label{eq:lin}
\end{equation}

Each crossing muon will produce scintillation photons that are detected by the PDS, with the amount detected varying with how the muon crosses the detector. Each crossing muon provides one equation in the form of \eqref{eq:lin}, and for many selected events, these equations can be assembled into the linear system

\begin{equation}
Ax = b
\label{eq:lin_full}
\end{equation}

where each row of $A$ corresponds to one crossing muon event and each column corresponds to one voxel. The solution vector $x$ contains the unknown light yield values for all voxels, and the vector $b$ contains the total detected light for each event. Solving this system therefore gives a voxelized 3D map of the detector light yield.

\subsection{Voxelization and Distance Calculation}
\label{sec:voxel}

The volume can be discretized into an array of voxels, with a larger number of voxels corresponding to a higher system resolution. In its simplest form, the muon track can be estimated by a start point and an end point, either from two CRT panels or from TPC reconstruction. The energy deposited in each voxel can be calculated by multiplying the muon energy loss in liquid argon by the distance the muon traverses within that voxel. This creates a need for an efficient method to determine these distances. For this work, the Liang-Barsky algorithm was chosen, which is a parametric line clipping algorithm commonly used in computer graphics \cite{liang_barsky_1984}. 

The reconstruction presented in this study uses only straight line muon trajectories defined by two crossing points, but the voxelization and path length calculation can be extended to more complicated tracks made up of many piecewise straight line paths, such as those from TPC charge reconstruction. By applying the Liang-Barsky algorithm described above to each piecewise segment, the total path length through the boundary is then obtained by summing the contributions from all clipped segments, and from this, the energy deposited in each voxel can be found following the same procedure as described above.

\subsection{Linear Solving}
\label{sec:linear}

The linear system is created from the voxel path lengths (converted to deposited energy) and total photon counts, as shown \eqref{eq:lin}. In this formulation, each element of the solution vector corresponds to the light yield in a voxel. An unconstrained least squares solution can therefore return negative light yield values, which are nonphysical. To avoid this, the reconstruction was formulated as a nonnegative least squares problem (NNLS),

\begin{equation}
\label{eq:nnls_cond}
    \min_{x \ge 0} \|Ax-b\|_2^2
\end{equation}

where $A$ is the track-coverage matrix, $x$ is the vector of voxel light yields, and $b$ is the vector of detected photon counts for the selected events.

For this work, two NNLS solving methods available in the Python library SciPy \cite{scipy_2020} were tested. The first was \texttt{scipy.optimize.nnls}, referred to here as the NNLS solver. This implementation uses the Lawson-Hanson active-set method \cite{nnls_lh_1995}, which solves the Karush-Kuhn-Tucker (KKT) conditions for the nonnegative least squares problem. The second method was \texttt{scipy.optimize.lsq\_linear} using the trust-region reflective algorithm, which is selected with \texttt{method='trf'} \cite{nnls_tr_1999}. This function solves a more general bounded linear least squares problem, which was set with lower bounds fixed to zero and upper bounds left unconstrained, making it solve the same nonnegative constrained objective. A direct comparison between these two methods is done in section~\ref{sec:solver_results}.

One practical difference between the \texttt{nnls} method and \texttt{lsq\_linear} method implementation in SciPy is that the \texttt{nnls} function requires a dense matrix, while the \texttt{lsq\_linear} method works with dense and sparse matrices. To achieve higher resolution light yield maps, the number of voxels and therefore number of variables in the linear system has a cubic dependence on the number of subdivisions along each spatial axis. Increasing the resolution can quickly create a dense A matrix of over a billion values, which for double data types can put significant memory strain on machines. Because straight line muon tracks only traverse through a small number of the total voxels in the system, this system is a good fit for sparse matrix methods, which reduces the memory concern issues that dense matrix methods pose.

\subsection{Regularization}
\label{sec:reg_tech}

The nonnegative least squares solutions described in section~\ref{sec:linear} can produce sparse or locally fluctuating light yield maps, especially in weakly constrained regions of the voxel grid. Regularization is a well-studied method that provides a way to stabilize the inverse problem by adding prior information about the expected spatial behavior of the solution. In this study, an $L_2$ smoothness regularization was used to penalize large differences between neighboring voxel light yield values \cite{hansen_inverse_2010}. Unlike a direct ridge penalty on the solution magnitude, which can shift the overall light yield normalization, this form of regularization acts on local voxel-to-voxel variations. The regularized nonnegative least squares problem is written as

\begin{equation}
\label{eq}
\min_{x \ge 0}
\left[
\frac{1}{2}\|Ax-b\|_2^2
+
\frac{\lambda}{2}
\sum_{\langle i,j\rangle}
\left(x_i-x_j\right)^2
\right],
\end{equation}

where the sum runs once over all nearest-neighbor voxel pairs, and $\lambda$ controls the strength of the smoothness penalty. Larger values of $\lambda$ increasingly suppress local voxel-to-voxel fluctuations, while $\lambda=0$ recovers the unregularized nonnegative least squares problem.

Results showing the unregularized maps to the smoothness regularized maps are presented in section~\ref{sec:ideal_compare}. Discussion of the observed sparsity of the solutions and how regularization can be utilized to overcome this effect is shown in section~\ref{sec:reg_results}. The method of selecting the best $\lambda$ is also outlined in section~\ref{sec:reg_results}, along with results for how the best $\lambda$ changes with resolution and event count.

\section{ProtoDUNE-VD Simulation Study}
\label{sec:study}

To study and develop this light yield mapping technique in a realistic detector environment, the ProtoDUNE-VD experiment was selected. ProtoDUNE-VD is a large-scale LArTPC with a sophisticated PDS, serving as a favorable platform for evaluating the feasibility and performance of this method. In addition to its relevance as a DUNE prototype, the experiment offers a mature simulation infrastructure within the LArSoft framework \cite{larsoft_2017}, allowing the detector geometry, particle propagation, energy deposition, scintillation light production, and photon detection response to be modeled accurately. The following sections describe the ProtoDUNE-VD detector and its PDS, and then the simulation chain and modeled detector effects used in this study are outlined.

\subsection{ProtoDUNE-VD Experiment}
\label{sec:pdvd}

\begin{figure}[htbp]
\centering
\includegraphics[width=.4\textwidth]{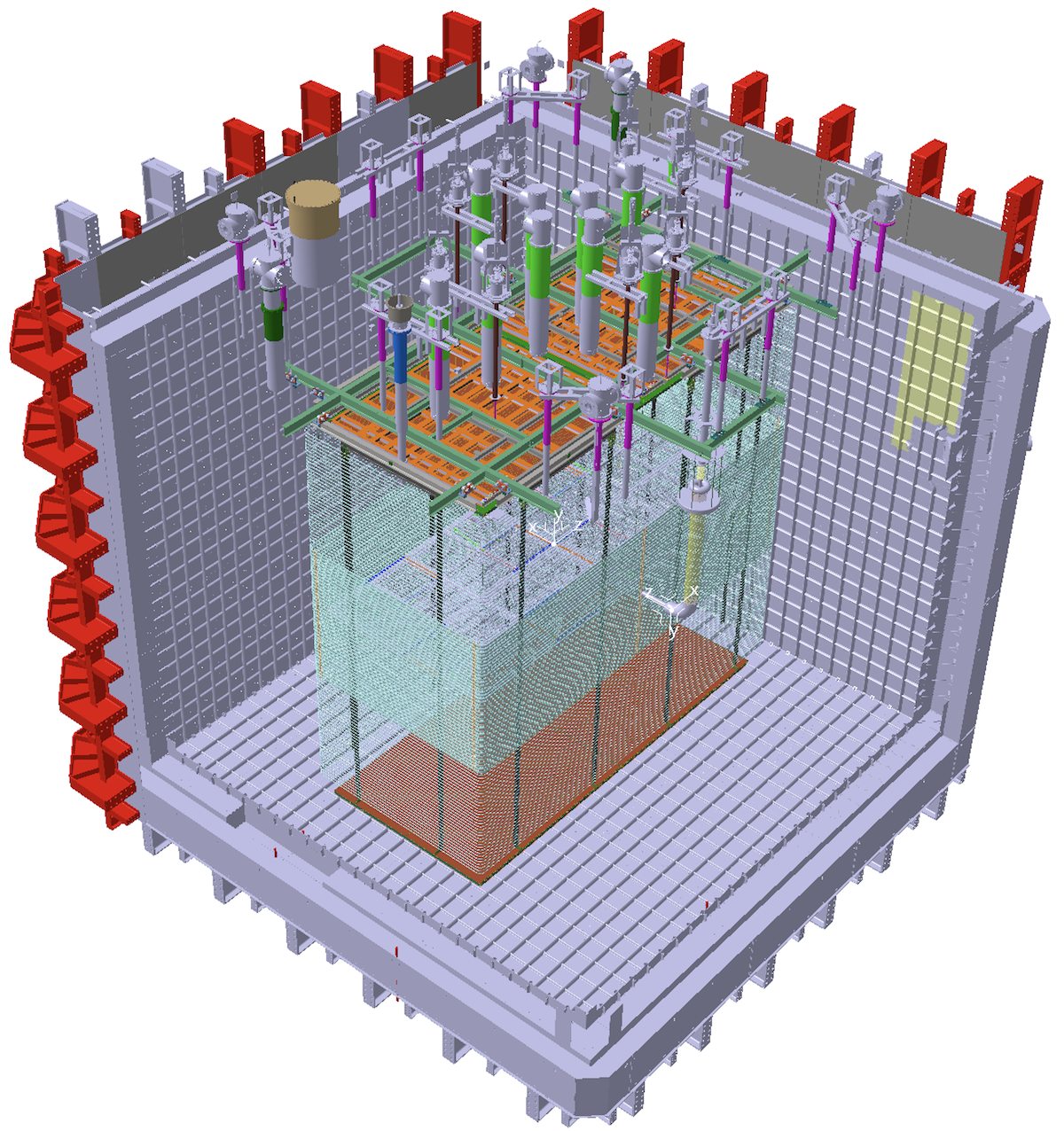}
\hspace{0.06\textwidth}
\includegraphics[width=.42\textwidth]{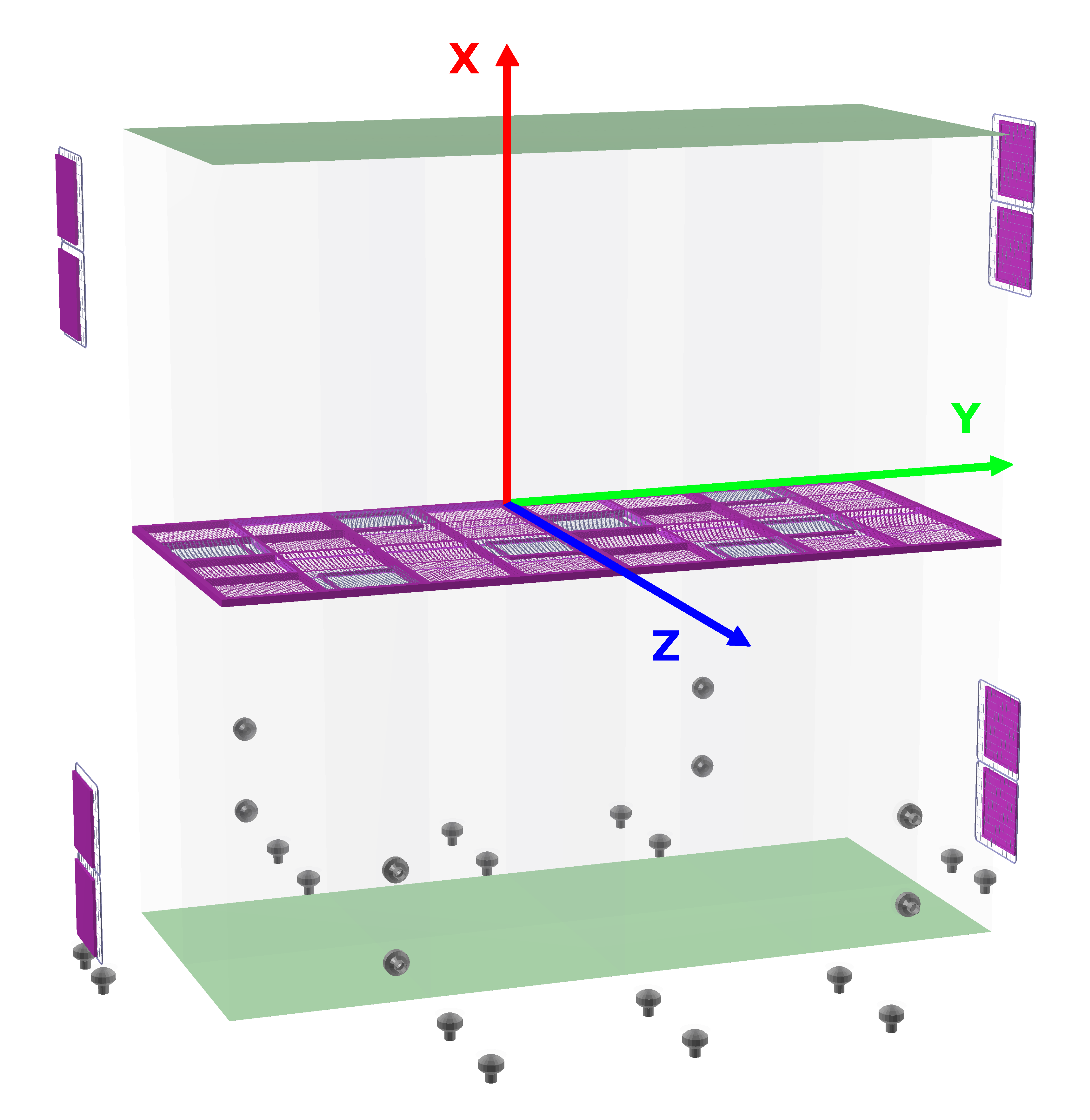}
\caption{A schematic of the ProtoDUNE-VD detector inside the NP02 cryostat at the CERN Neutrino Platform \cite{dune_vd_tdr} (left) and a cross-sectional view of the v4 PDS geometry with the defined coordinate axes (right). The X-ARAPUCAs are visible on the cathode plane and membrane walls, with the PMTs lining the lower drift volume. The axes in the right figure are what are used for all results presented in this study.}
\label{fig:pds_layout}
\end{figure}

ProtoDUNE-VD serves as a large-scale prototype of the vertical drift LArTPC design \cite{dune_vd_tdr} that will be used alongside the horizontal drift design as the first two DUNE far detector modules. Located at the Neutrino Platform at CERN, ProtoDUNE-VD uses the NP02 cryostat ($\sim$ 800 ton LAr) and is intended to be a full system integration: verifying that the charge readout, cathode, high voltage, photon detection system, supports, cabling, grounding, and installation procedure are understood before the 17-kton FD-VD module is assembled. A 3D rendering of the ProtoDUNE-VD detector in the NP02 cryostat is shown in figure~\ref{fig:pds_layout}.

ProtoDUNE-VD uses a vertical drift architecture, with the upper and lower drift regions read out by two charge readout planes (CRPs) \cite{crp_2018}, which are a development of the traditional wire plane readout design. Unlike previous LArTPCs, the FD-VD design uses only X-ARAPUCA photon detectors \cite{Machado_2018} for its PDS. Because the CRPs are opaque to light, it is not possible to have photon detectors behind the anode plane. This leaves the options to mount photon detectors inside the cathode plane and on the cryostat membrane walls behind the field cage. Because of the high-voltage environment, a nonconductive power and readout system is required for the cathode-mounted photon detectors, which is accomplished using power over fiber (PoF) technology \cite{pof_2024}. To monitor and calibrate the PDS response over time, ProtoDUNE-VD also includes a UV LED pulsing system that delivers light into the detector through optical fibers terminated with diffusers \cite{uv_calib_2026}. ProtoDUNE-VD tests both of these photon detector locations, including 8 cathode- and 8 membrane-mounted X-ARAPUCAs, as shown in figure~\ref{fig:pds_layout}. There are also 22 PMTs leftover from ProtoDUNE-DP~\cite{ProtoDUNE-DP} that have been used for precision light studies and crosschecks, but these are not part of the final FD-VD design.

\subsection{Visibility-Based Truth Light Yield Map}
\label{sec:ideal}

As an independent cross-check, the light yield in simulation can be estimated as

\begin{equation}
\text{LY}= Y_{\gamma}^{\mathrm{LAr}} \cdot V \cdot \varepsilon_{\mathrm{PD}},
\end{equation}

where $Y_{\gamma}^{\mathrm{LAr}}$ is the LAr scintillation yield, $V$ the visibility (fraction of emitted photons reaching the PD surfaces), and $\varepsilon_{\mathrm{PD}}$ the PD detection efficiency. For MIP energy deposition, the scintillation yield is taken as $25{,}000$ photons/MeV in the active volume for a nominal drift field of 500 V/cm and $44{,}000$ photons/MeV outside it, where the stray electric field is substantially lower~\cite{kubota1978recombination, bonivento_2024, doke1990estimation}. These scintillation yields are consistent with the LArQL model~\cite{LArQL, LArQL_update} used in the cosmic muon simulation detailed in section~\ref{sec:sim}. The PD efficiencies are 3.0\% for X-ARAPUCAs \cite{xarapuca_eff_2025}, 3.6\% for PEN-coated PMTs, and 12.0\% for TPB-coated PMTs \cite{PDDP_2022}, accounting for the production of photoelectrons per impinging photon. The X-ARAPUCA value corresponds to the FD-VD target photon detection efficiency, while the PMT values are chosen as effective efficiencies for the wavelength shifter and PMT response.

The photon visibility $V$ predicts photon detection probabilities directly from the scintillation vertex position. This prediction is obtained using a one-dimensional generative neural network (1D Genn) fast optical model \cite{Mu_2022}, implemented as a computable graph and trained on Geant4 \cite{geant4_2003} samples generated with the full simulation chain, including photon transport and material optical properties, that associate each scintillation vertex with the corresponding detected photon distribution. The ProtoDUNE-VD volume was divided into 80 planes along each spatial axis, and the mean visibility projection was computed for each coordinate.

This approach provides an idealized estimate of the detector response because the visibility is evaluated directly from interaction vertices sampled on the computable graph, rather than from reconstructed particle events. Consequently, the detector response can be probed at arbitrary spatial positions, yielding higher spatial resolution than analyses based on simulated crossing muons or experimental data, which are limited by event statistics and reconstruction uncertainties. Additionally, this light yield estimation does not include the reconstruction phase to obtain waveforms; thus, the electronics response is estimated simply by multiplying visibility and efficiency.

\begin{figure}[htbp]
\centering
\includegraphics[width=.32\textwidth]{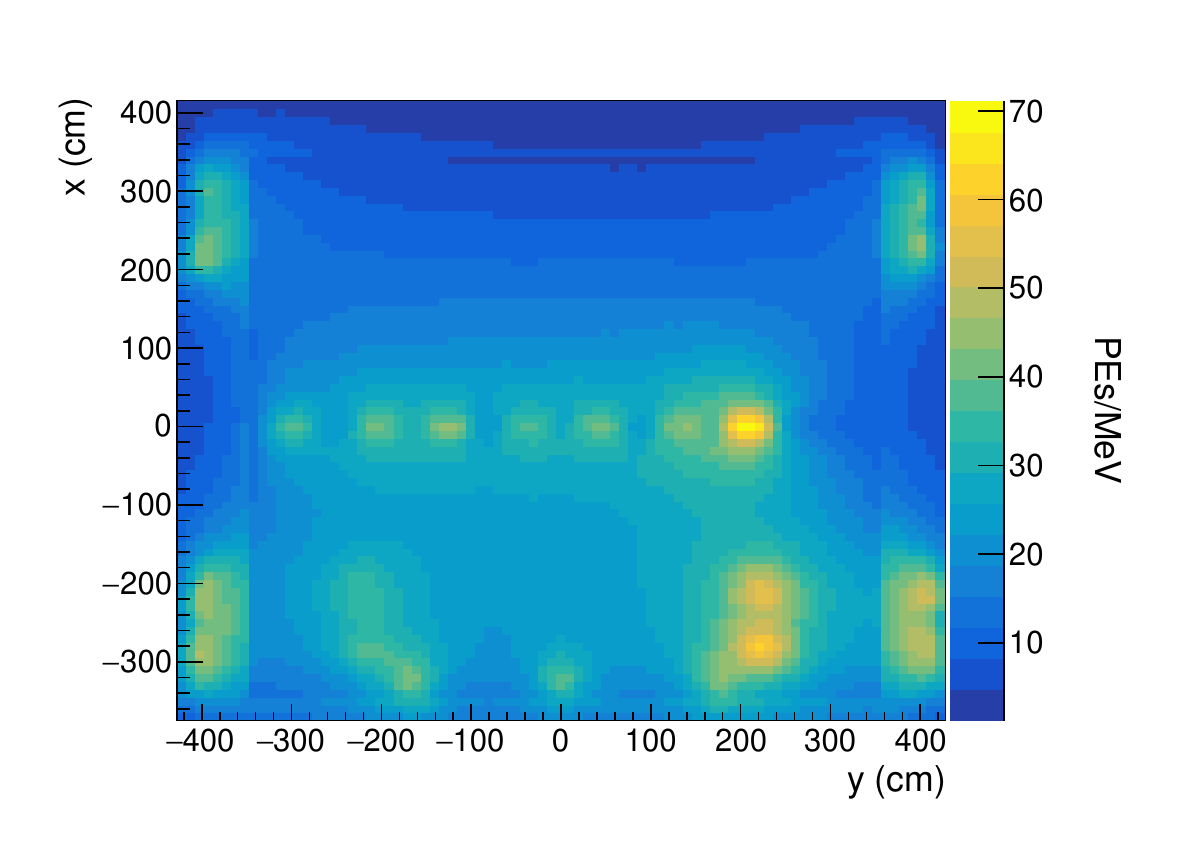}
\includegraphics[width=.32\textwidth]{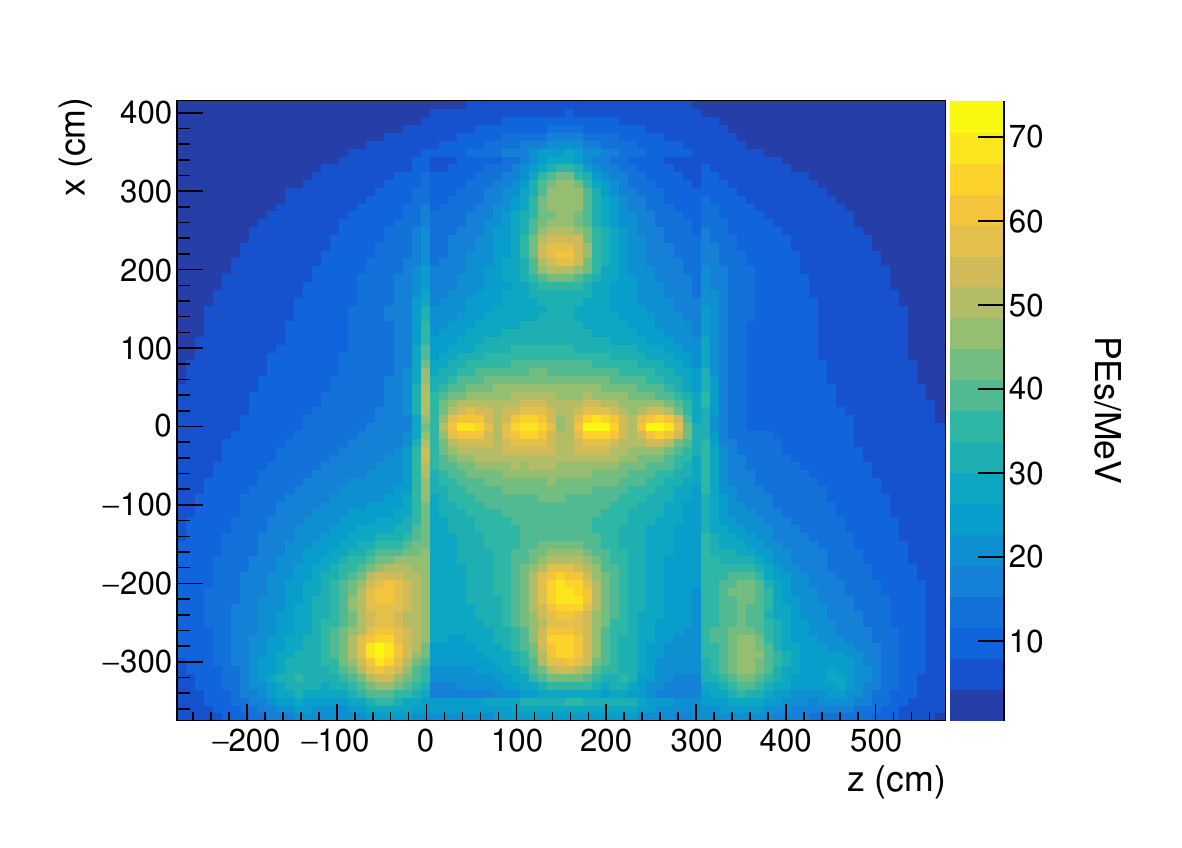}
\includegraphics[width=.32\textwidth]{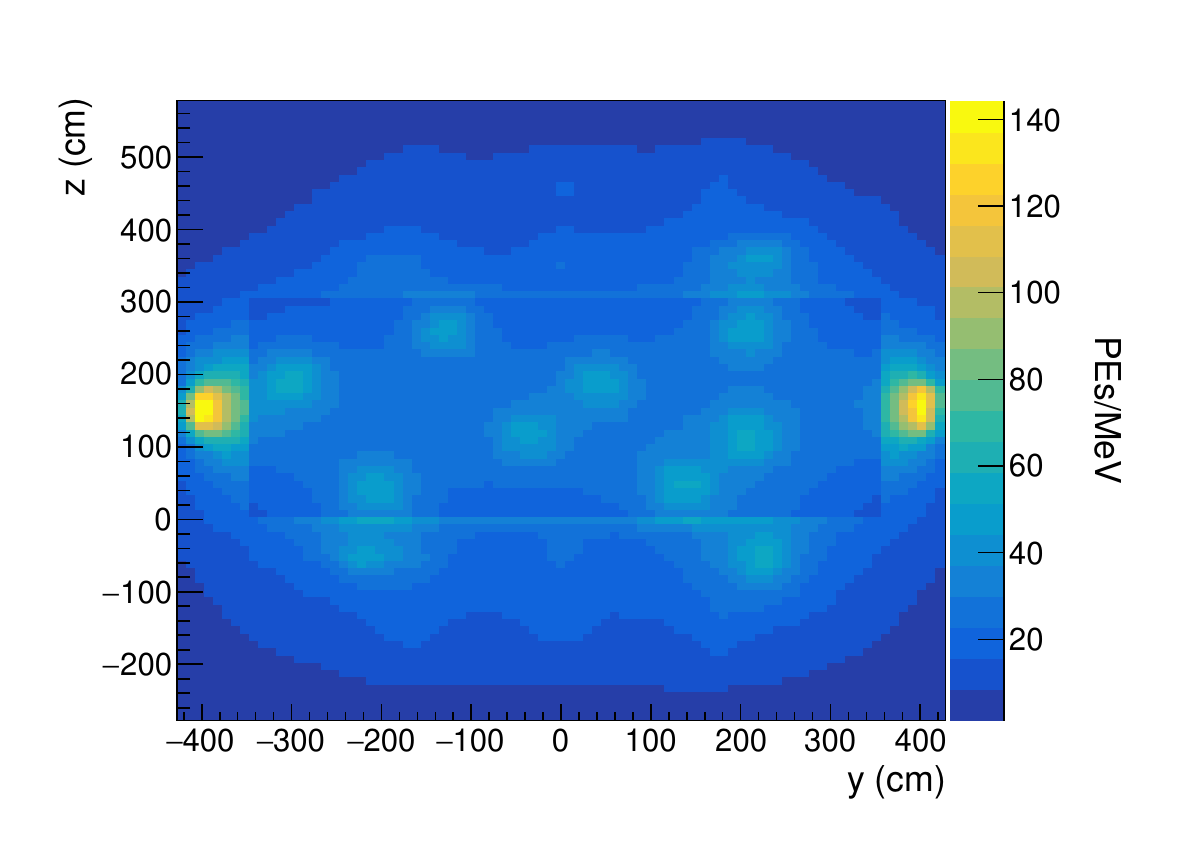}
\includegraphics[width=.32\textwidth]{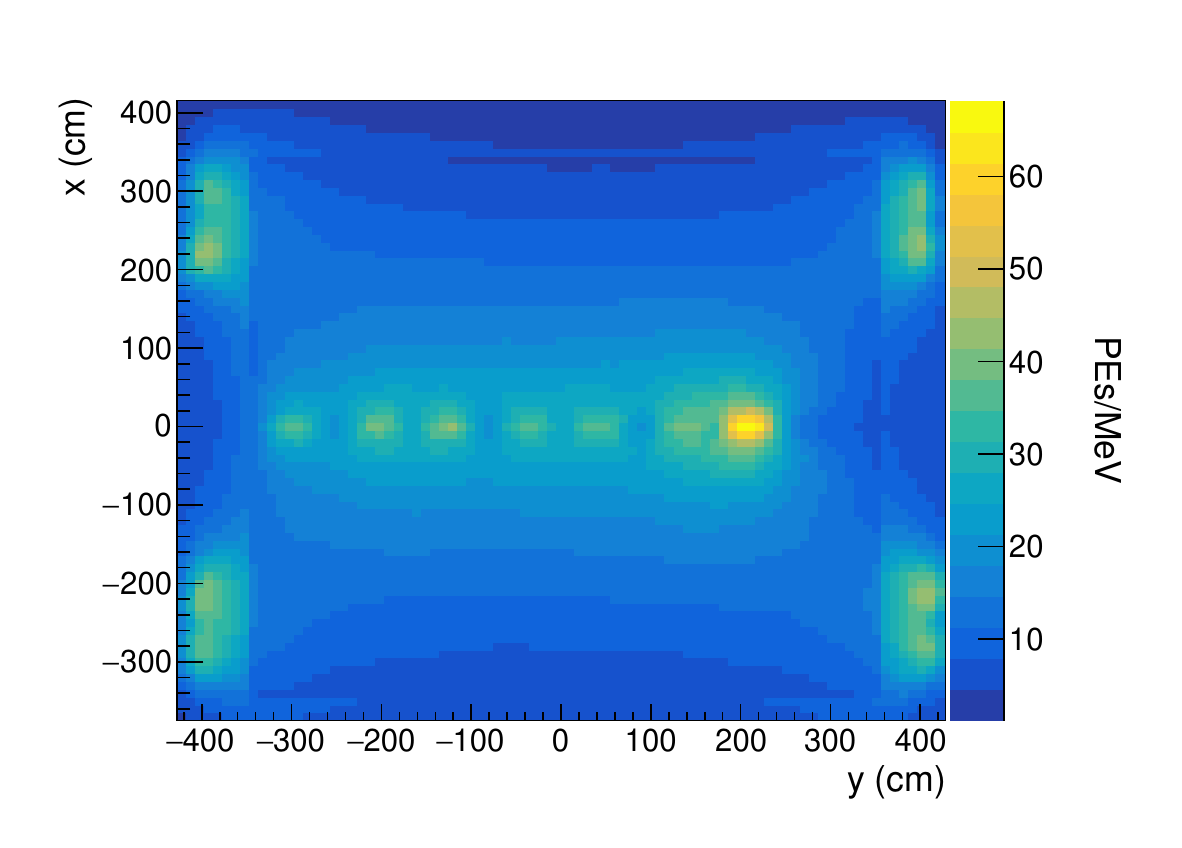}
\includegraphics[width=.32\textwidth]{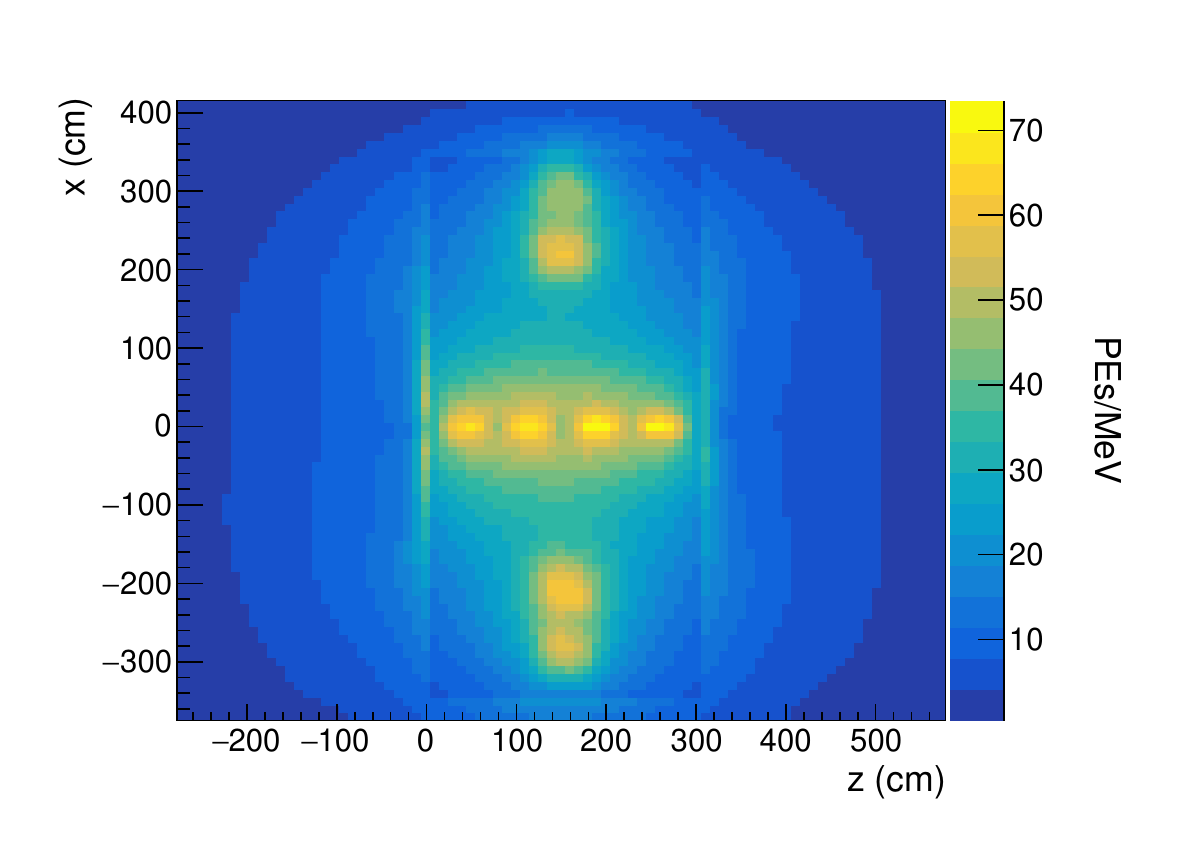}
\includegraphics[width=.32\textwidth]{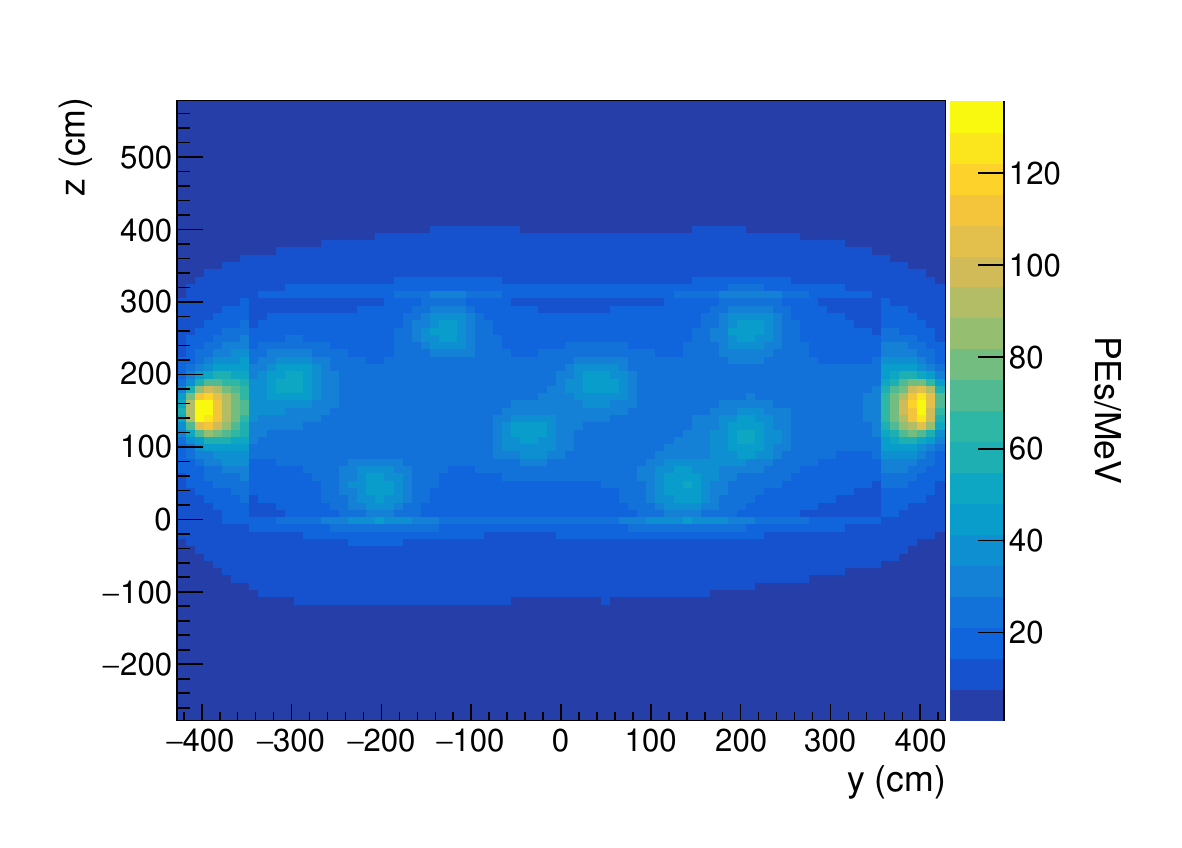}
\caption{Visibility-based truth light yield projections for the ProtoDUNE-VD v4 geometry. The left column shows the YX projection averaged over $z$, the middle column shows the ZX projection averaged over $y$, and the right column shows the YZ projection averaged over $x$. The top row includes PMT and X-ARAPUCA contributions, while the bottom row includes only the X-ARAPUCA contribution.\label{fig:ideal}}
\end{figure}

The resulting visibility-based true light yield projections are shown in figure~\ref{fig:ideal}. In this study, these maps are treated as the visibility-based truth reference for comparisons with the crossing muon reconstructions presented in section~\ref{sec:ideal_compare}. The truth maps show the expected spatial nonuniformity of the detector response, with enhanced light yield near the photon detector locations and reduced response farther from optically active regions. The X-ARAPUCA-only maps also exhibit an asymmetry in the cathode-plane response. This asymmetry arises from the specific arrangement of the X-ARAPUCA modules, including regions where modules are positioned adjacent to one another rather than forming a fully symmetric diagonal pattern (cf. figure~\ref{fig:pds_layout} right plot). When the PMT contribution is included, the response changes primarily in the lower detector region where the PMTs are located. The PMT contribution is also asymmetric because the PMT system contains different wavelength-shifter configurations, with the more efficient TPB-coated PMTs and the less efficient PEN-coated PMTs.

\subsection{Cosmic Muon Simulation Setup}
\label{sec:sim}

The simulated crossing muon sample used in this work was produced with the ProtoDUNE-VD simulation geometry version 4 (v4) in the LArSoft simulation framework \cite{larsoft_2017}. Muon kinetic energies were sampled from a CORSIKA-derived surface cosmic-muon energy distribution \cite{corsika}, rather than from a fixed or uniform energy range. The input distribution spans approximately $2$--$393~\mathrm{GeV}$ and is strongly weighted toward lower energies, with a mean kinetic energy of $9.9~\mathrm{GeV}$ and a median of $5.4~\mathrm{GeV}$. Particle transport was simulated using the Geant4 toolkit, while LArSoft was used to record the energy deposition and the corresponding ionization and scintillation production in the detector volume. The correlation between ionization charge and scintillation light is taken into account in the LArSoft simulation using the LArQL model. The optical response to each scintillation energy deposition was evaluated using the same 1D Genn fast optical model described in section~\ref{sec:ideal}. 

The detailed PDS response, including the conversion of detected photons into digitized waveforms, was not simulated. Instead, a constant photon detection efficiency of $3\%$ was assumed for each photon detector. For this study, the expected number of detected photons was calculated by multiplying the number of photons incident on the optical detector surfaces by the assumed detection efficiency. The resulting detected light count for each event was used as the corresponding entry of the vector $b$ in the linear system described in section~\ref{sec:technique}. The 3\% efficiency is then corrected in the analysis to the efficiencies for PMTs with different coatings outlined in section~\ref{sec:pdvd}.

\begin{figure}[htbp]
\centering
\includegraphics[width=.49\textwidth]{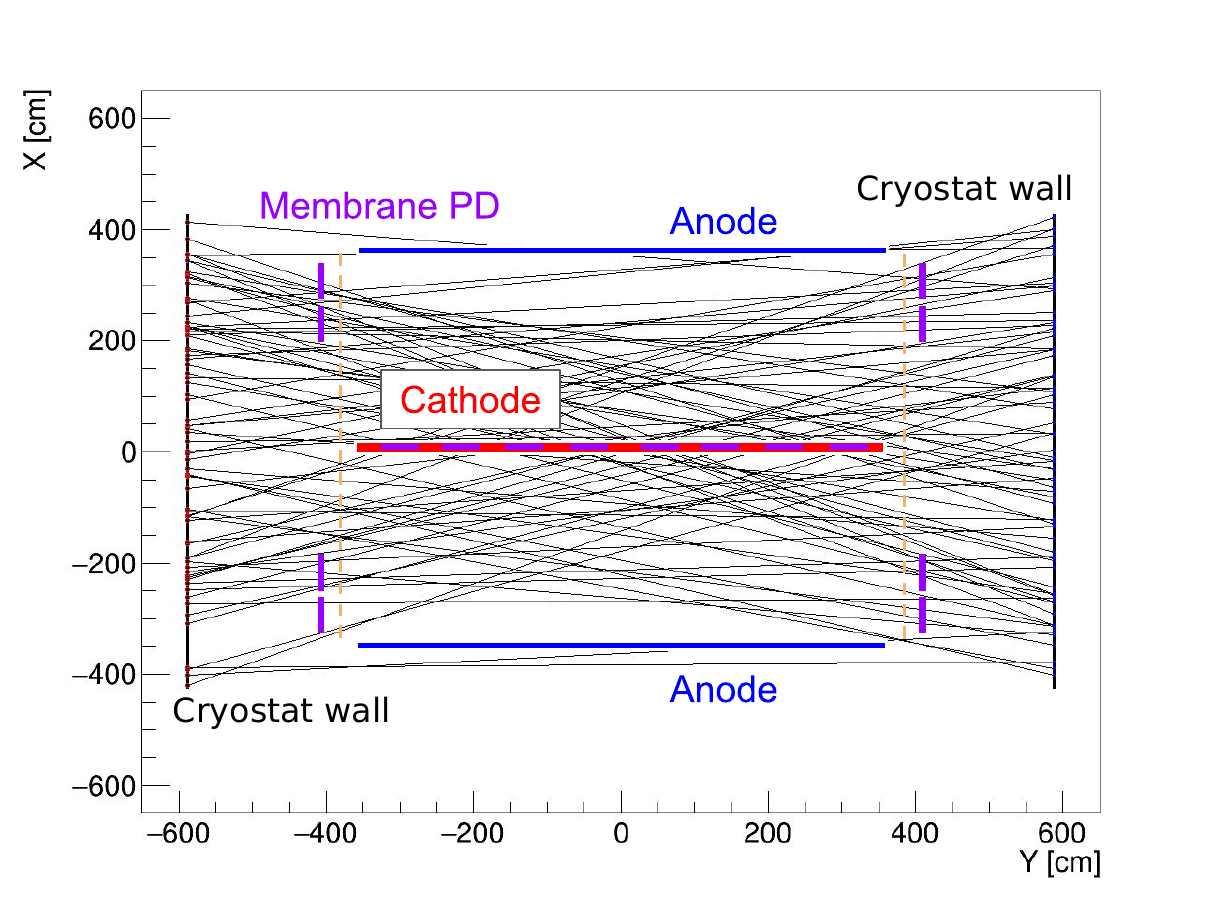}
\qquad
\includegraphics[width=.42\textwidth]{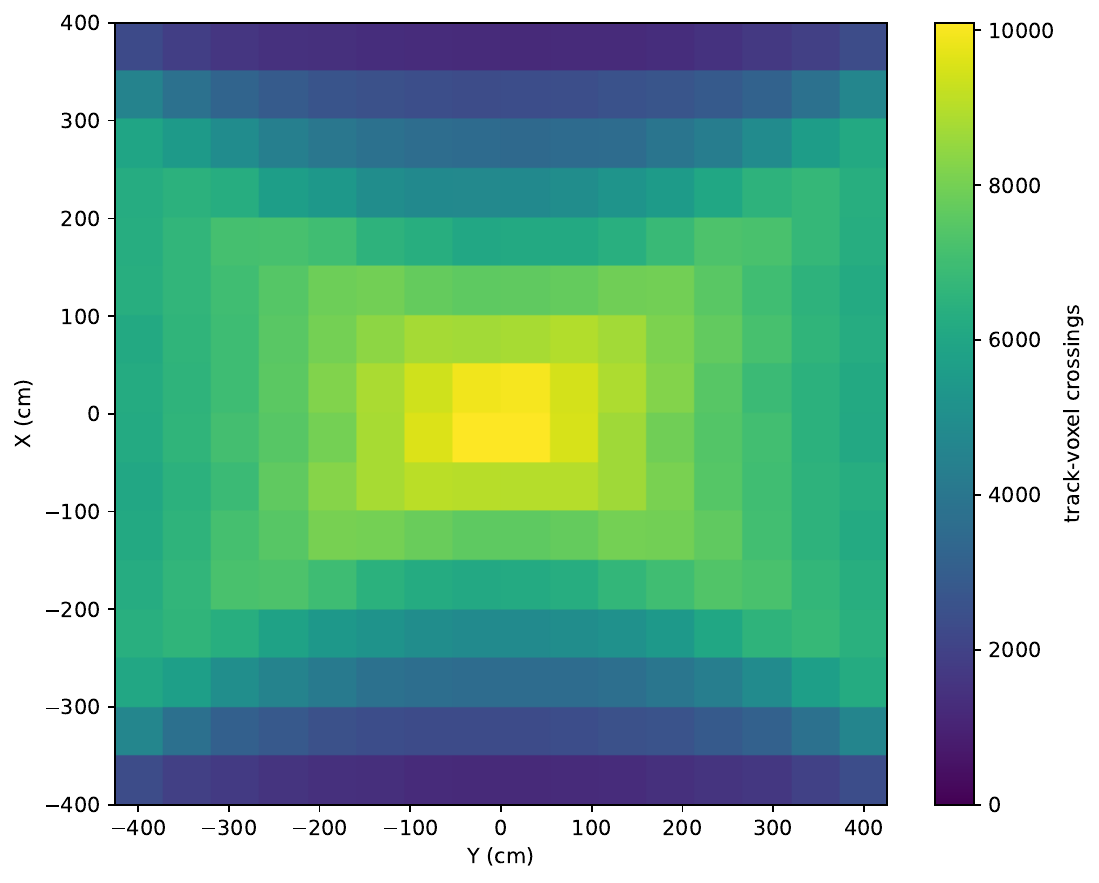}
\caption{Diagram of the ProtoDUNE-VD detector geometry and generated
crossing-muon trajectories (left). YX projection of the total
track--voxel crossing count for the 60,000 event sample, summed over the
$Z$ direction (right).\label{fig:crt_full}}
\end{figure}

The muon selection boundaries were defined using opposite outer cryostat walls, providing an idealized crossing muon sample that spans the full liquid argon volume and gives track coverage throughout the region used for the light yield reconstruction, as shown in figure~\ref{fig:crt_full}. The coordinate axes and detector orientation used in this figure, and throughout the remainder of the paper, follow the convention defined in figure~\ref{fig:pds_layout}. Only muons intersecting both selected cryostat walls were retained. Muons intersecting only one wall were rejected because they do not provide the two spatial points required to define the straight line track approximation used in this study.

\section{Results}
\label{sec:results}

Following the reconstruction technique outlined in section~\ref{sec:technique} and using the ProtoDUNE-VD simulated data described in section~\ref{sec:sim}, this section describes various studies done to demonstrate and characterize the effectiveness of this technique. 

\subsection{Comparison to Truth Reference}
\label{sec:ideal_compare}

The primary validation of the reconstruction method is a comparison between the light yield map inferred from crossing muons and the truth reference map obtained from the visibility-based calculation shown in figure~\ref{fig:ideal}. The truth map represents the expected detector response to scintillation light produced at fixed spatial positions, with the response calculated directly from the photon visibility and photon detector efficiency. In contrast, the crossing-muon reconstruction infers the voxelized light yield indirectly from the integrated photon counts associated with extended muon trajectories. For a direct comparison, the truth reference map is rebinned to the voxel configuration used by the reconstruction. Agreement between the two approaches would therefore indicate that the linear inversion recovers the underlying spatial response encoded in the optical simulation.

In addition to the unregularized nonnegative least squares solution, a reconstruction incorporating the smoothness regularization outlined in section~\ref{sec:reg_tech} is also considered. This regularization penalizes large differences between neighboring voxels, thereby suppressing spatial variations that are weakly constrained by the available muon trajectories. The regularization strength used for the results in this section is selected using the procedure described in the following section.

\begin{figure}[htbp]
\centering
\includegraphics[width=.32\textwidth]{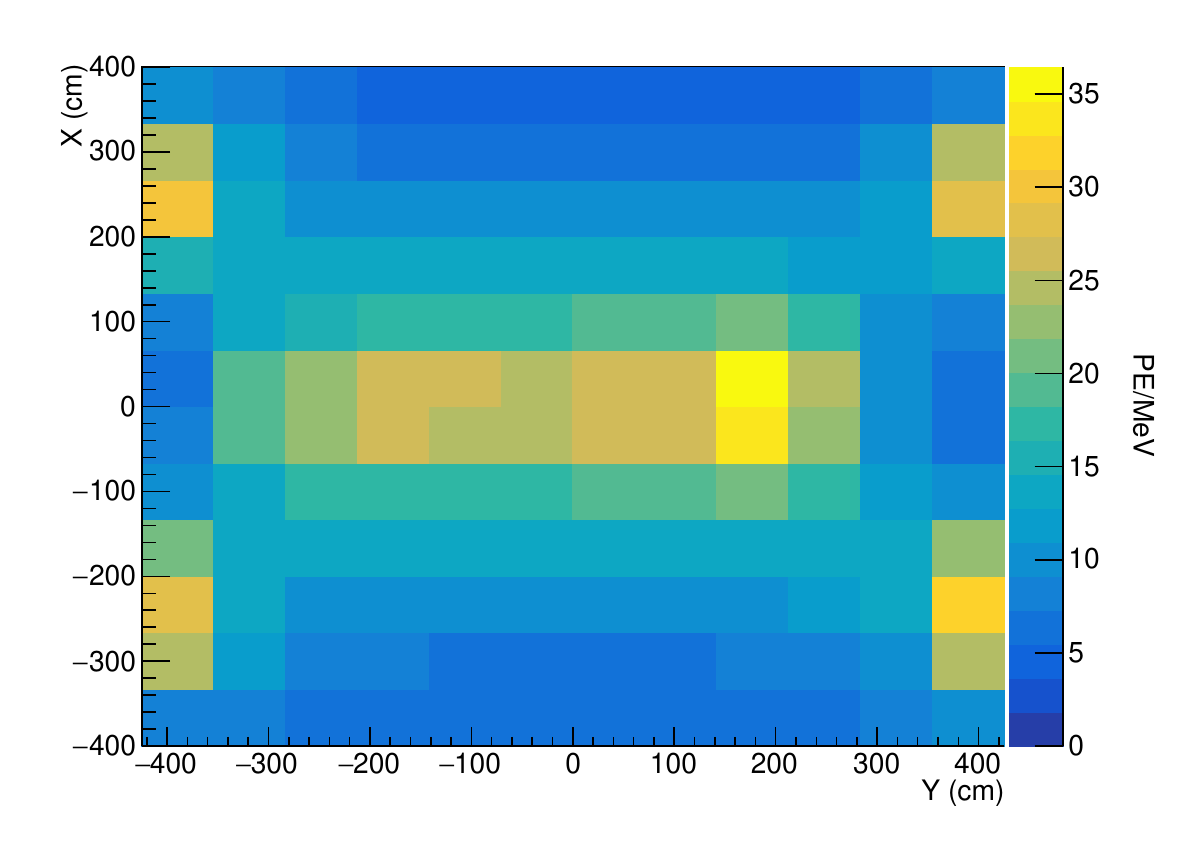}
\includegraphics[width=.3\textwidth]{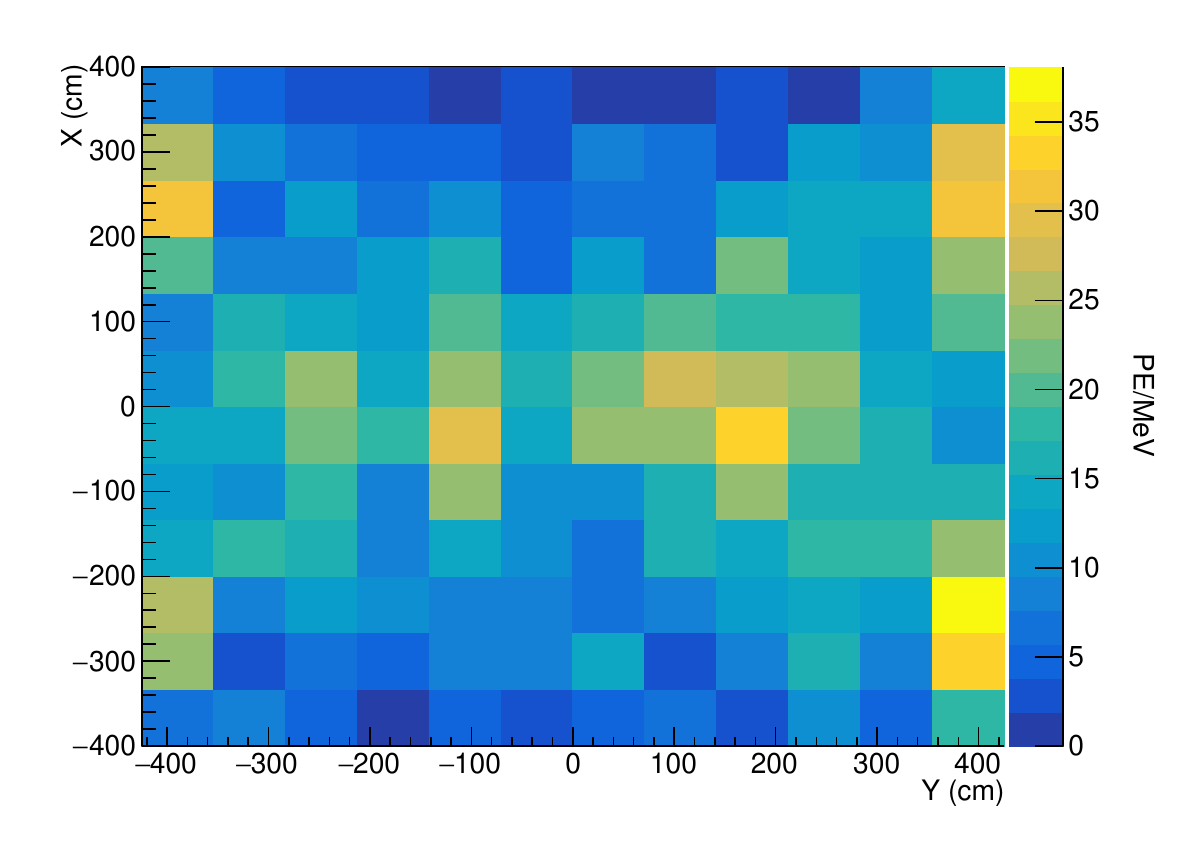}
\includegraphics[width=.3\textwidth]{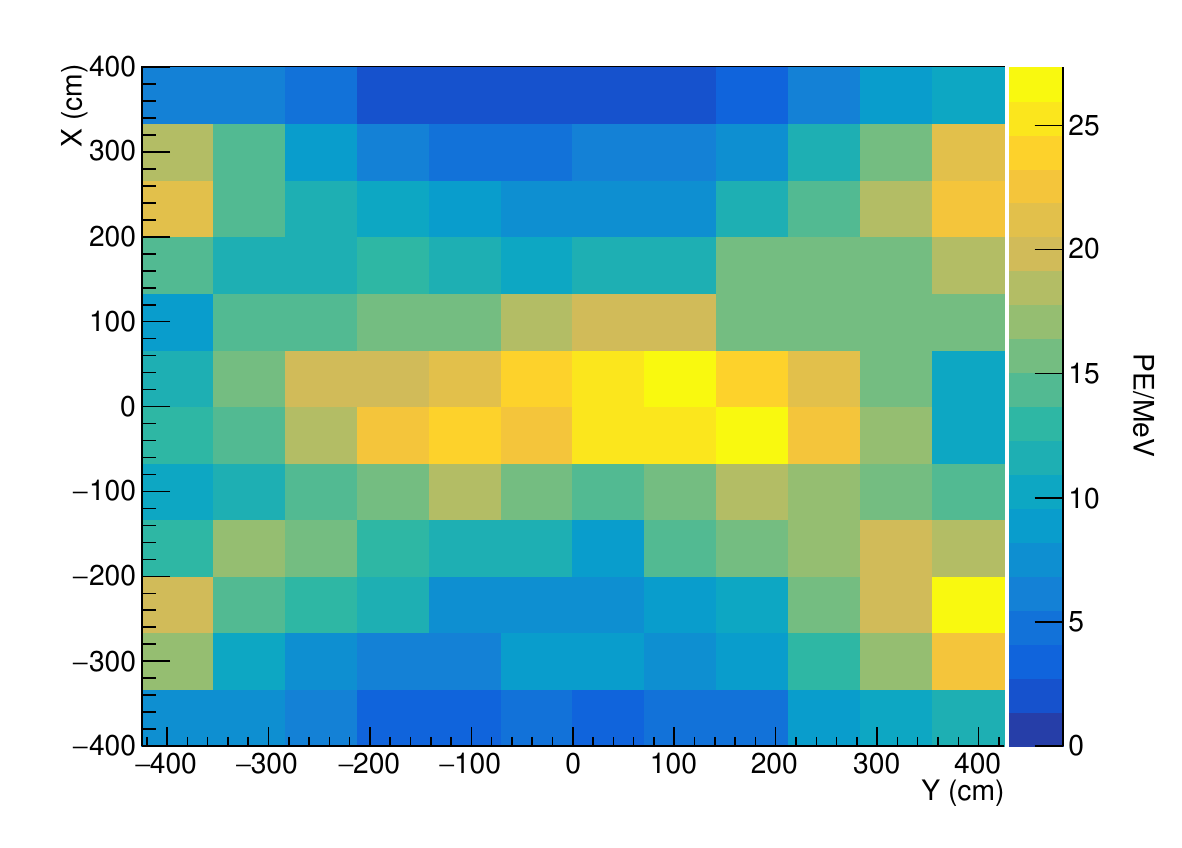}
\includegraphics[width=.32\textwidth]{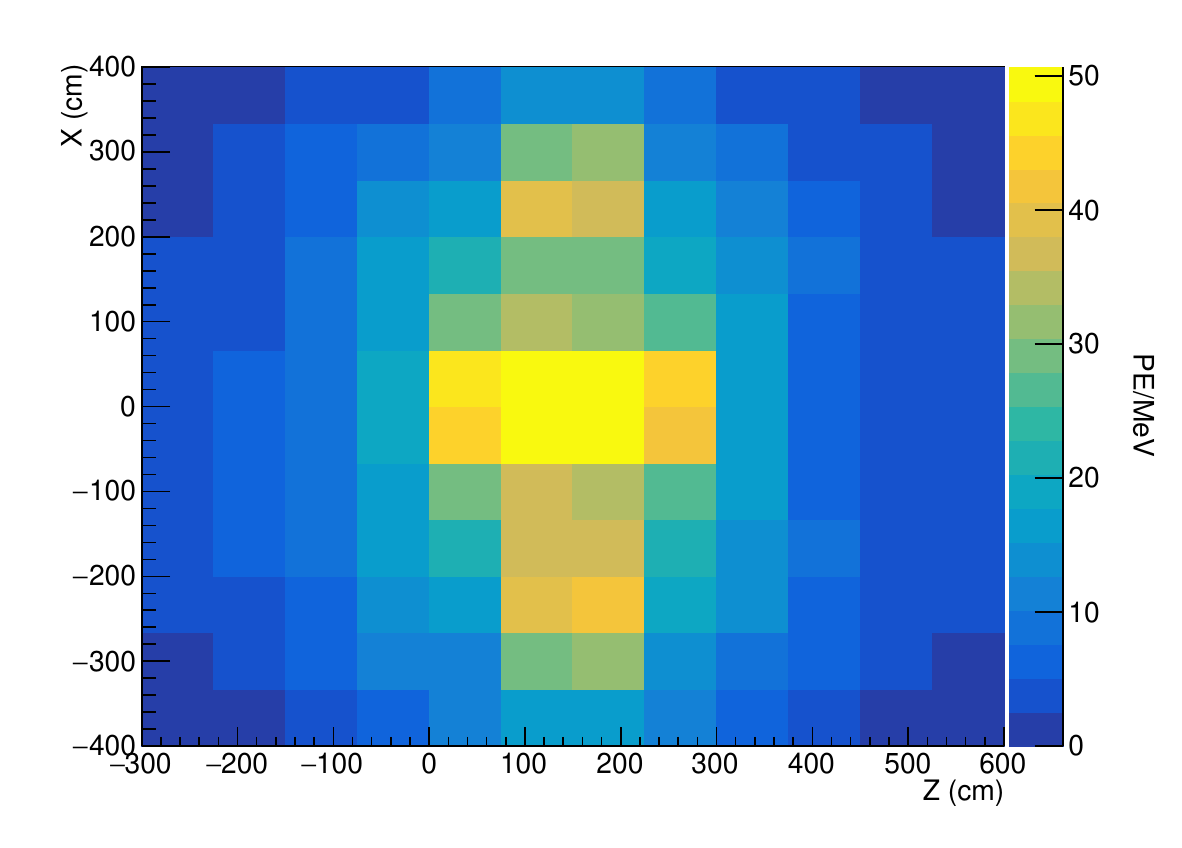}
\includegraphics[width=.3\textwidth]{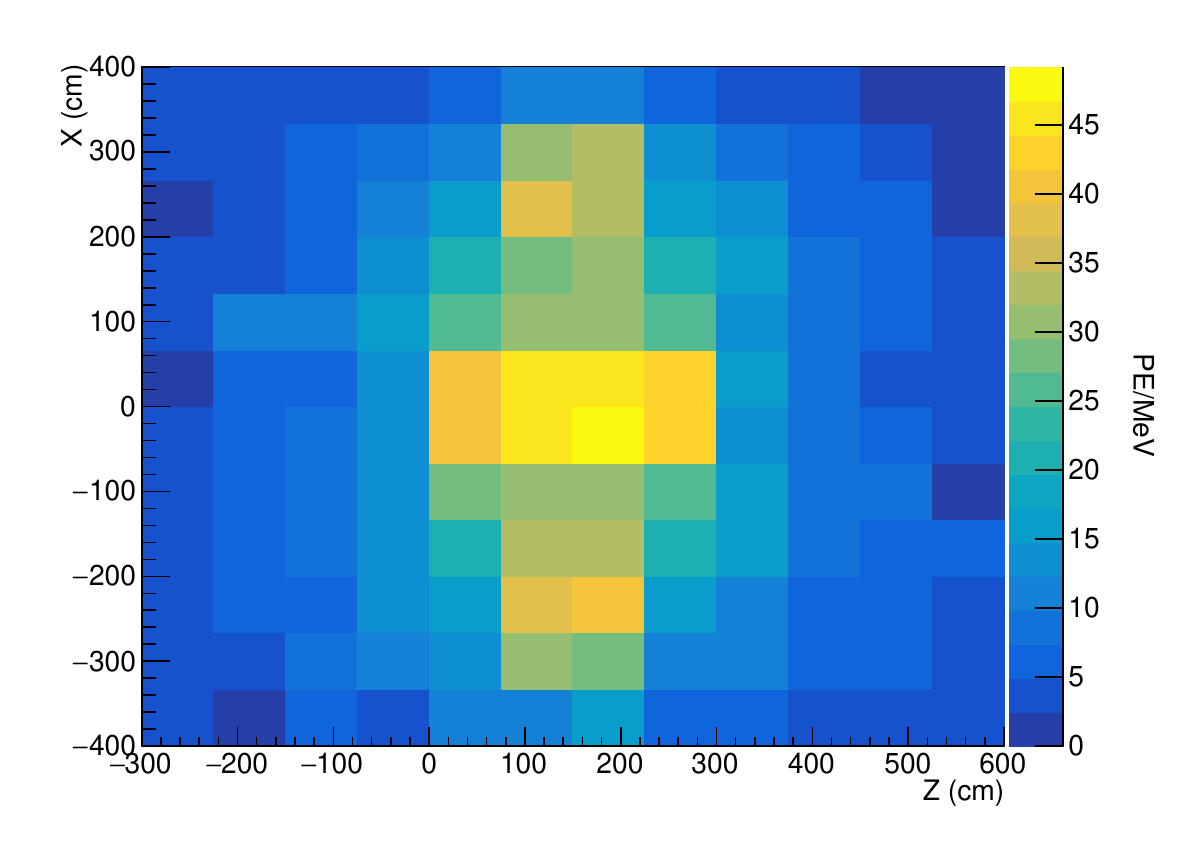}
\includegraphics[width=.3\textwidth]{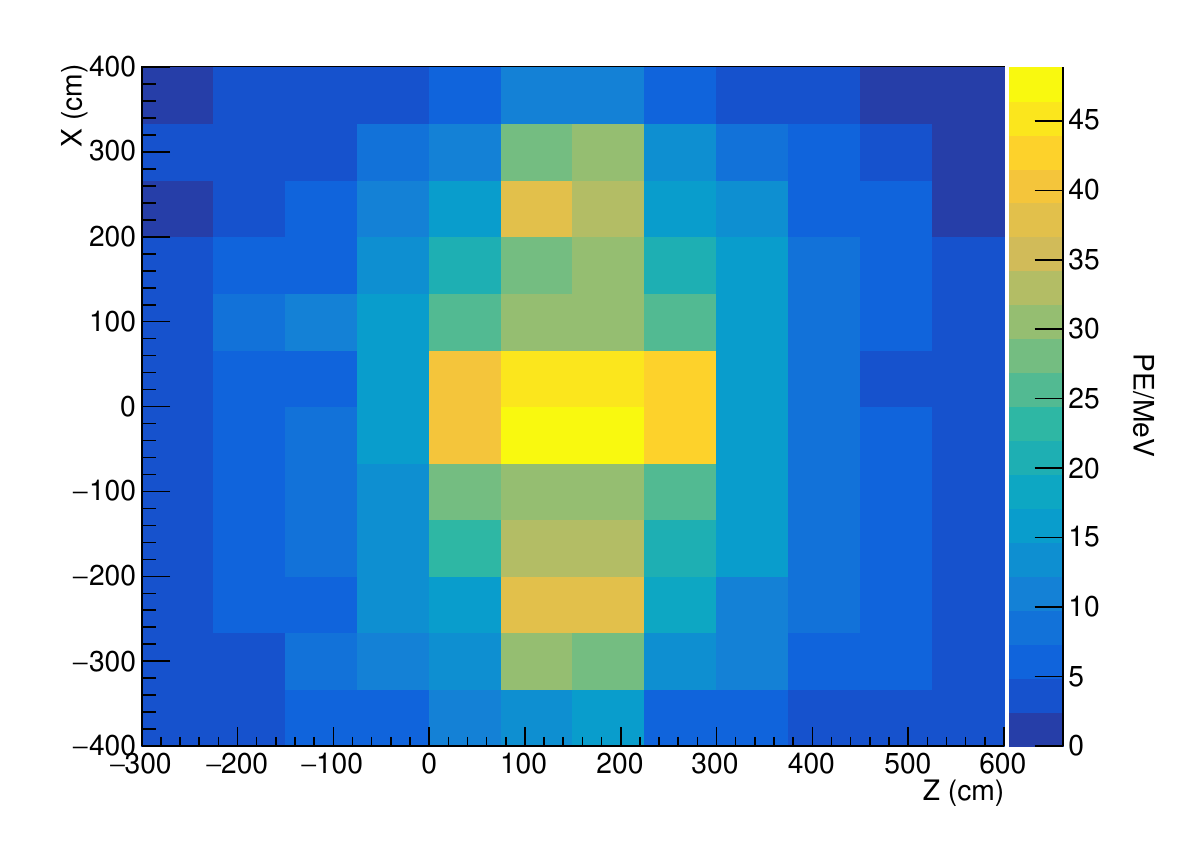}
\caption{Comparison of the visibility-based truth map (left), the unregularized reconstruction (center), and the smoothness regularized reconstruction (right) for a $12\times12\times12$ voxel map with the PMT contributions removed. The top and bottom rows show the average YX and ZX projections, respectively.\label{fig:ideal_compare}}
\end{figure}

Both reconstructed maps reproduce the main qualitative features expected from the detector geometry, as seen in figure~\ref{fig:ideal_compare}. Regions closer to the photon detectors show larger reconstructed light yield, while regions farther from the optically active surfaces show reduced response. This spatial behavior is consistent with the visibility-based expectation and demonstrates that the crossing muon method is sensitive to the large-scale optical structure of the detector. However, the unregularized reconstruction contains stronger voxel-to-voxel fluctuations and isolated regions of anomalously high or low light yield. These deviations are not present to the same extent in the truth map and arise because some spatial modes of the inversion are only weakly constrained by the available track sample. Smoothness regularization suppresses these localized fluctuations and produces a more continuous spatial distribution that more closely resembles the truth map.

A comparison of the detector volume mean light yield is given in table~\ref{tab:ideal_table}. With the PMTs included, the unregularized reconstruction has a mean of $18.5$~PE/MeV, compared with $19.5$~PE/MeV for the truth map. This corresponds to an absolute difference of $1.0$~PE/MeV, or approximately $5.1\%$ relative to the truth value. With the PMT contributions removed, the reconstructed mean is $13.2$~PE/MeV compared with $13.5$~PE/MeV, corresponding to a difference of $0.3$~PE/MeV, or approximately $2.2\%$ relative to the truth result. The regularized reconstruction gives mean values of $18.5$ and $13.2$~PE/MeV for the two photon detector selections respectively. The detector volume mean is therefore essentially unchanged by the smoothness constraint, showing that the absolute normalization of the system is unchanged by regularization. This behavior is expected because the smoothness term penalizes differences between neighboring voxels rather than the absolute magnitude of the solution, thereby redistributing light yield spatially while largely preserving the detector-average value.

\begin{table}[htbp]
\centering
\caption{Mean light yield for the visibility-based truth and crossing-muon reconstructed maps. The quoted uncertainties are estimated using the square root of the reported value. Results are shown for the full PDS configuration and with the PMT contributions removed.\label{tab:ideal_table}}
\smallskip
\begin{tabular}{l|cc}
\hline
Mapping & With PMTs [PE/MeV] & Without PMTs [PE/MeV]\\
\hline
 Visibility-based truth& $19.5 \pm 4.4$ & $13.5 \pm 3.7$\\
Unregularized reconstruction & $18.5 \pm 4.3$ & $13.2 \pm 3.6$\\
Smoothness regularized reconstruction & $18.5 \pm 4.3$ & $13.2 \pm 3.6$\\
\hline
\end{tabular}
\end{table}

\begin{figure}[htbp]
\centering
\includegraphics[width=.5\textwidth]{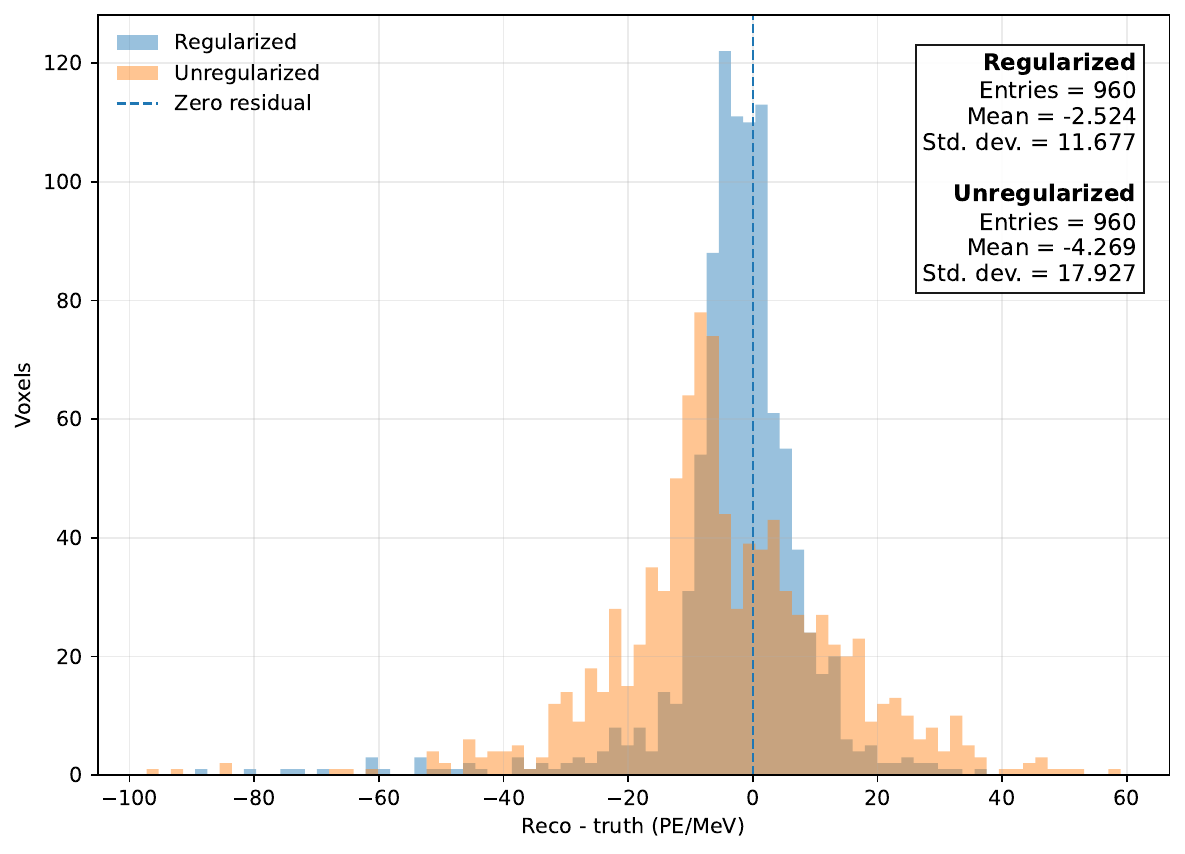}
\caption{Overlaid distributions of the voxel-by-voxel signed difference between the reconstructed and truth light yields for the unregularized and smoothness-regularized reconstructions of a $16\times16\times16$ solution. Both distributions are evaluated over the same 960 voxels in the active region, and the dashed vertical line indicates zero residual. The regularized reconstruction has a narrower residual distribution, with a standard deviation of 11.7~PE/MeV compared to 17.9~PE/MeV for the unregularized reconstruction, indicating a reduction in the spread of the voxel-level differences. The mean residuals are negative for both cases, with values of -2.5 PE/MeV for the regularized reconstruction and -4.3 PE/MeV for the unregularized reconstruction.\label{fig:reg_1d_compare}}
\end{figure}

The effect of regularization is more apparent in a voxel-by-voxel comparison than in the detector volume mean. Figure~\ref{fig:reg_1d_compare} shows the distribution of the signed residual,

\begin{equation}
    \Delta x = x_{\mathrm{reco}}-x_{\mathrm{truth}},
\end{equation}

where $x_{\mathrm{reco}}$ and $x_{\mathrm{truth}}$ are the reconstructed and truth light yields in the same voxel. Both distributions are evaluated using the same set of 960 voxels for the active volume of the detector. The unregularized residual distribution is broad and contains long tails produced by voxels with large reconstruction errors. Its mean residual is $-4.3$~PE/MeV, with a standard deviation of $17.9$~PE/MeV. After smoothness regularization, the distribution becomes substantially narrower, with a mean residual of $-2.5$~PE/MeV and a standard deviation of $11.7$~PE/MeV. Regularization therefore reduces the voxel-to-voxel spread of the residuals by approximately $35\%$, and the average negative offset from the truth map also decreases.

The root mean square error (RMSE) can be obtained from the mean and standard deviation of the residual distribution according to

\begin{equation}
    \mathrm{RMSE}
    =
    \sqrt{\left\langle \Delta x^{2}\right\rangle}
    =
    \sqrt{\mu_{\Delta x}^{2}+\sigma_{\Delta x}^{2}}.
\end{equation}

The displayed residual distributions correspond to an RMSE of approximately $18.4$~PE/MeV for the unregularized reconstruction and $11.9$~PE/MeV for the regularized reconstruction. Smoothness regularization therefore reduces the voxel-level RMSE by approximately $35\%$. This improvement is driven primarily by the suppression of large local deviations rather than by a change in the detector volume mean.

\begin{figure}[htbp]
\centering
\includegraphics[width=.4\textwidth]{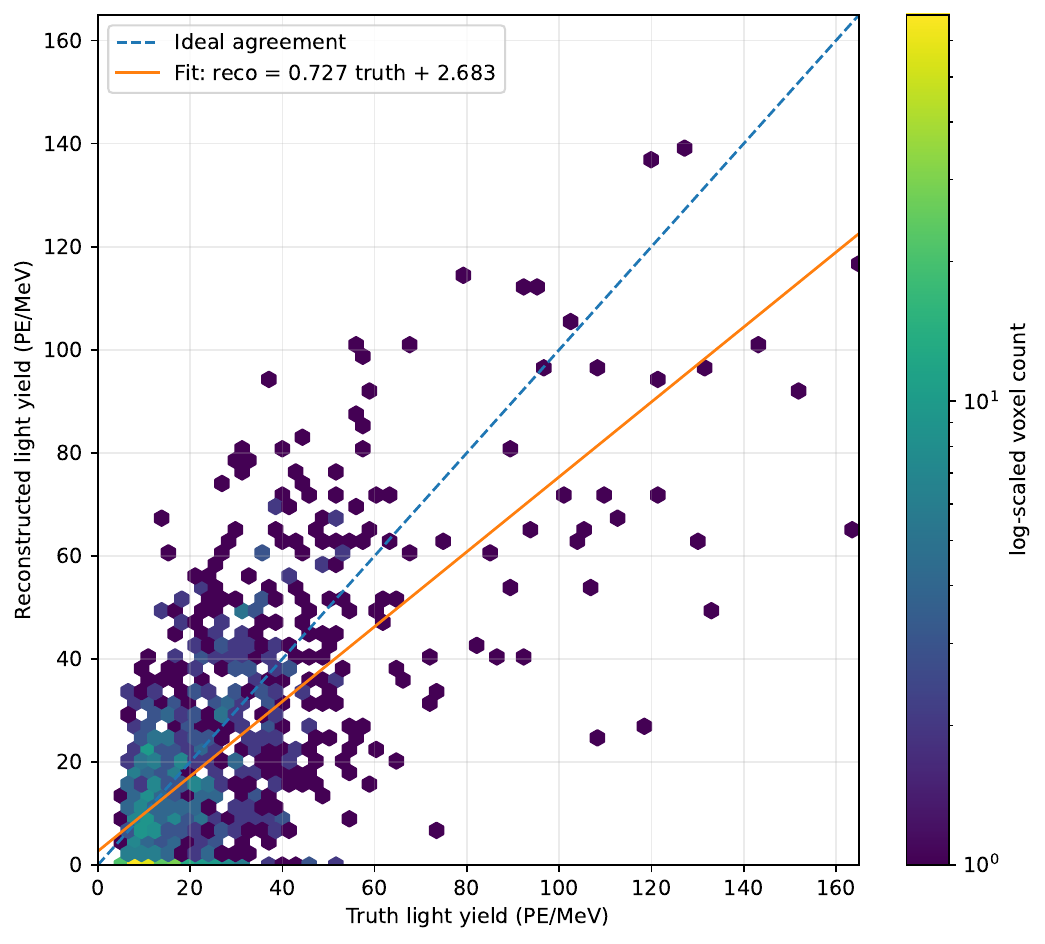}
\qquad
\includegraphics[width=.4\textwidth]{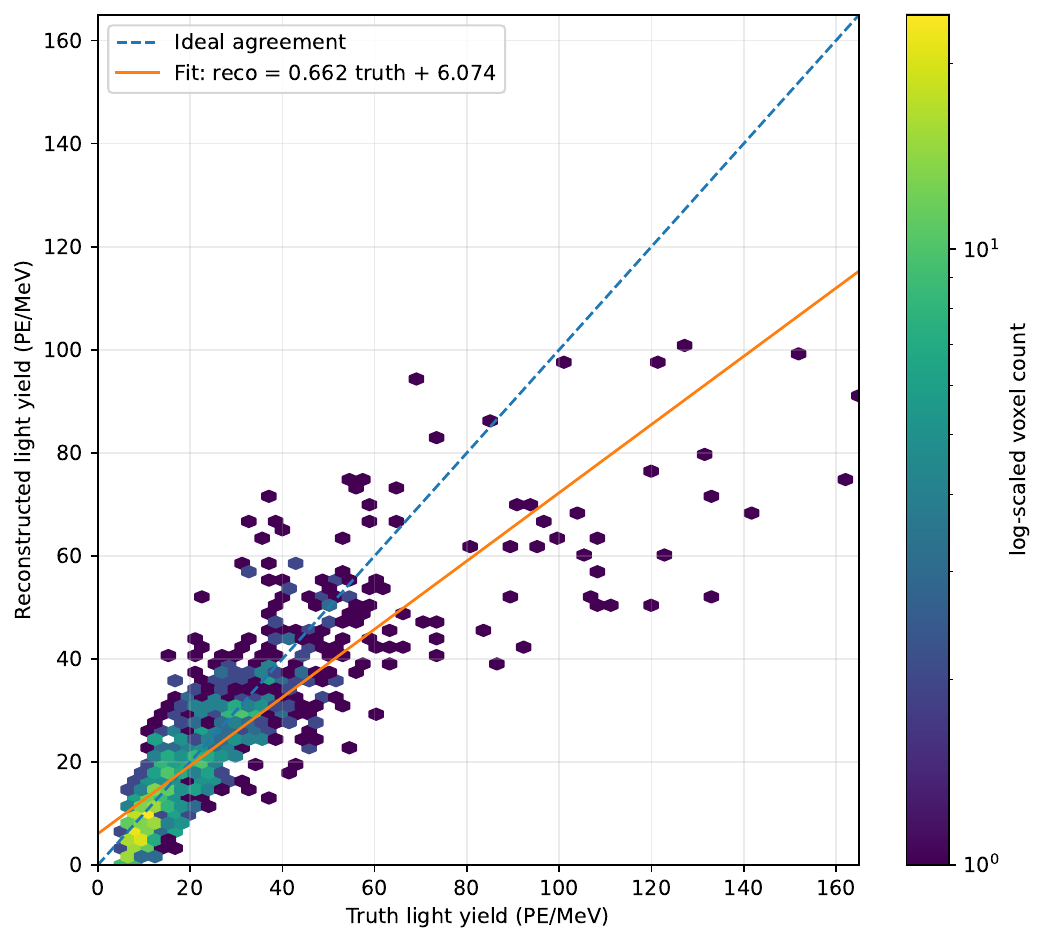}
\caption{Voxel-by-voxel comparison of reconstructed and truth light yield for the unregularized reconstruction (left) and the smoothness-regularized reconstruction (right). The dashed line represents ideal agreement, while the solid line is a linear fit to the voxel values. The color scale gives the voxel count on a logarithmic scale. Regularization reduces the scatter and extreme positive deviations.\label{fig:reg_2d_compare}}
\end{figure}

The same improvement is visible in the 2D voxel comparisons shown in figure~\ref{fig:reg_2d_compare}. The unregularized reconstruction exhibits substantial scatter relative to ideal agreement, including several voxels for which the reconstructed light yield is much larger than the corresponding truth value. These large deviations are consistent with the extended tails observed in the unregularized residual distribution.

The regularized reconstruction is more tightly concentrated around ideal agreement and contains fewer extreme voxel-level deviations, as shown by the lower standard deviation and RMSE. The remaining spread indicates that regularization does not completely remove the differences between the reconstructed and truth maps. Instead, it suppresses the largest localized fluctuations while retaining the principal spatial variation in the detector response.

\begin{figure}[htbp]
\centering
\includegraphics[width=.4\textwidth]{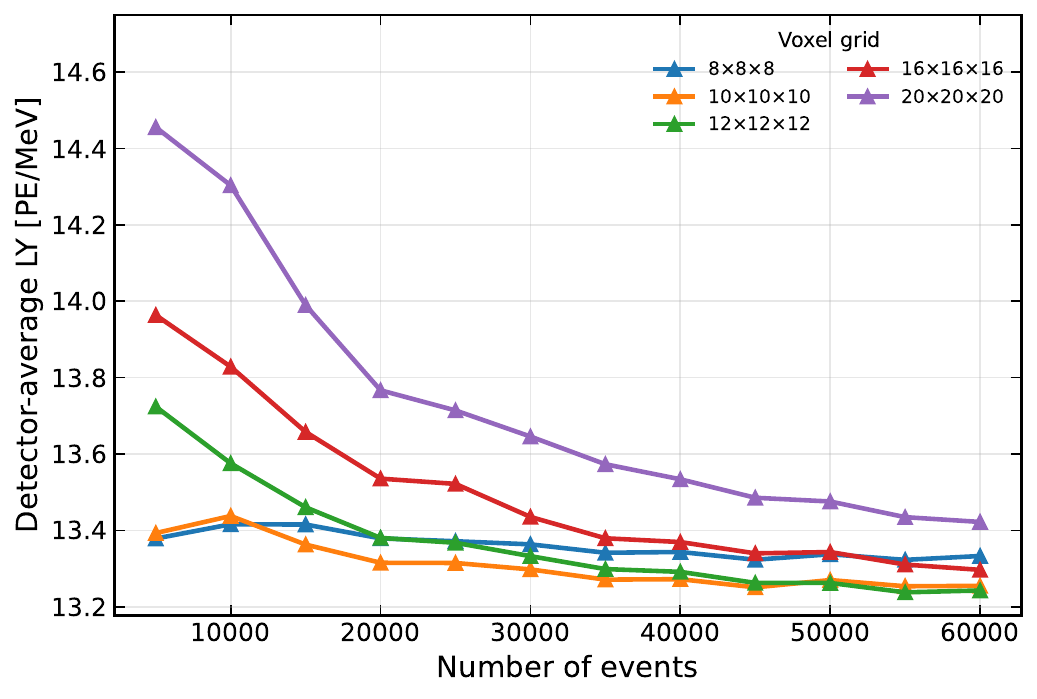}
\qquad
\includegraphics[width=.4\textwidth]{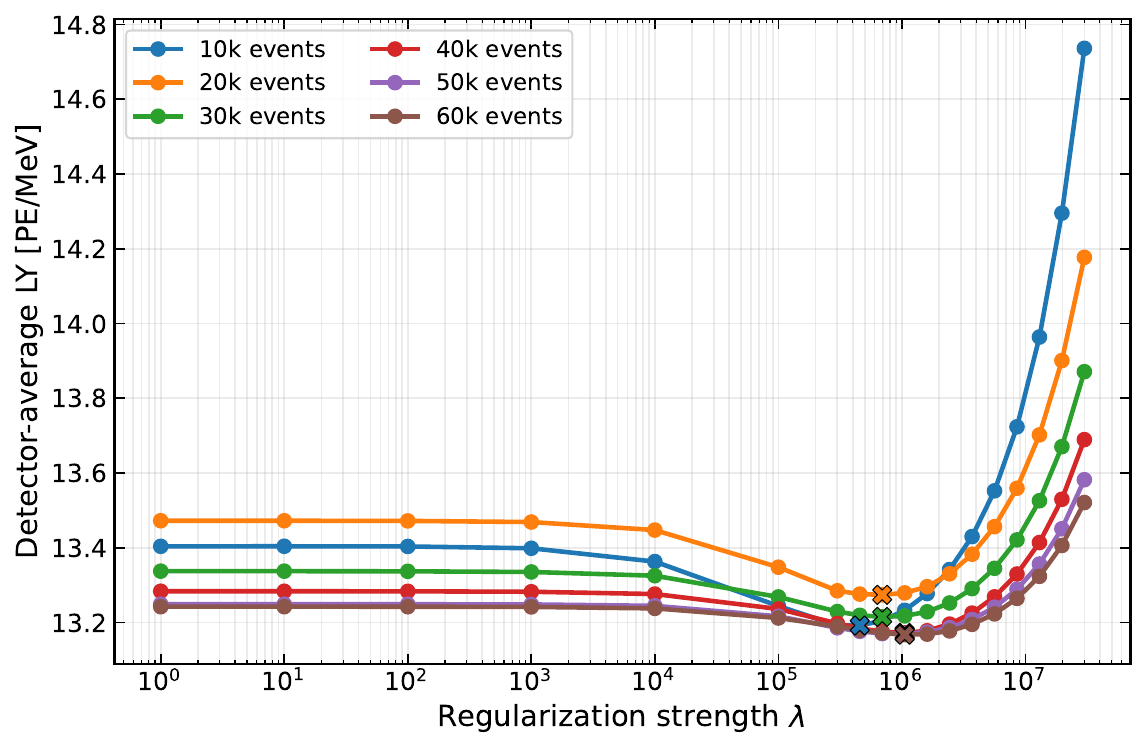}
\caption{Dependence of the detector volume average light yield on event statistics and regularization strength. The left panel shows the unregularized detector-average light yield as a function of event count for several voxelizations for the \texttt{lsq\_linear} solver. The right panel shows the detector-average light yield for the $12\times12\times12$ reconstruction as a function of $\lambda$ for several event samples, with the cross marks indicating the lowest sampled means. Increasing the event count drives the unregularized solutions toward a stable plateau, while the selected regularization strengths leave the detector-average light yield approximately unchanged.\label{fig:avg_vs_events}}
\end{figure}

The reconstruction also depends on the number of crossing muon events used to construct the linear system. The left panel of figure~\ref{fig:avg_vs_events} shows the detector volume average light yield for the unregularized solutions as a function of event count. At low statistics, the reconstructed average is larger and depends more strongly on the voxel resolution. This effect is most pronounced for the finer voxelizations, for which the available track information is distributed among a larger number of fitted parameters. As the event count increases, the reconstructed averages decrease and approach a common high-statistics plateau.

For the geometry and event selection studied here, the largest change in the detector-average light yield occurs at low statistics, with the variation becoming much smaller above around 20,000 events. To quantify this behavior, the residual event count dependence was evaluated using the fractional range of the detector-average light yield over the high-statistics portion of the scan. Over the range $N\geq20{,}000$, the detector-average light yield varies by 0.42--2.54\% across the voxelizations studied. Restricting the comparison to $N\geq40{,}000$, this range decreases to 0.15--0.83\%. Equivalently, using the 60,000 event value as a reference, all voxelizations are within 2\% of the final detector-average light yield by 30,000 events, within 1\% by 40,000 events, and within 0.5\% by 45,000 events. These should not be interpreted as universal minimum event requirements, since the convergence rate could depend on the detector geometry, voxelization, and event selection. A similar event count scan should therefore be performed for each detector configuration to verify that the available statistics are sufficient to produce a stable reconstruction at the required precision.

The right panel of figure~\ref{fig:avg_vs_events} shows the detector-average light yield for the $12\times12\times12$ reconstruction as a function of the regularization strength. Over the range of selected values of $\lambda$, the detector-average light yield changes only modestly, consistent with the comparison in table~\ref{tab:ideal_table}. At sufficiently large $\lambda$, however, the average begins to increase, indicating that excessive regularization can alter both the spatial structure and overall normalization of the solution. The selected values therefore lie in an intermediate region where local fluctuations are reduced without producing a large change in the detector-average response.

Several effects may contribute to the remaining differences between the truth and reconstructed maps. The truth map is evaluated using point-like scintillation vertices sampled directly in the visibility model, whereas the reconstructed map is inferred from integrated light along extended muon tracks. The reconstruction is therefore likely influenced by the available track coverage and finite event statistics.

\subsection{Effect of Smoothness Regularization}
\label{sec:reg_results}

The comparison with the truth map demonstrates that the unregularized reconstruction recovers the dominant optical response but can contain substantial voxel scale fluctuations. A related feature is the presence of voxels fitted at or near the nonnegative lower bound. In practice, the \texttt{nnls} solver assigns exact zeros, while the \texttt{lsq\_linear} solver returns small floating point values instead of exact zeros. For the purpose of comparing the two solvers then, voxels with reconstructed light yield below 0.1~PE/MeV are classified as zero. Since the true optical response is not expected to be zero at points near the photon detectors, these zero-valued voxels are interpreted as a reconstruction effect rather than as regions with no physical light collection.

Each crossing muon constrains only the voxels lying along its trajectory. If the detector is divided into $N$ bins along each dimension, the number of fitted parameters grows as $N^3$, while a single track crosses only approximately $N$ voxels. At fixed event count, finer voxelizations therefore provide fewer independent track constraints per fitted voxel. The nonnegative solution can consequently place weakly constrained voxels at the lower bound while concentrating light yield into other voxels along the same trajectories.

\begin{figure}[htbp]
\centering
\includegraphics[width=.4\textwidth]{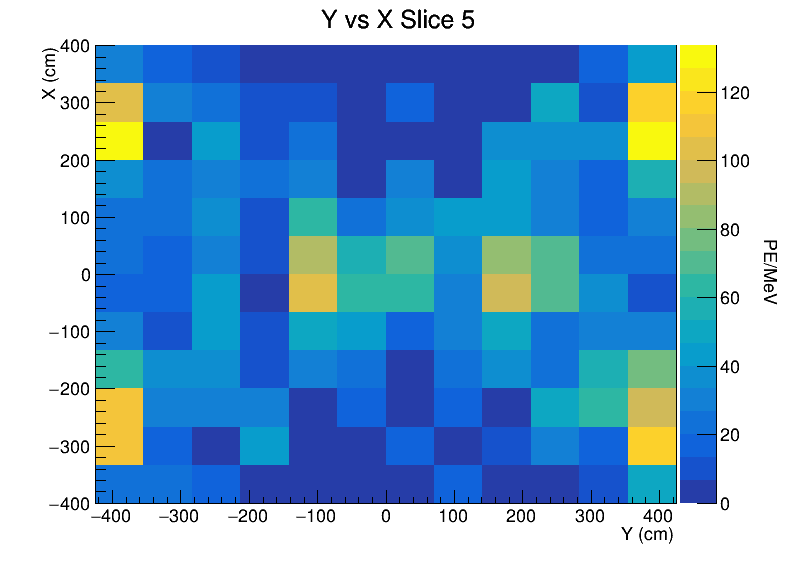}
\qquad
\includegraphics[width=.4\textwidth]{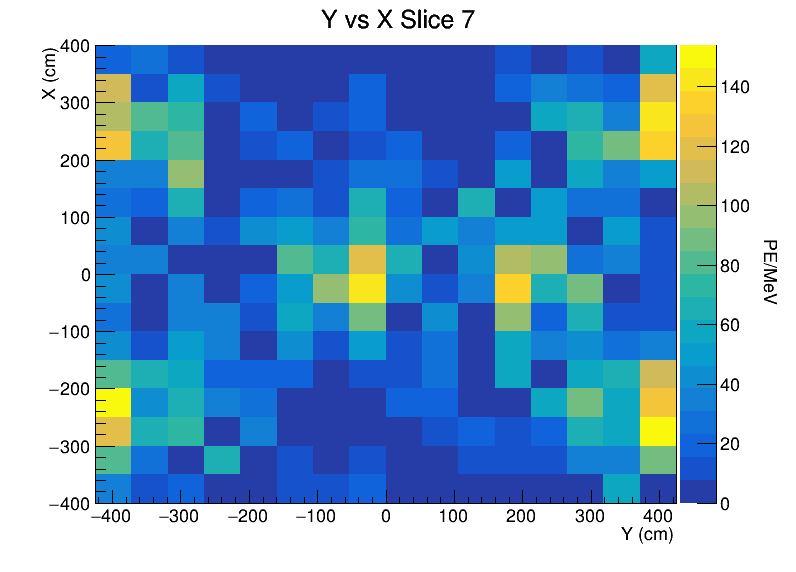}
\includegraphics[width=.4\textwidth]{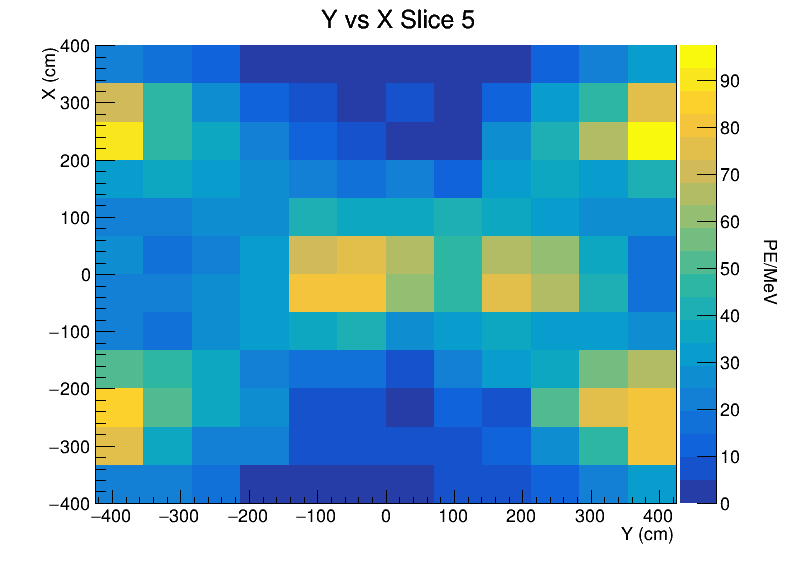}
\qquad
\includegraphics[width=.4\textwidth]{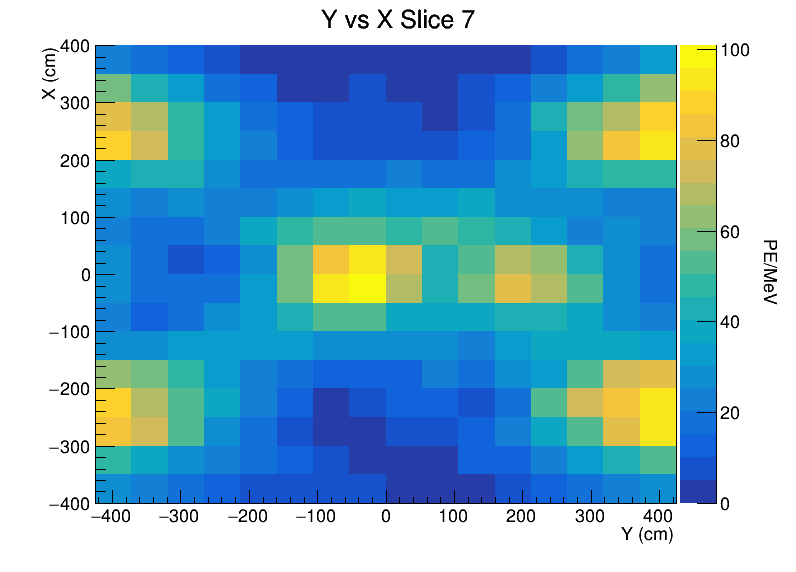}
\caption{Representative middle YX slices for the $12\times12\times12$ reconstruction (left) and the $16\times16\times16$ reconstruction (right), both without PMT contributions. The top row shows the unregularized reconstructions, while the bottom row shows the corresponding smoothness-regularized reconstructions. In the unregularized case, increasing the voxel resolution increases the number of fitted parameters and produces a larger fraction of zero-valued voxels for the same track sample. This resolution-dependent sparsity is not visible in the regularized reconstructions, where the smoothness penalty suppresses large voxel-to-voxel variations.\label{fig:sparse_compare}}
\end{figure}

This behavior is visible in the unregularized slices in the top row of figure~\ref{fig:sparse_compare}. The $16\times16\times16$ reconstruction contains more zero-valued voxels and stronger local fluctuations than the $12\times12\times12$ reconstruction produced from the same event sample. The smoothness-regularized slices in the bottom row are more continuous and do not exhibit the same increase in sparsity with resolution. The regularization penalty couples neighboring voxel values, reducing the freedom to place individual weakly constrained voxels at zero while assigning large values to nearby voxels.

\begin{figure}[htbp]
\centering
\includegraphics[width=.4\textwidth]{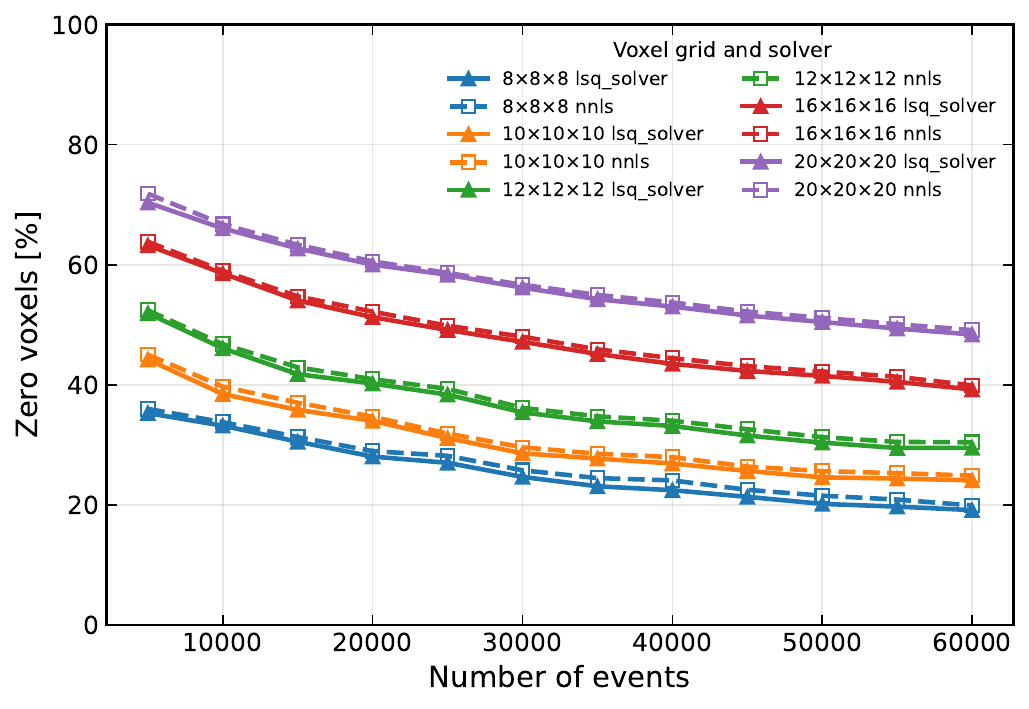}
\qquad
\includegraphics[width=.4\textwidth]{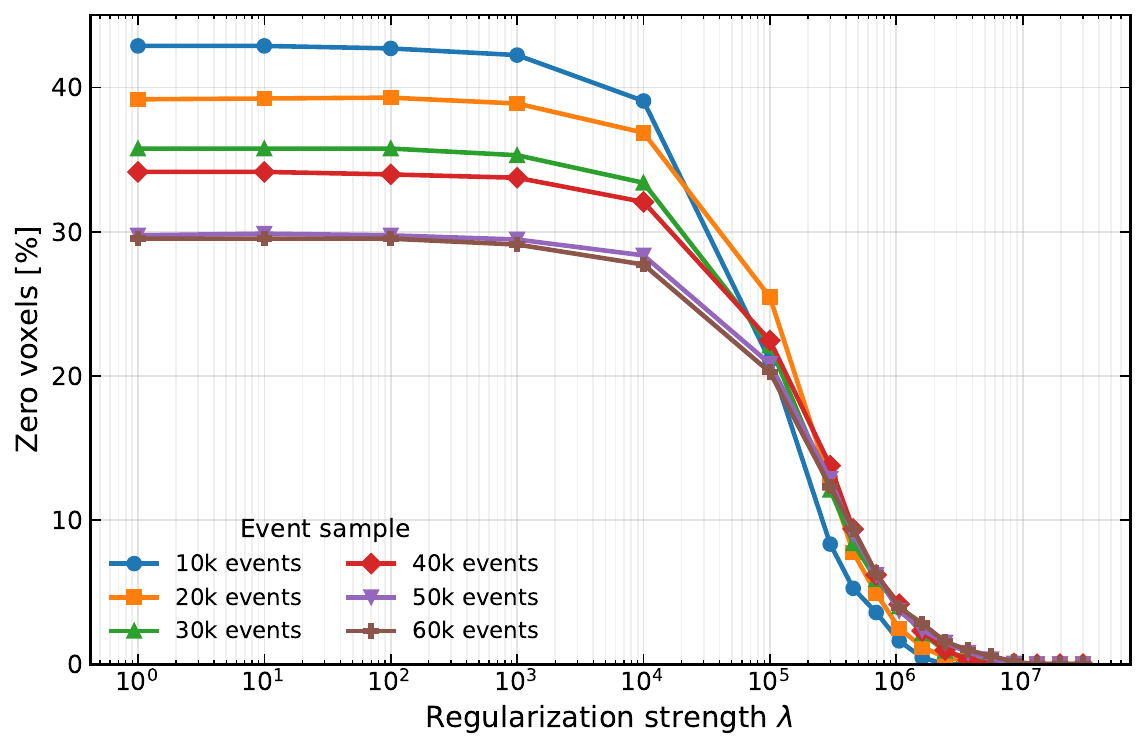}
\caption{Fraction of voxels classified as zero, defined as voxels with a
reconstructed light yield below $0.1~\mathrm{PE/MeV}$. The left panel shows the zero voxel fraction for unregularized reconstructions as a function of event count for several voxel resolutions. The right panel shows the zero voxel fraction for the $12\times12\times12$ reconstruction as a function of the smoothness regularization strength $\lambda$.\label{fig:sparse_vs_events}}
\end{figure}

The dependence of this sparsity on event count and regularization strength is shown in figure~\ref{fig:sparse_vs_events}. In the unregularized solutions, the zero voxel fraction decreases as the event count is increased because additional tracks provide new constraints on previously weakly sampled regions. Nevertheless, a substantial zero voxel fraction remains at high statistics, particularly for the finer voxelizations. Increasing the number of events therefore improves the constraint coverage but does not fully remove the sparsity associated with the nonnegative inverse problem. The sparsity effect is also consistent between the \texttt{nnls} and \texttt{lsq\_linear} solvers. A more detailed comparison between the outputs of the \texttt{nnls} and \texttt{lsq\_linear} solvers is presented in section~\ref{sec:solver_results}.

For the $12\times12\times12$ reconstruction, the zero voxel fraction also decreases rapidly as the regularization strength is increased. At small $\lambda$, the result approaches the unregularized solution and retains a large fraction of zero-valued voxels. As the smoothness penalty becomes stronger, neighboring voxel values become coupled and the zero voxel fraction falls toward zero. This behavior, together with the improved voxel-wise agreement shown in figures~\ref{fig:reg_1d_compare} and~\ref{fig:reg_2d_compare}, motivates the use of smoothness regularization in the final reconstructed maps.

Although regularization reduces both sparsity and local reconstruction error, its strength must be chosen carefully. If $\lambda$ is too small, the solution remains close to the unregularized result and retains weakly constrained voxel-scale fluctuations. If $\lambda$ is too large, genuine spatial variations in the detector response are suppressed and the reconstructed map becomes oversmoothed. As shown in the right panel of figure~\ref{fig:avg_vs_events}, sufficiently large values can also alter the detector-average light yield.

To determine an appropriate value, logarithmically spaced scans in $\lambda$ were performed independently for each voxel resolution and event count. For every point in a scan, the reconstructed map was compared with the visibility-based truth map rebinned to the same voxelization, and the voxel-wise RMSE was calculated. A smooth local fit was then applied near the lowest sampled RMSE, and the minimum of this fit was used as the optimal $\lambda$. Using the fitted minimum avoids restricting the result to the discrete values included in the scan.

\begin{figure}[htbp]
\centering
\includegraphics[width=.41\textwidth]{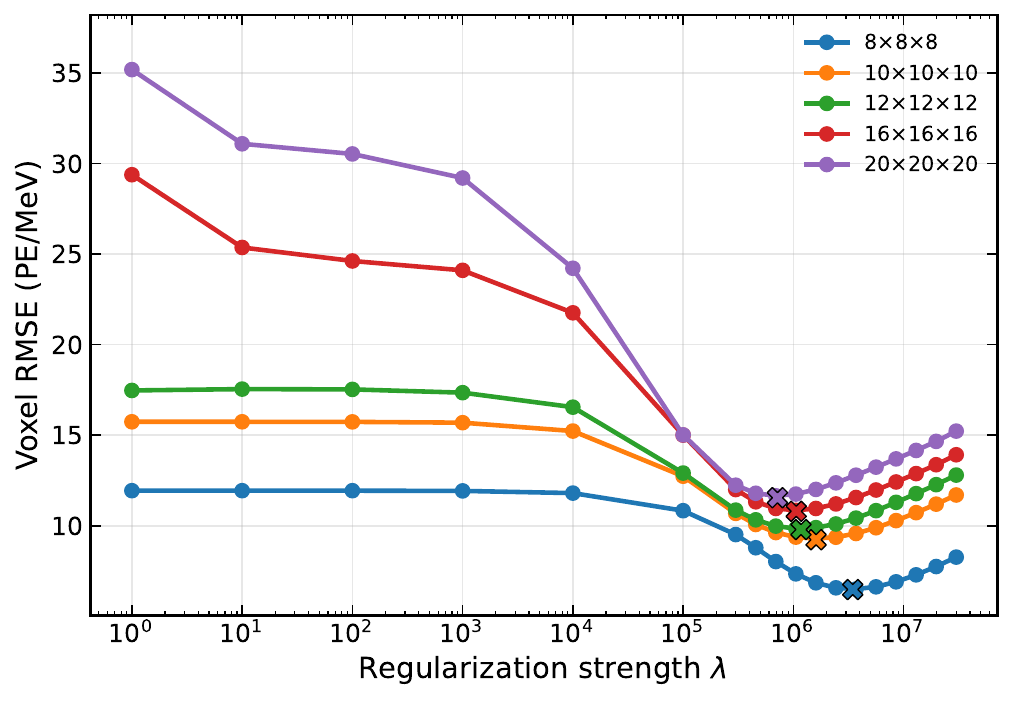}
\qquad
\includegraphics[width=.46\textwidth]{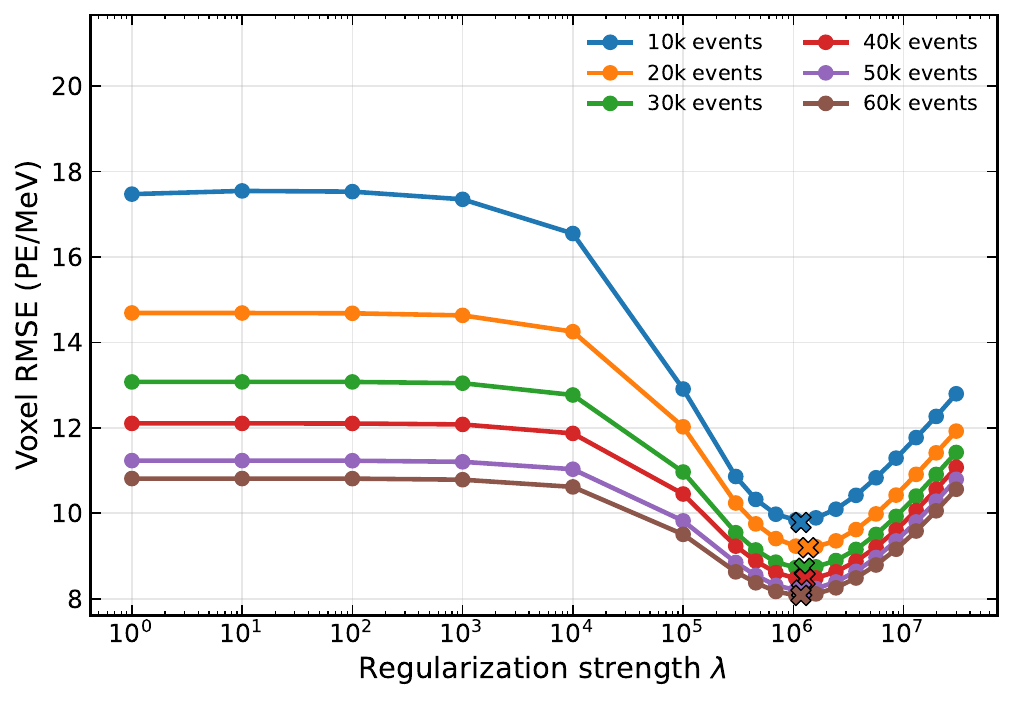}
\caption{Voxel-wise RMSE between the reconstructed and truth maps as a function of the smoothness regularization strength $\lambda$. The left panel compares the studied voxel resolutions solved with 10,000 events, while the right panel compares different event counts for a $12\times12\times12$ resolution. The markers indicate the minima obtained by interpolating the sampled RMSE curves, but the interpolation curves are not shown. At low $\lambda$, the RMSE is dominated by unstable voxel scale variations, while at high $\lambda$, the RMSE increases because the reconstructed spatial response becomes oversmoothed.\label{fig:reg_lambda_example}}
\end{figure}

Examples of these scans are shown in figure~\ref{fig:reg_lambda_example}. The RMSE curves exhibit a broad minimum separating the under-regularized and over-regularized regimes. At low $\lambda$, increasing the regularization strength reduces the influence of poorly constrained voxel scale fluctuations. Beyond the minimum, the RMSE increases because the solution is smoothed more strongly than is supported by the spatial structure of the truth map. The marked points indicate the fitted minima used in the subsequent comparisons.

\begin{figure}[htbp]
\centering
\includegraphics[width=.46\textwidth]{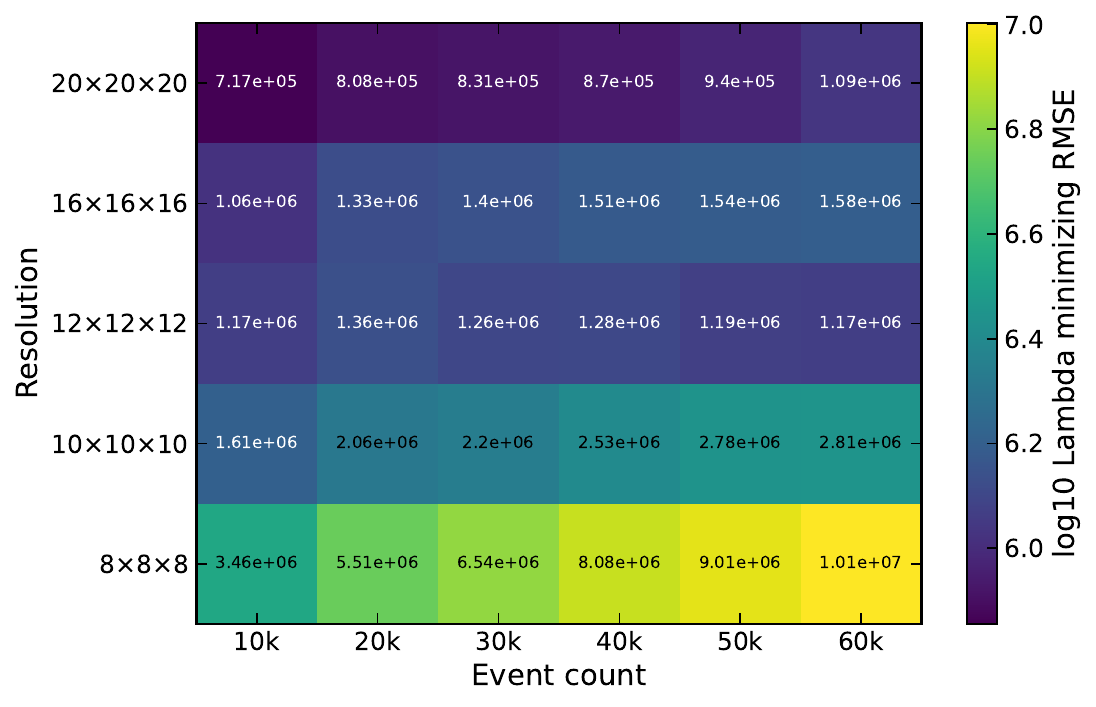}
\qquad
\includegraphics[width=.4\textwidth]{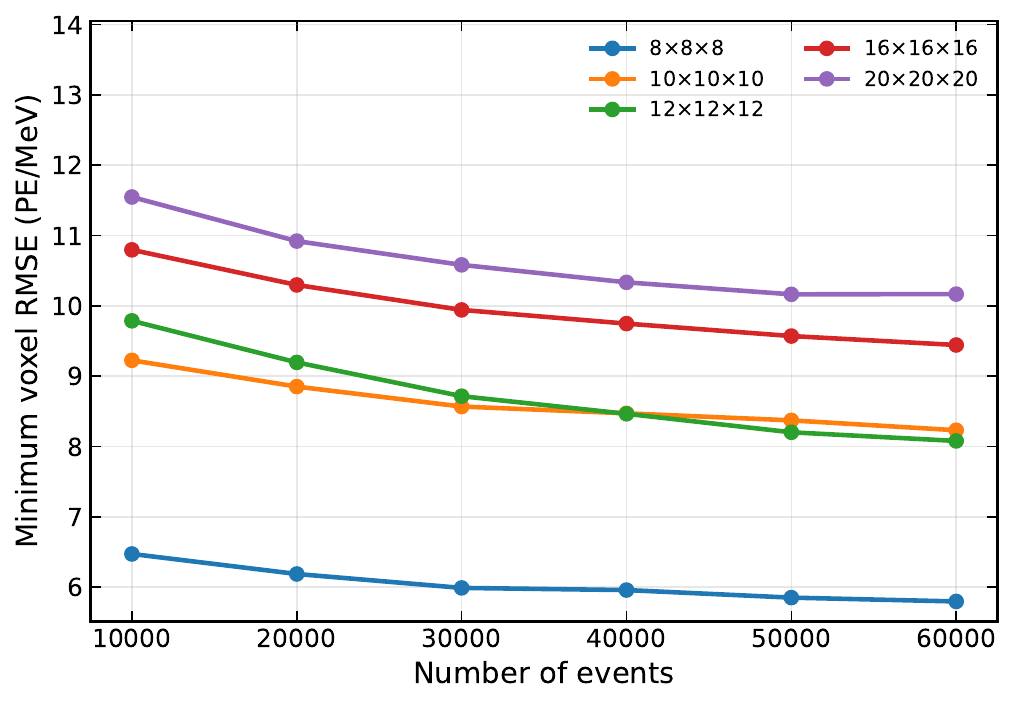}
\caption{The fitted regularization strength, $\lambda$, that minimizes the
voxel-wise RMSE between the reconstructed and truth maps for each voxelization
and event count (left), and the corresponding minimum RMSE as a function of
event count for each voxelization (right).\label{fig:reg_lambda_select}}
\end{figure}

The RMSE-minimizing values obtained for all resolutions and event counts are summarized in figure~\ref{fig:reg_lambda_select}. For a fixed resolution, the preferred $\lambda$ increases with event count for the coarsest and finest grids and is flat within its uncertainty for the 12×12×12 grid.

The minimum RMSE generally decreases as more events are included, with the largest improvement occurring between 10,000 and 20,000 events. Beyond around 20,000 events, the improvement becomes more gradual, consistent with the convergence of the detector-average light yield shown in figure~\ref{fig:avg_vs_events}. Additional events continue to improve the voxel-level reconstruction, but with diminishing gains once the principal spatial modes are sufficiently constrained. The decrease is not strictly monotonic for every
voxelization. A larger
sample may provide improved overall coverage while changing the relative
constraint on individual voxels, producing small fluctuations in the
voxel-wise agreement with the fixed truth map.

Finer voxelizations have a larger minimum RMSE than coarser voxelizations at every event count except for $10\times10\times10$ and $12\times12\times12$ beyond 40,000 events, where they cross. Increasing the resolution introduces more independent voxel values while reducing the average track coverage within each voxel, making the inversion more weakly constrained. Smoothness regularization reduces the resulting fluctuations but cannot fully compensate for the loss of information per voxel.

\subsection{Sensitivity to Geometry}
\label{sec:geo_sens}

The PDS configuration used in this study is shown in section~\ref{sec:pdvd}, and includes both X-ARAPUCA photon detectors and PMTs. Since the reconstructed light yield map is intended to describe the detector response for a specific optical configuration, a useful validation of the method is to test whether it is sensitive to controlled changes in the photon detector geometry. In particular, removing a subset of photon detectors should produce a corresponding change in the reconstructed spatial response, especially near the region where those detectors were located. An example of this can be seen with the visibility-based truth maps shown in figure~\ref{fig:ideal}, where the top row of maps have PMT contributions included, while the bottom row maps have the PMTs excluded.

\begin{figure}[htbp]
\centering
\includegraphics[width=.4\textwidth]{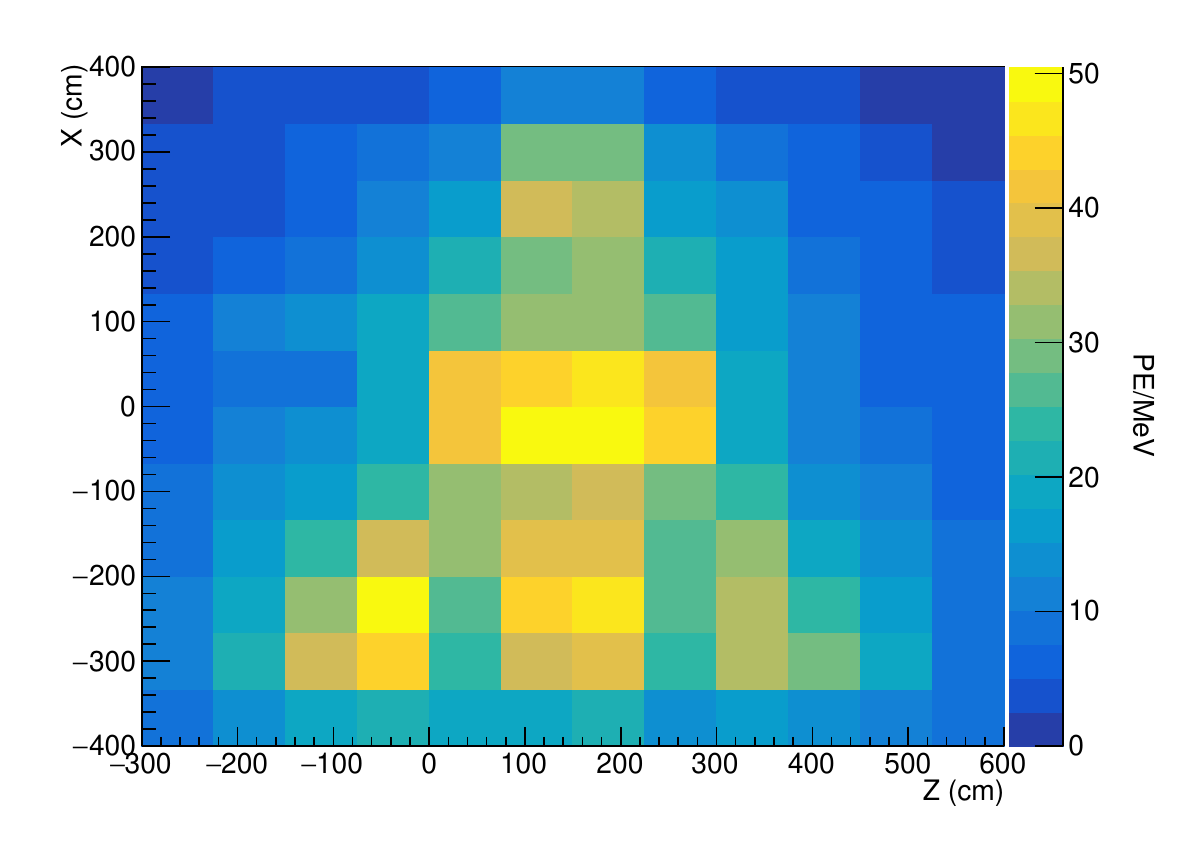}
\qquad
\includegraphics[width=.4\textwidth]{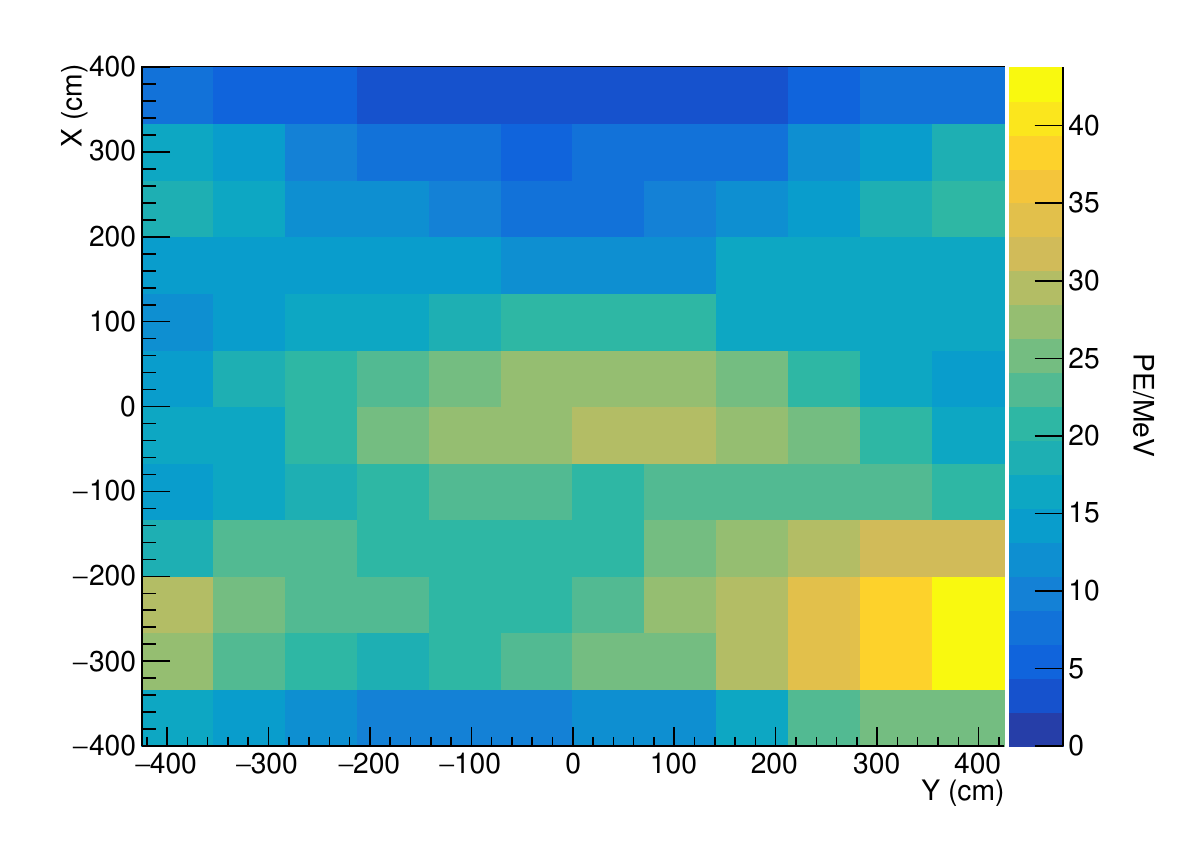}
\caption{Smoothness-regularized average light-yield projections for the $12\times12\times12$ reconstruction with PMT contributions included. The left and right panels show the ZX and YX projections respectively. The reconstruction uses the same event sample, voxelization, and regularization procedure as the PMT-excluded reconstruction shown in figure~\ref{fig:ideal_compare}.\label{fig:pmt_compare}}
\end{figure}

To test this, two otherwise identical reconstructions were performed: one including the full PDS response, and one excluding the PMT contribution. In both cases, the same regularized reconstruction procedure was applied, so differences between the maps can be attributed to the change in photon detector configuration rather than to changes in the voxelization or track sample. The reconstruction with PMTs included is shown in figure~\ref{fig:pmt_compare}
and can be compared with the PMT-excluded reconstruction in
figure~\ref{fig:ideal_compare}.

When the PMTs are included, the average projection contains an enhanced light yield contribution near the lower portion of the detector, corresponding to the additional photon collection from the PMT array. When the PMT contribution is removed, this enhancement is reduced, and the remaining spatial structure is dominated by the X-ARAPUCA response, demonstrating the expected reconstruction behavior. The ability to remove their contribution and recover a different spatial response indicates that the same reconstruction framework can be applied to different detector layouts or channel selections.

\subsection{Spatial Resolution}
\label{sec:resolution}

A solved light yield map is generated for a fixed voxelization of the detector volume, where the voxelization is defined by the map boundaries and by the number of divisions along each spatial axis. Increasing the number of divisions improves the nominal spatial resolution of the map, but it also increases the number of fitted light yield parameters. This makes the reconstruction more sensitive to the available track statistics and to the uniformity of the muon coverage.
 
For this study, only maps with the same number of divisions along each dimension were considered. In principle, different numbers of divisions could be used along the three axes if finer resolution were desired in one direction, and this would only change the voxelization and distance calculation step described in section~\ref{sec:voxel}. Because the map boundaries used in this study are not equal along all three dimensions (for consistency with truth maps used in section~\ref{sec:ideal}), the resulting voxels are rectangular cuboids rather than cubes.

\begin{figure}[htbp]
\centering
\includegraphics[width=0.32\textwidth]{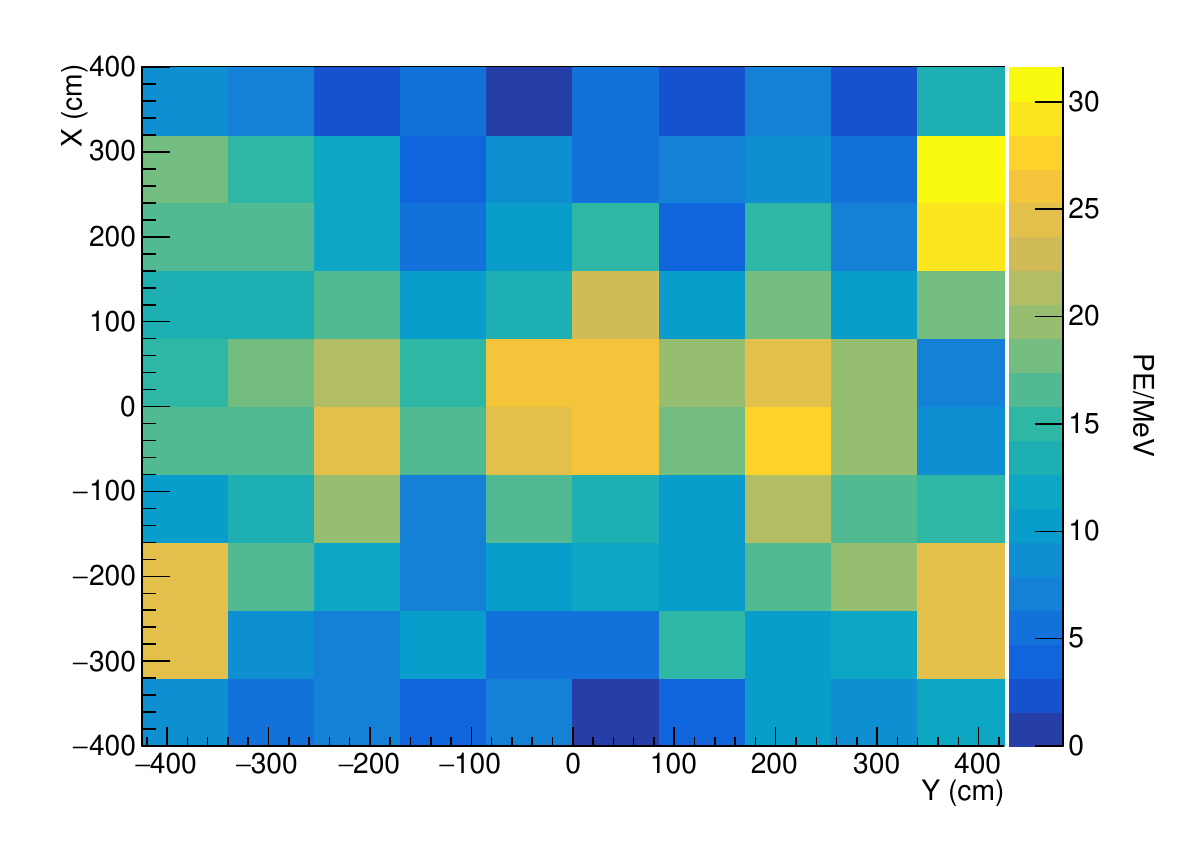}
\includegraphics[width=0.32\textwidth]{figs/avg_h_yx_12x12_nnls_nopmts.pdf}
\includegraphics[width=0.32\textwidth]{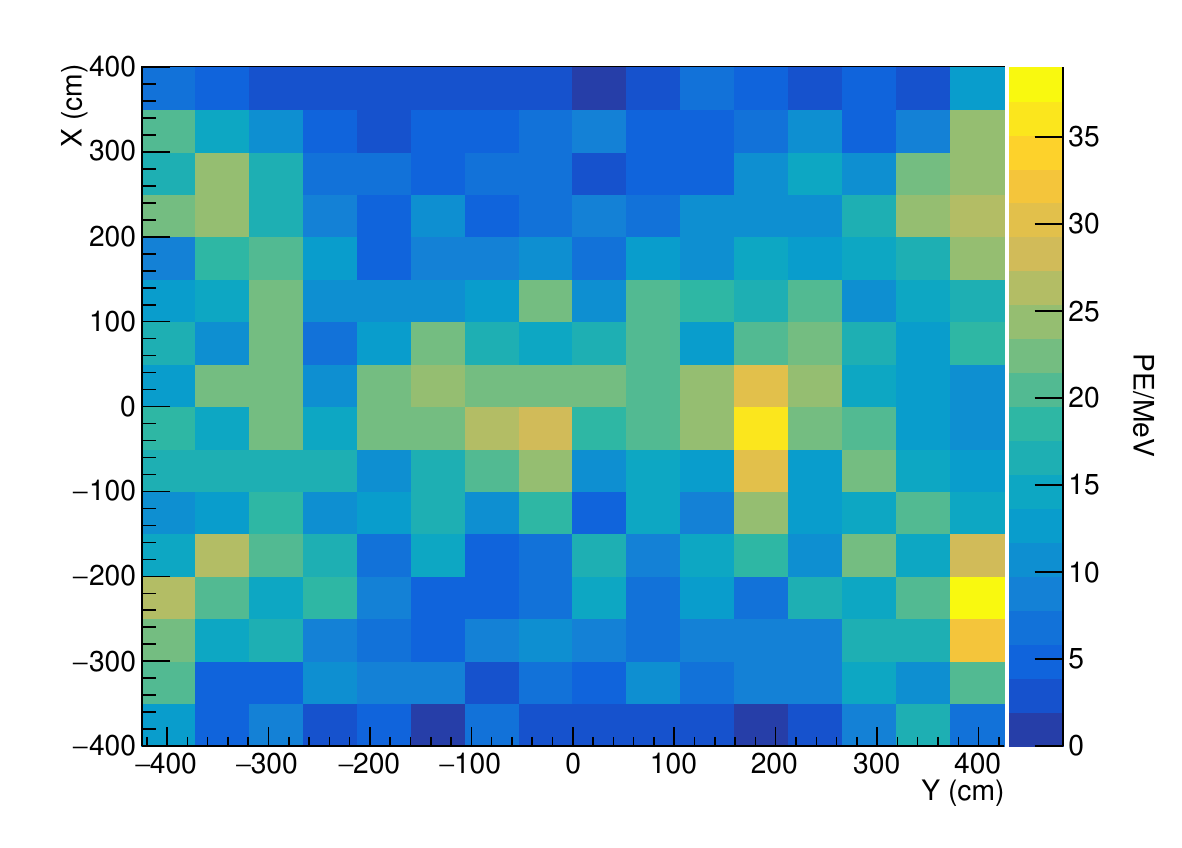}
\includegraphics[width=0.32\textwidth]{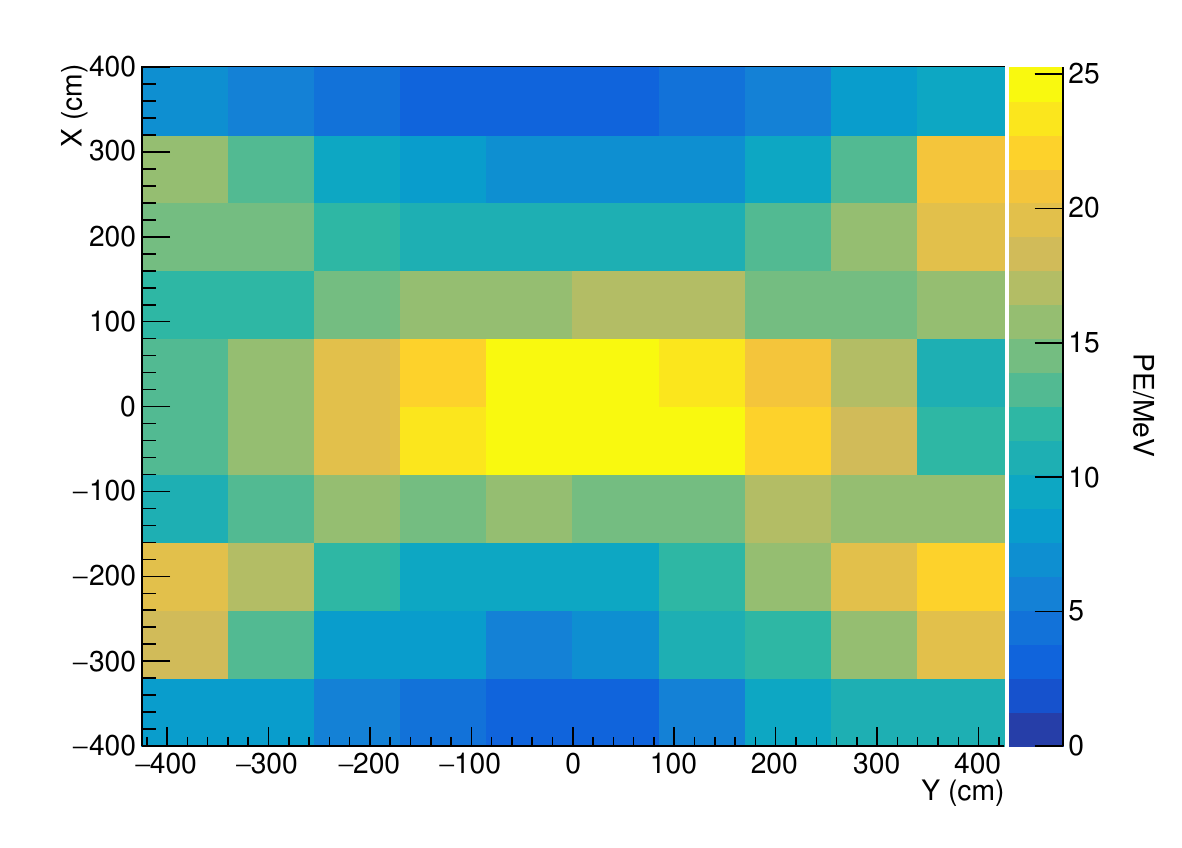}
\includegraphics[width=0.32\textwidth]{figs/avg_h_yx_12x12_nnls_nopmts_reg.pdf}
\includegraphics[width=0.32\textwidth]{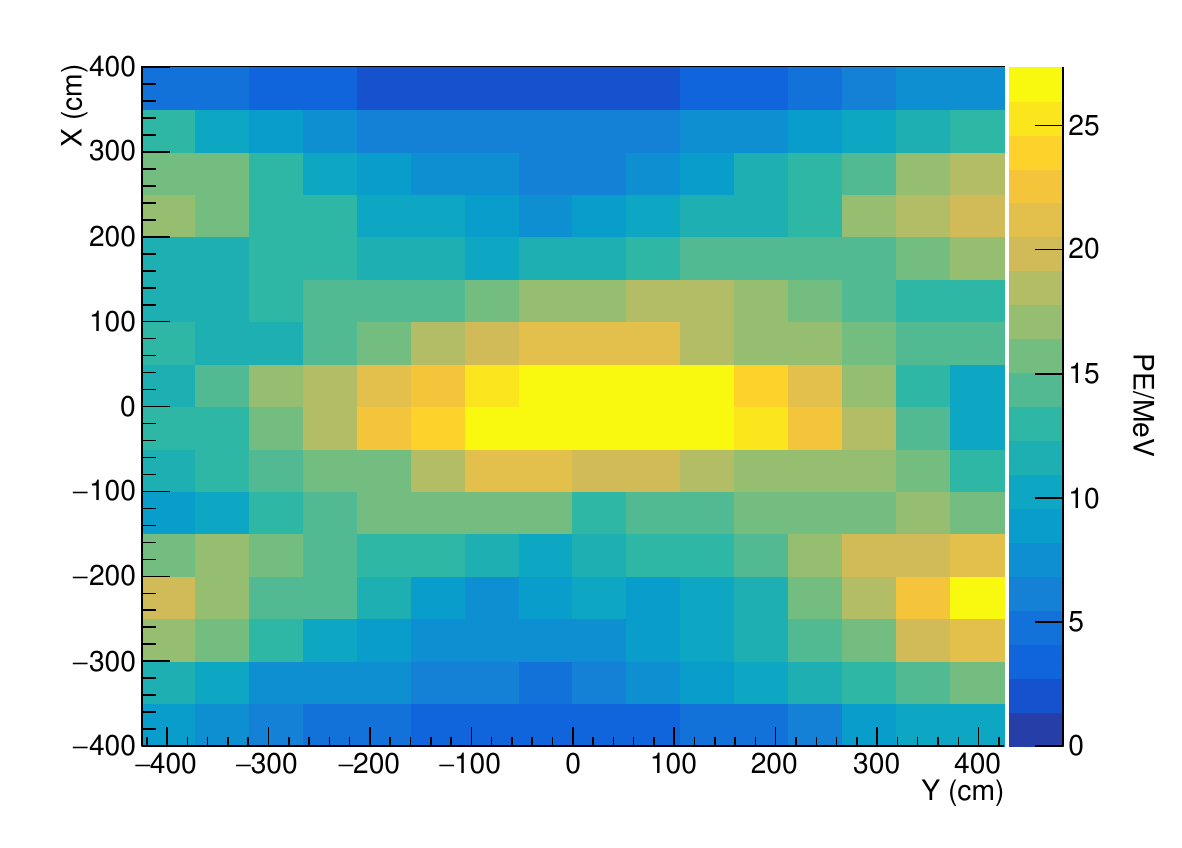}
\caption{Comparison of various YX average projections, without (top) and with regularization (bottom), and with increasing resolution; $10\times10\times10$ (left), $12\times12\times12$ (center), $16\times16\times16$ (right).\label{fig:res_compare}}
\end{figure}

Figure~\ref{fig:res_compare} shows the YX average projections for three voxelizations using the configuration without PMT contributions. The large-scale spatial structure of the reconstructed light yield map is stable as the resolution is increased, indicating that the dominant detector response features are not artifacts of a particular voxel size. At lower resolution, the unregularized reconstructed map is smoother because each voxel averages over a larger detector volume and is crossed by a larger fraction of the muon sample. At higher resolution, finer spatial variations become visible, but individual voxels are constrained by fewer tracks. This increases the sensitivity to local coverage variations and can enhance voxel-to-voxel fluctuations, especially in regions where the track sample is less uniform. The regularized maps reduce these local fluctuations by penalizing large differences between neighboring voxels, allowing the large-scale spatial response to remain stable as the voxel resolution is increased.

Based on the voxel map bounds used for this study, table~\ref{tab:res_configs} shows the voxel sizes for the maps of increasing resolution shown in figure~\ref{fig:res_compare}.

\begin{table}[htbp]
\centering
\caption{Voxel dimensions for the increasing resolution maps in figure~\ref{fig:res_compare}.\label{tab:res_configs}}
\smallskip
\begin{tabular}{l|ccc}
\hline
Configuration&X (cm)&Y (cm)&Z (cm)\\
\hline
$10\times10\times10$ & 80 & 85 & 90\\
$12\times12\times12$ & 66.7 & 70.8 & 75\\
$16\times16\times16$ & 50 & 53.1 & 56.3\\
\hline
\end{tabular}
\end{table}

\subsection{Comparison of Linear Solvers}
\label{sec:solver_results}

The reconstruction procedure depends on solving a constrained linear inverse problem, so it is important to verify that the recovered light yield structure is not an artifact of the particular numerical solver used. To test this, the same simulated crossing muon sample, voxelization, detector bounds, and photon count inputs were reconstructed using both the \texttt{nnls} method and the \texttt{lsq\_linear} method described in section~\ref{sec:linear}. The resulting maps were then compared using average projections and residual difference maps, providing a numerical cross check of the reconstruction.

\begin{figure}[htbp]
\centering
\includegraphics[width=1\textwidth]{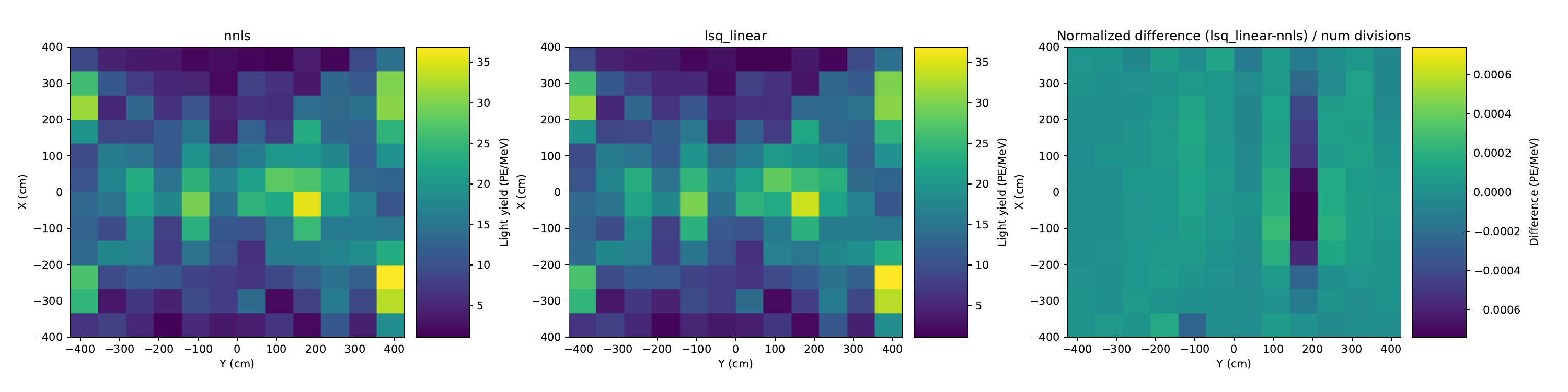}
\includegraphics[width=1\textwidth]{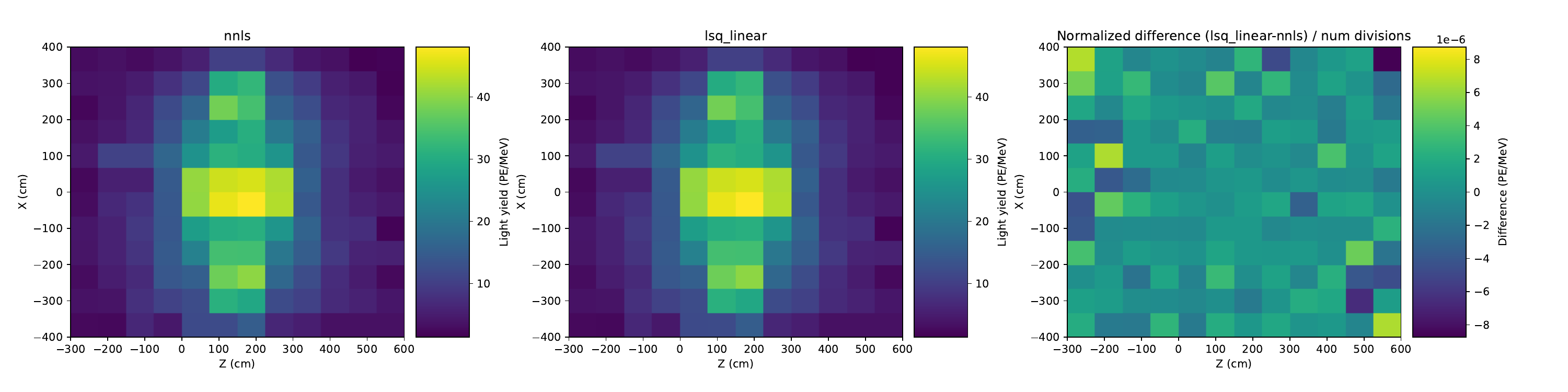}
\caption{Plots for the YX (top) and ZX (bottom) average projections using the \texttt{nnls} method (left) and the \texttt{lsq\_linear} method (center), with a normalized solver difference map between the two methods (right). The two methods produce nearly identical reconstructed light yield maps, with only small residual differences visible after subtracting the solutions.\label{fig:solver_compare}}
\end{figure}

The comparison of the solving methods in figure~\ref{fig:solver_compare} shows strong voxel-level agreement between the two reconstructions. The normalized residual difference maps are consistent with negligible solver dependent differences, indicating that both methods recover the same light yield structure for this problem. 

To quantify the agreement at the level of the original minimization problem, each reconstructed solution was inserted back into the linear system and compared to the input photon count vector. The event-level residual vector is defined as

\begin{equation}
r = Ax-b,
\end{equation}

where (Ax) is the predicted photon count vector obtained from the reconstructed light yield map and (b) is the simulated photon count input. The normalized residual norm is then defined as

\begin{equation}
R_{\mathrm{norm}} = \frac{\|r\|_2}{\|b\|_2}
= \frac{\|Ax-b\|_2}{\|b\|_2},
\end{equation}

which measures the total remaining event-level residual relative to the size of the input photon count vector. In addition, the total predicted to observed photon ratio is defined as

\begin{equation}
S =
\frac{\sum_i (Ax)_i}{\sum_i b_i}.
\end{equation}

This quantity tests whether the fitted reconstruction reproduces the overall photon count normalization of the sample. A value of (S=1) indicates that the total predicted photon count equals the total input photon count.

These two quantities are reported for both solvers in Table~\ref{tab:solver_fit}. The values of S are 1.001001 for \texttt{nnls} and 1.000990 for \texttt{lsq\_linear}, showing that both reconstructions reproduce the total photon count normalization to approximately (0.1\%). The normalized residual norms are also nearly identical: $(R_{\mathrm{norm}}=0.41677422)$ for \texttt{nnls} and $(R_{\mathrm{norm}}=0.41677452)$ for \texttt{lsq\_linear}. The absolute difference between the two normalized residuals is $3.0\times10^{-7}$, corresponding to a relative difference of around $7.2\times10^{-7}$. Therefore, both solvers not only produce visually consistent light yield maps, but also reach the same level of agreement with the original photon count data.

\begin{table}[htbp]
\centering
\caption{Fit quality comparison between the $12\times12\times12$ unregularized \texttt{nnls} and \texttt{lsq\_linear} reconstructions for 60,000 events with no PMT contributions. The residual is defined as $r=Ax-b$.\label{tab:solver_fit}}
\begin{tabular}{lccc}
\hline
Solver & $\sum Ax/\sum b$ & $\|Ax-b\|_2/\|b\|_2$ \\
\hline
\texttt{nnls}
& $1.001001$
& $0.41677422$ \\
\texttt{lsq\_linear}
& $1.000990$
& $0.41677452$ \\
\hline
\end{tabular}
\end{table}

The residual comparison indicates that the observed reconstruction is not solver dependent. The nearly identical residual metrics show that both solvers converge to the same minimum of the objective within the numerical precision relevant for this analysis. Since \texttt{lsq\_linear} can operate directly on sparse matrices, it is better suited for larger event samples and finer voxelizations, avoiding the memory limitations associated with constructing dense matrices for the \texttt{nnls} solver.

\subsection{Crossing Plane Variation}
\label{sec:vary_crossing}

\begin{figure}[htbp]
\centering
\includegraphics[width=.32\textwidth]{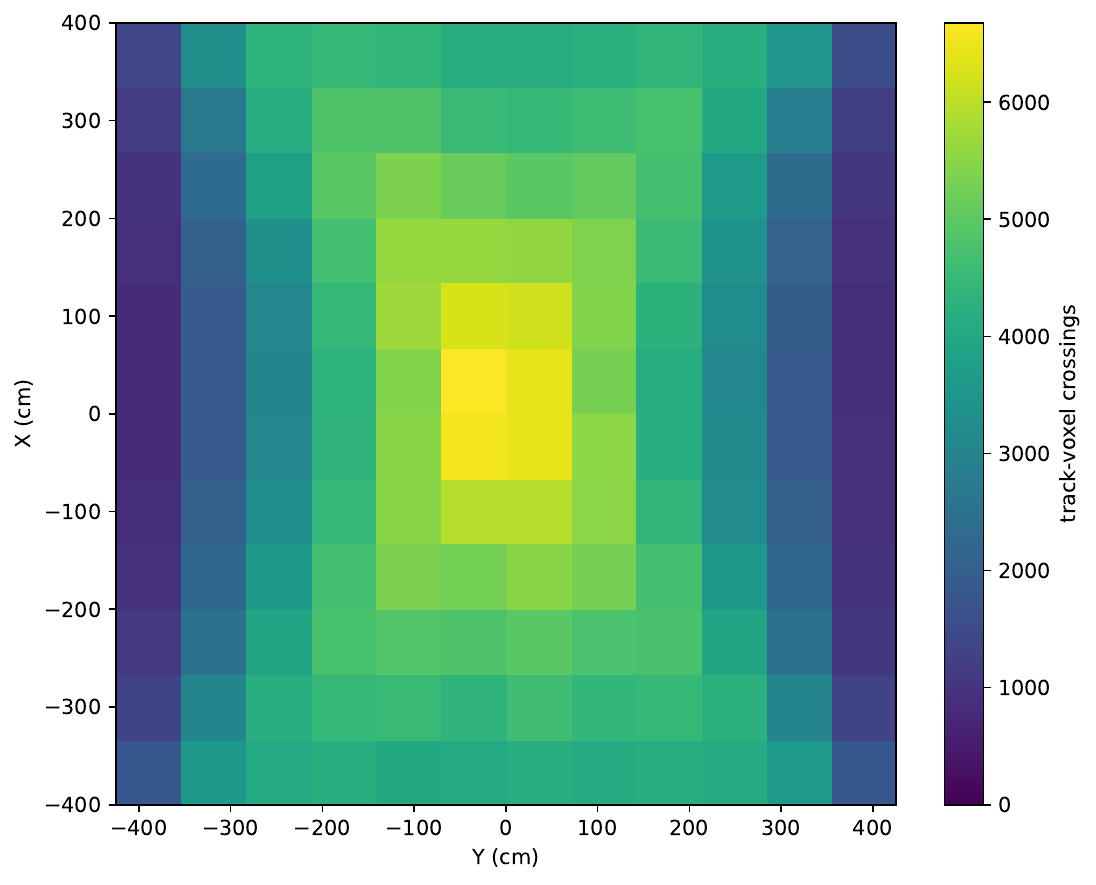}
\includegraphics[width=.32\textwidth]{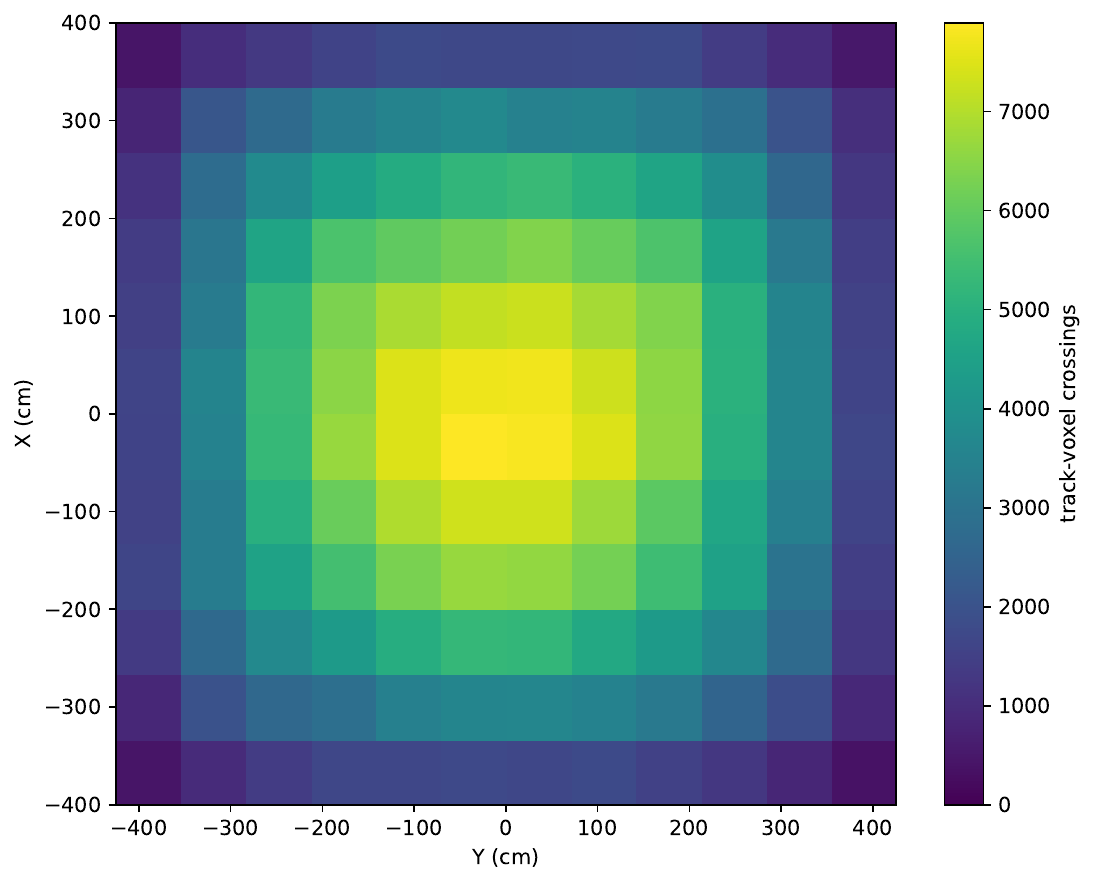}
\includegraphics[width=.32\textwidth]{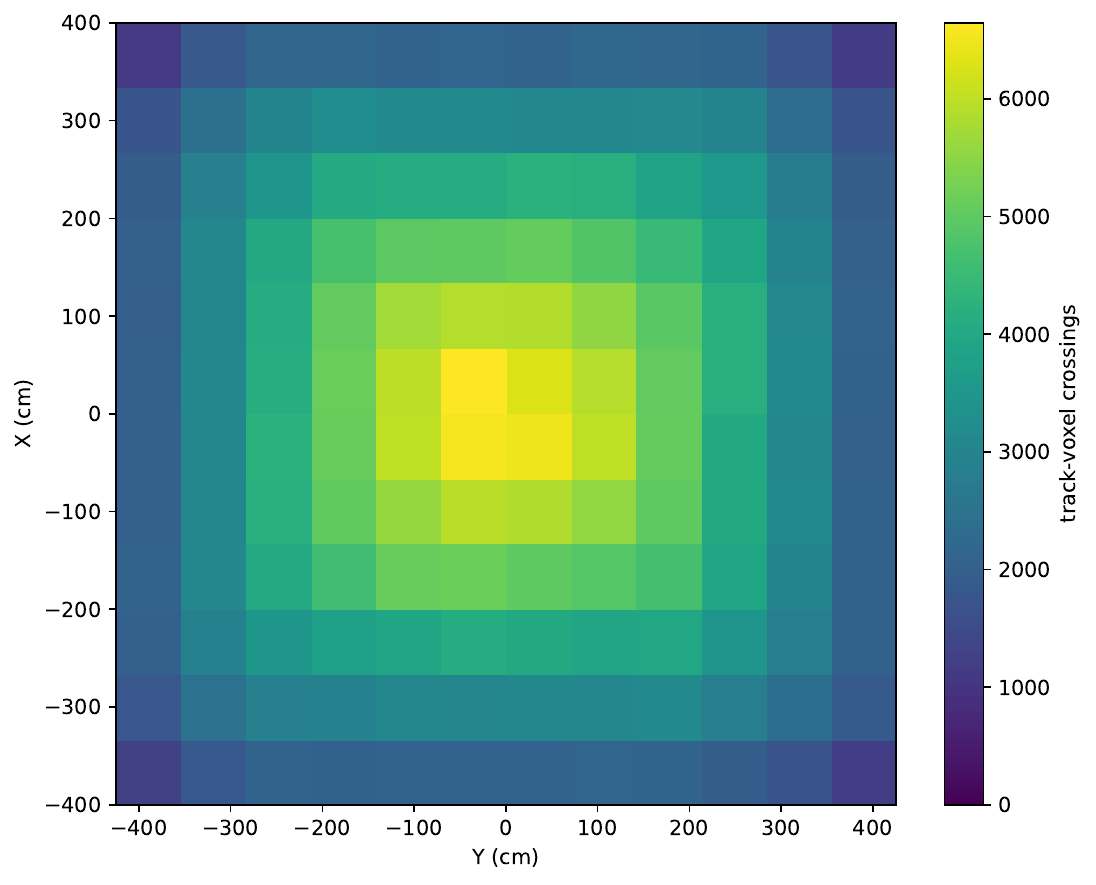}
\caption{The voxel occupancy plots for crossing muon generation through the X plane (left), the Z plane (center), and the combination of X, Y, and Z planes (right), all for 30,000 events. These plots can be compared to figure~\ref{fig:crt_full} (right), which shows the Y plane crossing data used in the rest of the paper.\label{fig:occupancy_compare}}
\end{figure}

The results presented so far have utilized crossing muons generated between a single pair of opposing planes, as shown in figure~\ref{fig:crt_full}. To test the effect of the crossing-muon direction, additional datasets were generated using opposing planes normal to each of the simulation's X, Y, and Z axes. The Y-plane configuration uses the same panels as the main sample. Each configuration contains 30,000 events, with starting and ending panel points sampled uniformly. The three samples were also reconstructed together, giving 90,000 events, and a random subset of 30,000 events was selected from the combined sample to compare mixed directions with a single direction at similar statistics. All configurations retain the event-generation settings and reconstruction procedure used for the main sample. The voxel occupancy plots for each of these new configurations are shown in figure~\ref{fig:occupancy_compare}.

\begin{table}[tbp]
\centering
\caption{Comparison of the regularized crossing-plane reconstructions generated with the same settings as the main sample. The mean light yield and voxel-by-voxel RMSE are evaluated over the full three-dimensional maps. The mean light yield of the simulation truth is 13.5~PE/MeV. The 30,000-event mixed sample is a subset of the 90,000-event sample.}
\label{tab:crossing_plane_rmse}
\small
\begin{tabular}{lrrr}
\hline
Configuration & Events & Mean LY (PE/MeV) & RMSE (PE/MeV) \\
\hline
$X$ planes & 30,000 & 12.5 & 9.8 \\
$Y$ planes & 30,000 & 13.2 & 8.4 \\
$Z$ planes & 30,000 & 12.7 & 14.3 \\
$X+Y+Z$ subset & 30,000 & 12.7 & 6.0 \\
$X+Y+Z$ combined & 90,000 & 12.5 & 4.8 \\
\hline
\end{tabular}
\end{table}

Table~\ref{tab:crossing_plane_rmse} compares the regularized reconstructions with the simulation truth. Among the single direction samples, the Y planes give the lowest RMSE, 8.4~PE/MeV, followed by the X planes at 9.8~PE/MeV, while the Z planes give the largest error of 14.3~PE/MeV. Combining the three directions reduces the RMSE to 4.8~PE/MeV, a reduction of approximately 43\% relative to the Y-plane sample. The mixed subset reaches 6.0~PE/MeV with only 30,000 events, approximately 29\% below the RMSE of the 30,000-event Y sample. The improvement therefore cannot be attributed to increased statistics alone. Despite the differences in RMSE, the mean reconstructed light yields span a narrow range, approximately 2--7\% below truth. Similar detector-average yields can therefore accompany substantial differences in the accuracy of the spatial reconstruction.

\begin{figure}[htbp]
\centering
\includegraphics[width=.32\textwidth]{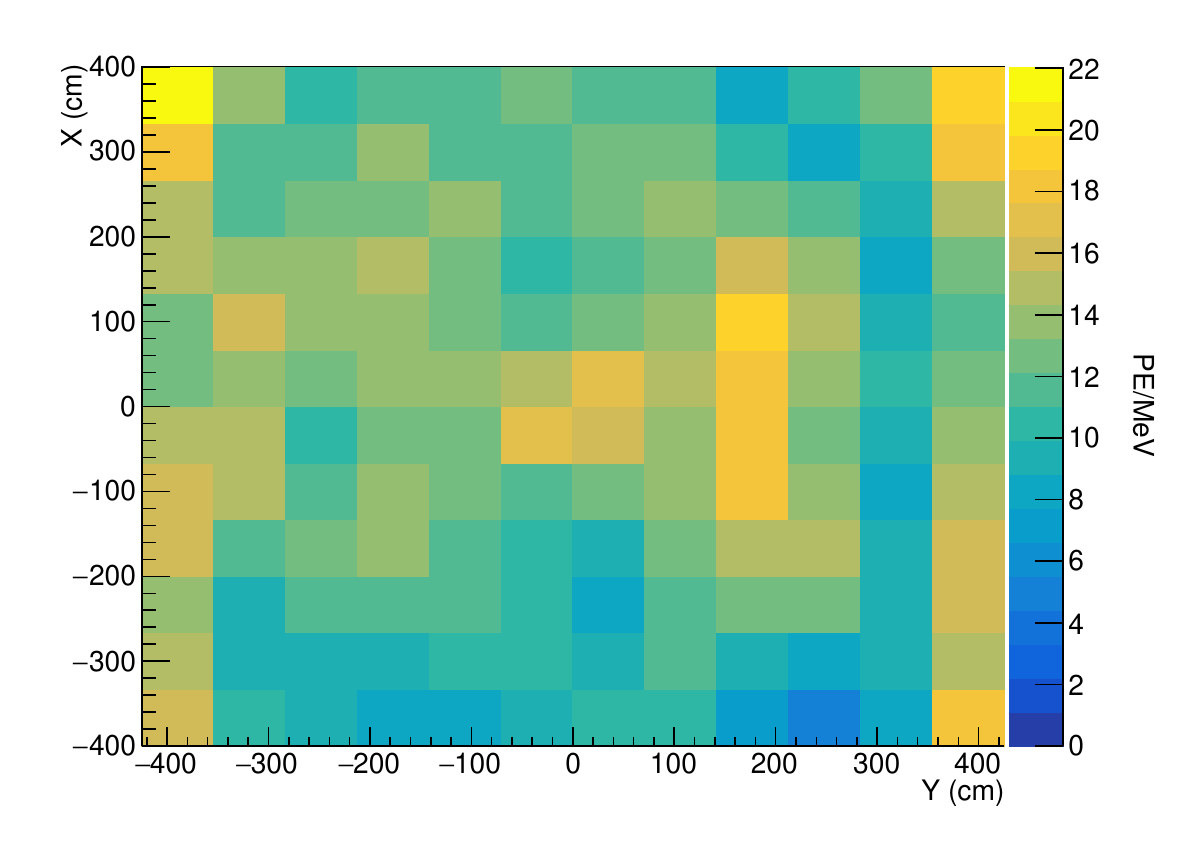}
\includegraphics[width=.32\textwidth]{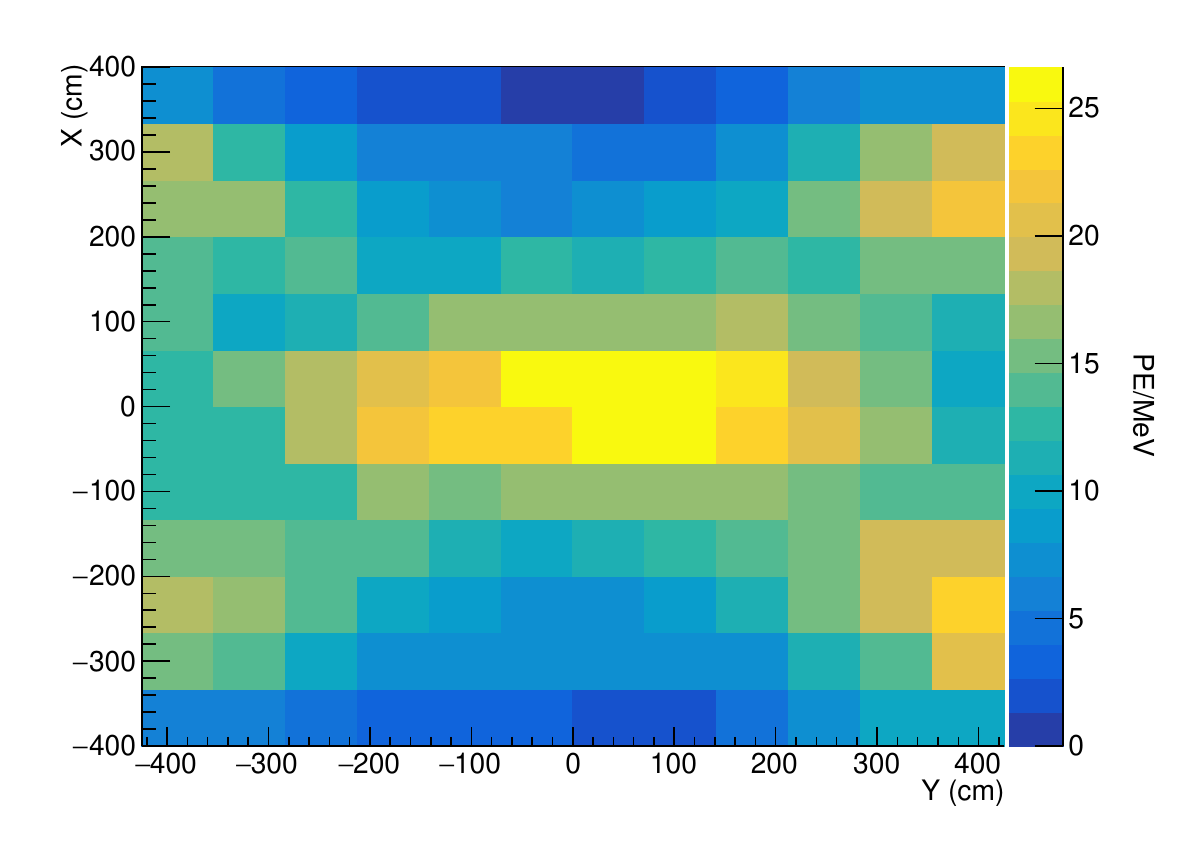}
\includegraphics[width=.32\textwidth]{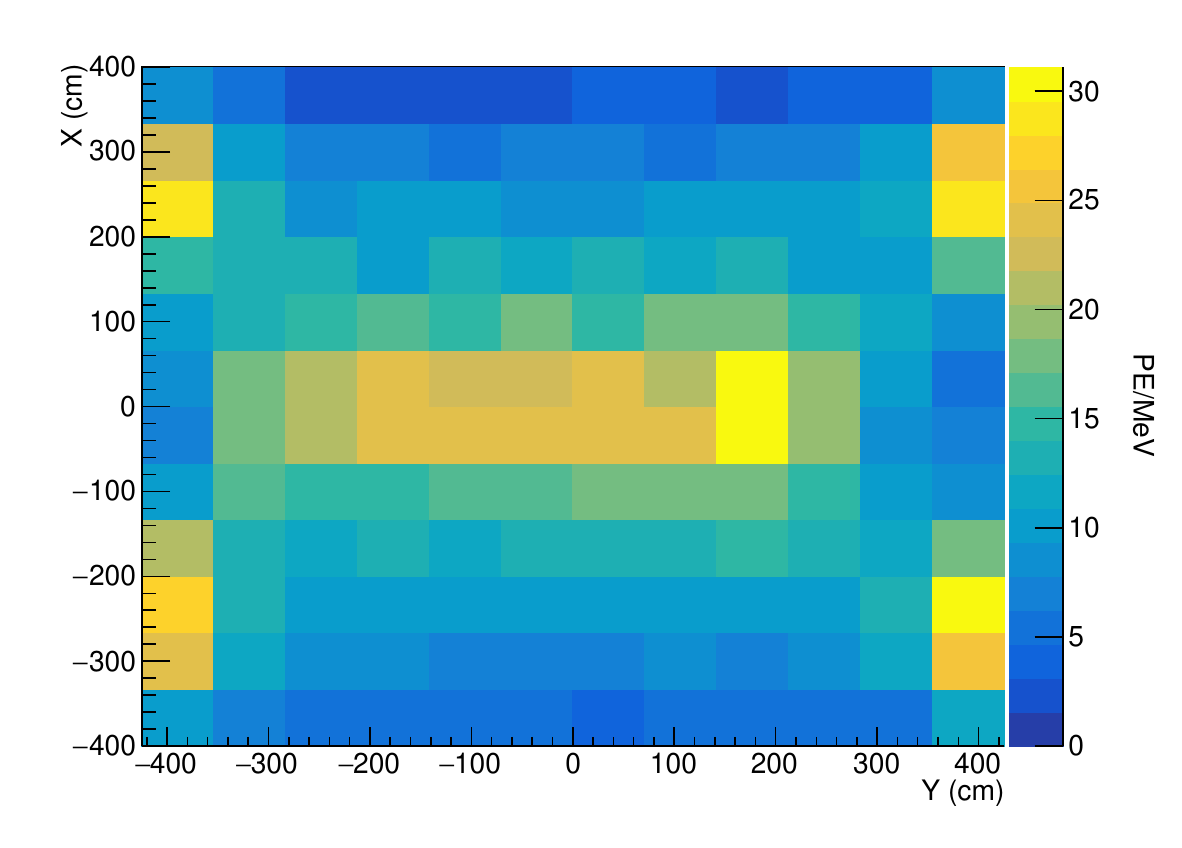}
\includegraphics[width=.32\textwidth]{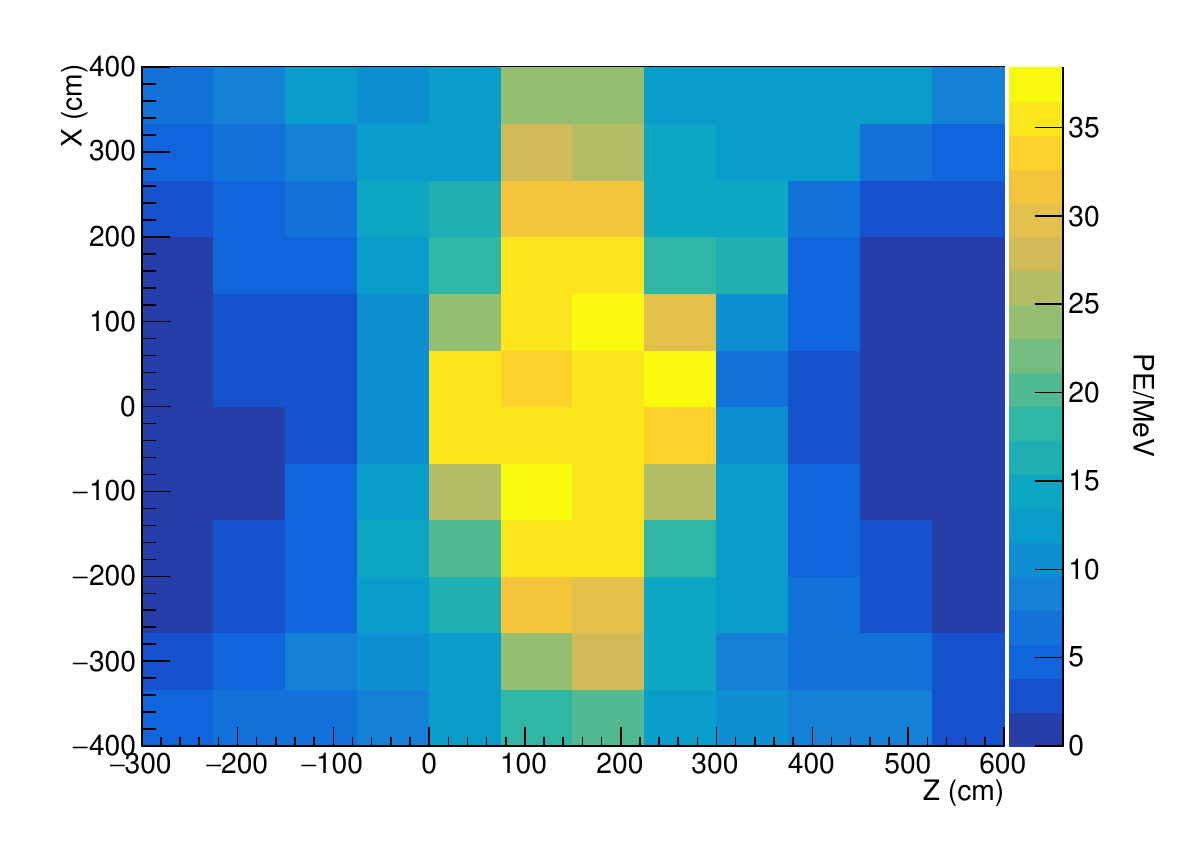}
\includegraphics[width=.32\textwidth]{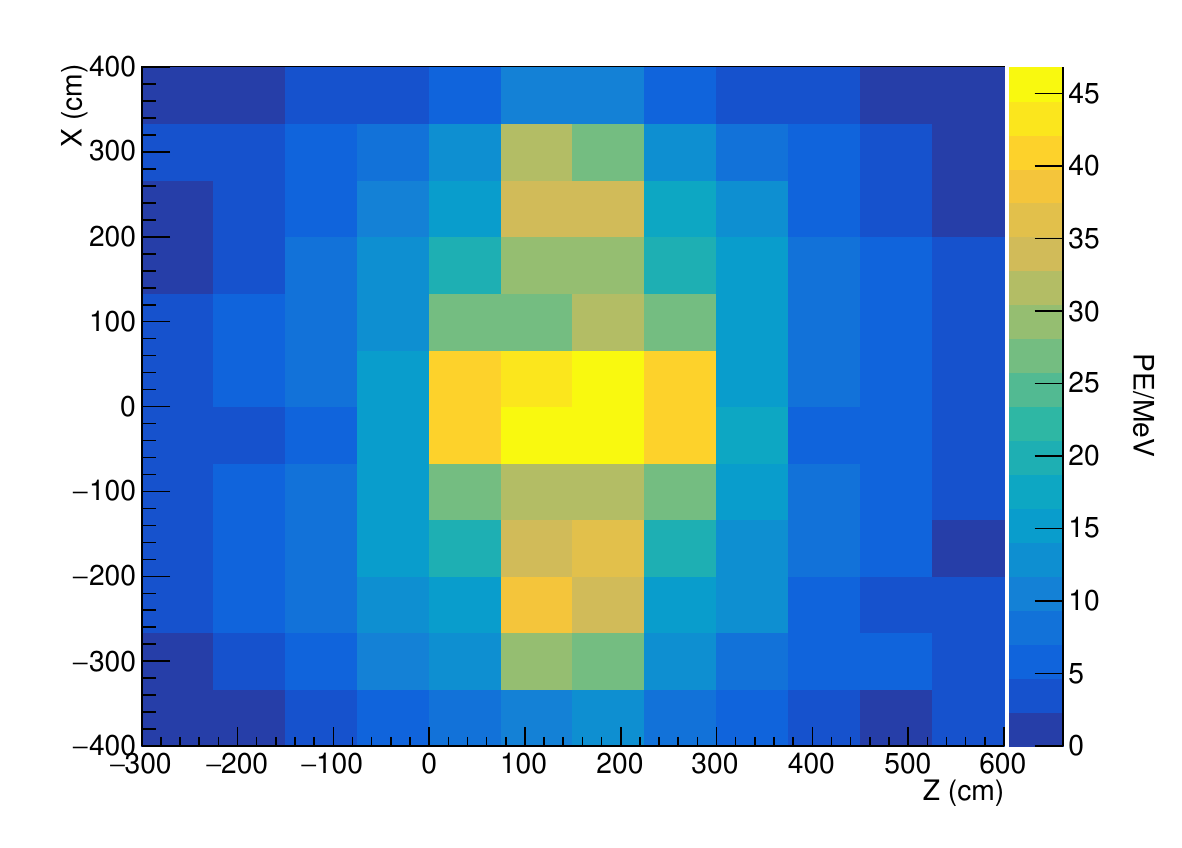}
\includegraphics[width=.32\textwidth]{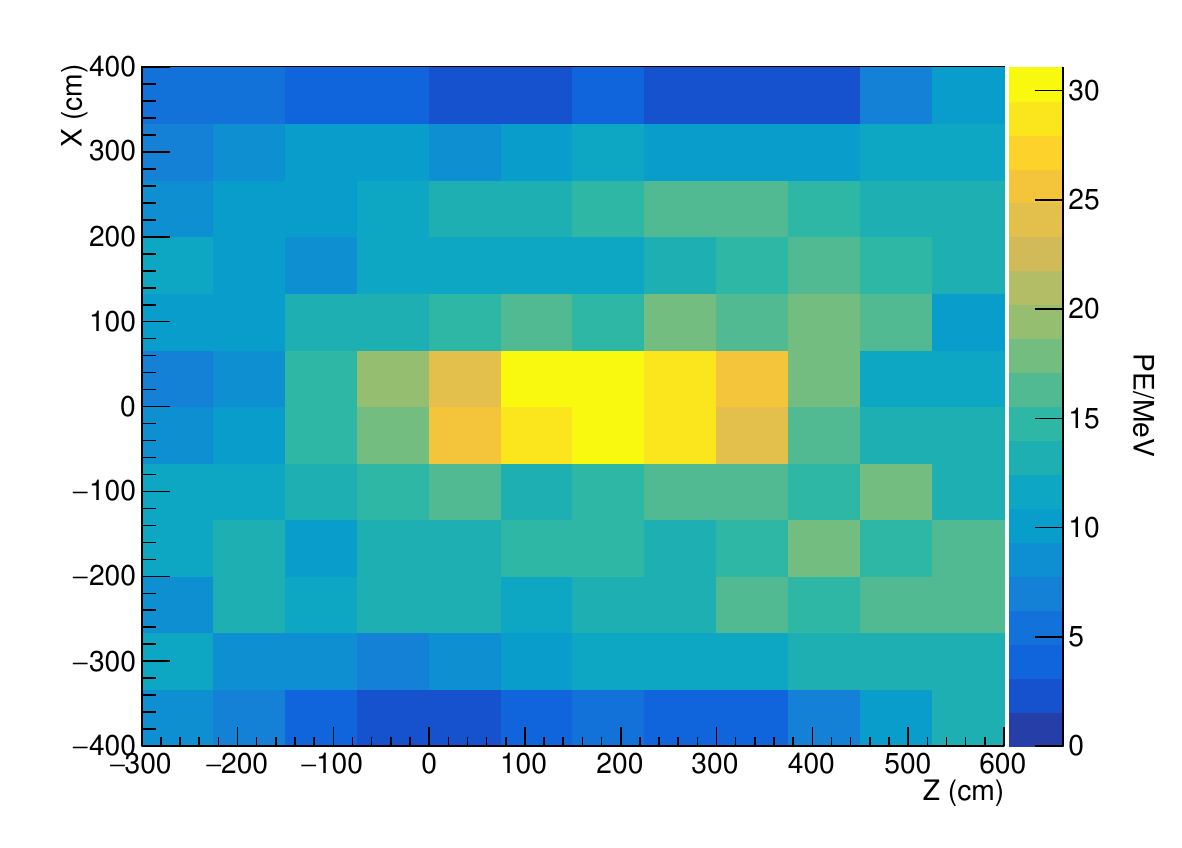}
\includegraphics[width=.32\textwidth]{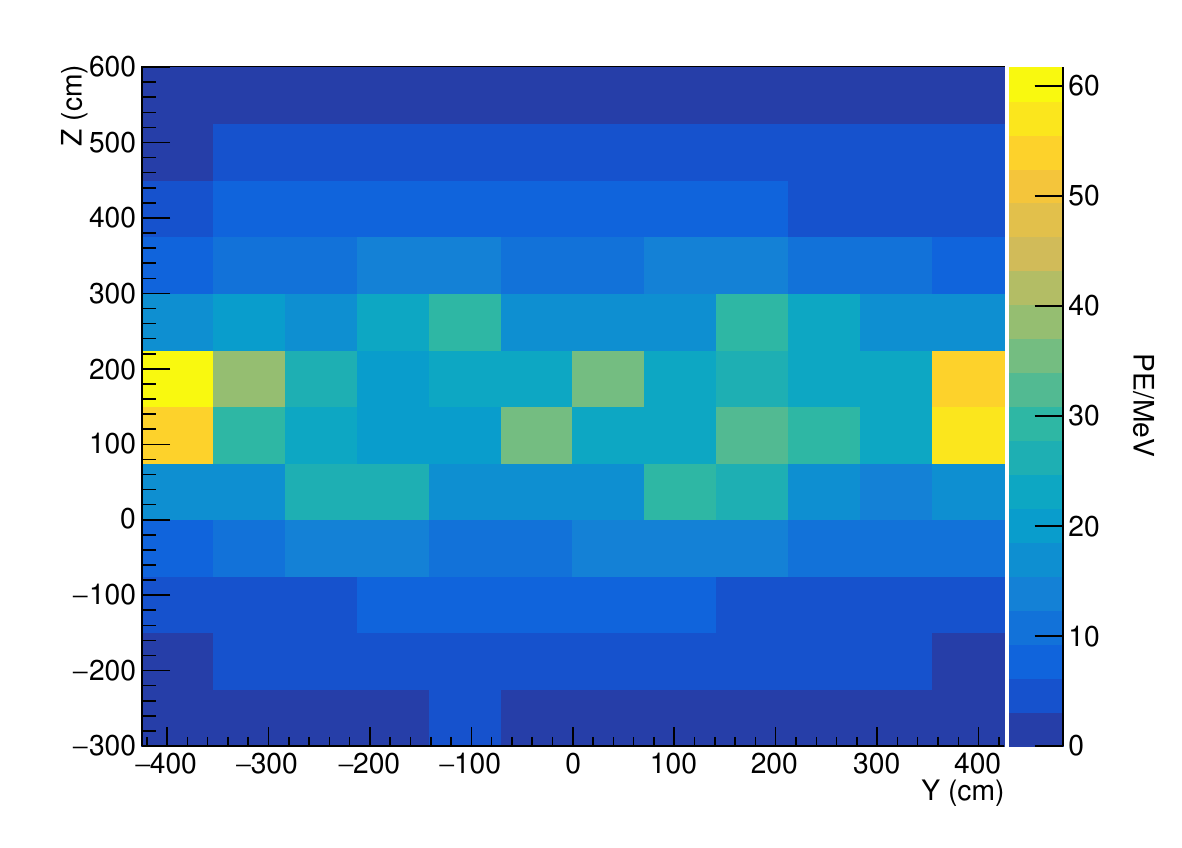}
\includegraphics[width=.32\textwidth]{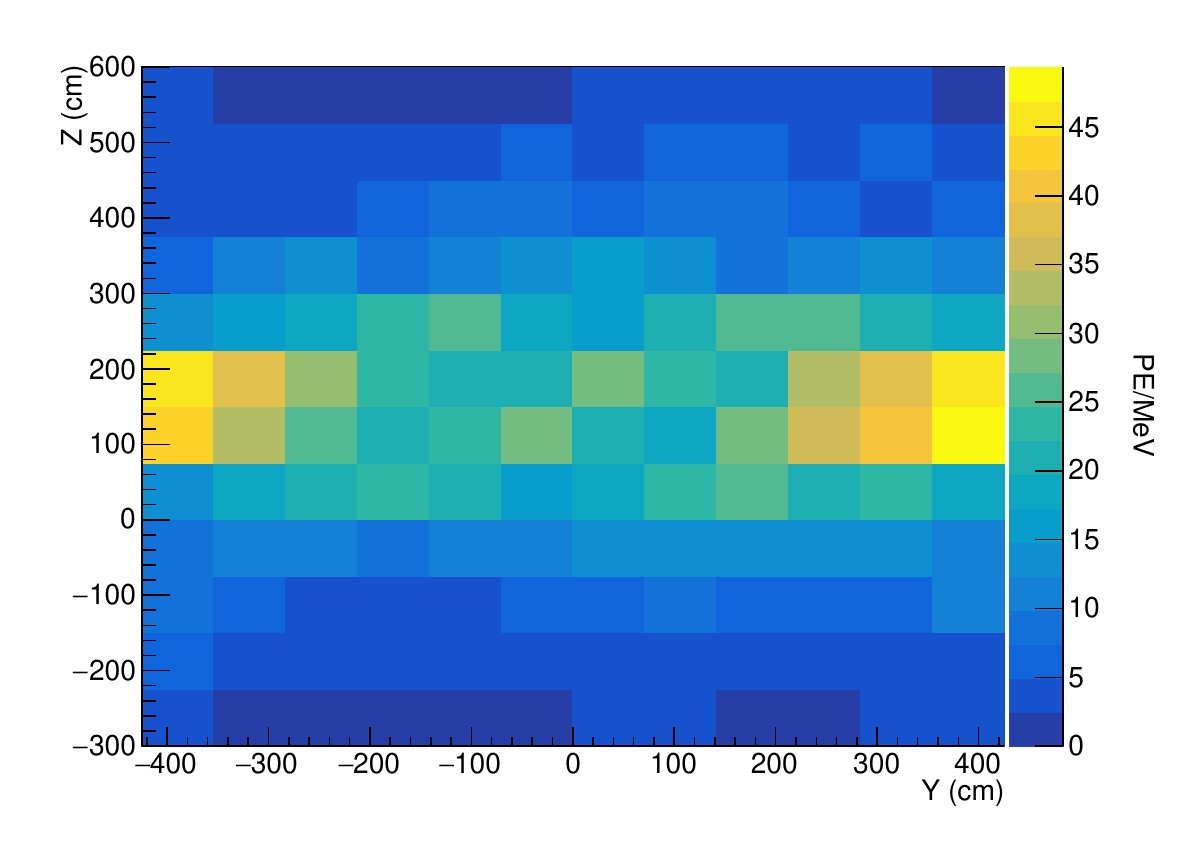}
\includegraphics[width=.32\textwidth]{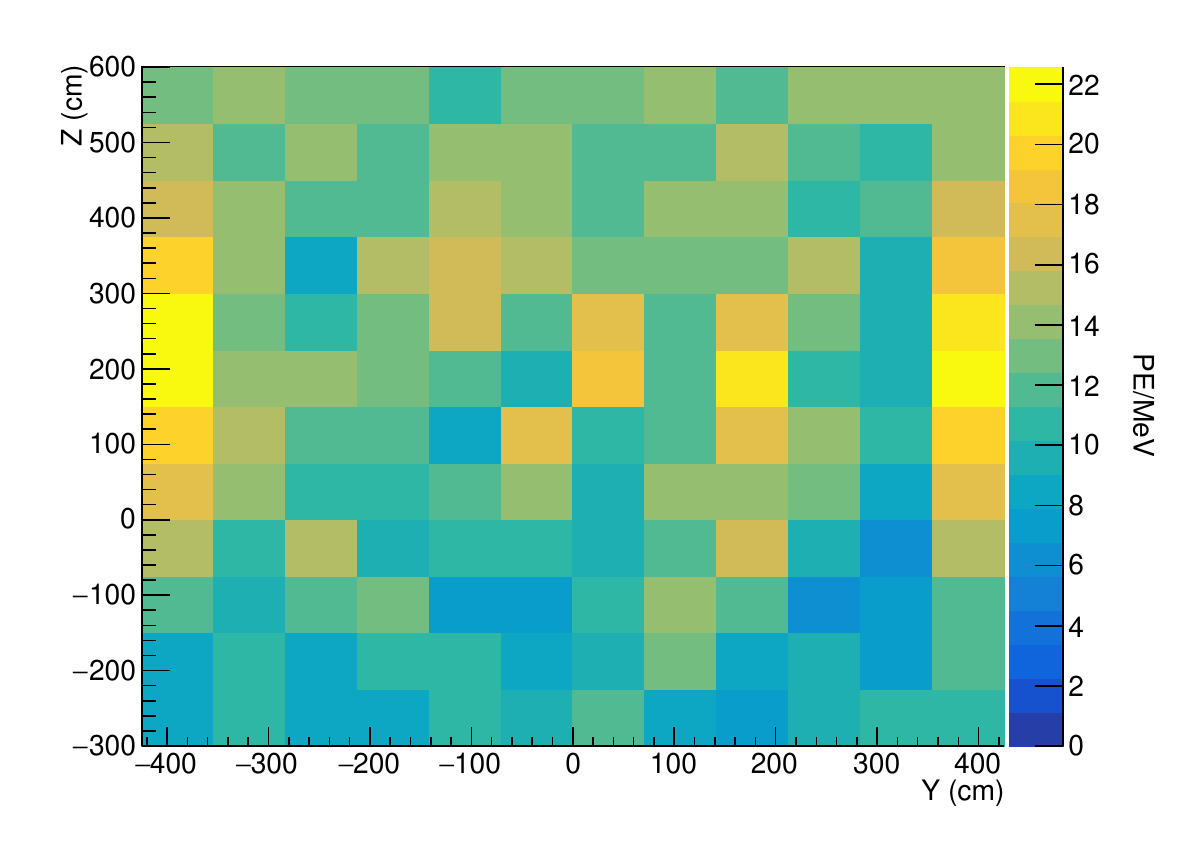}
\caption{Average projections of the regularized light yield maps reconstructed using individual crossing-plane configurations. From left to right, the columns correspond to the X, Y, and Z plane samples. The top row shows the displayed YX projection, the middle row shows the ZX projection, and the bottom row shows the displayed YZ projection, with each map averaged over the remaining coordinate.\label{fig:plane_compare}}
\end{figure}

\begin{figure}[htbp]
\centering
\includegraphics[width=.32\textwidth]{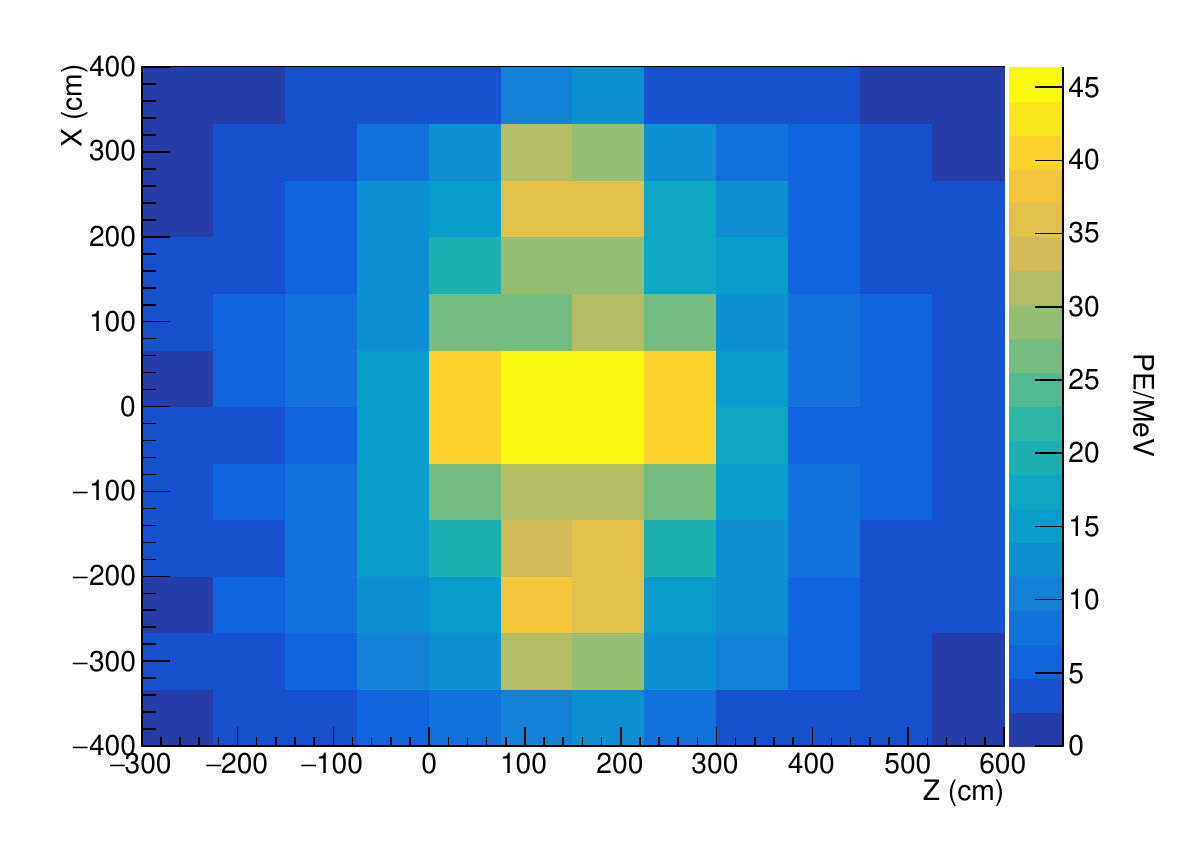}
\includegraphics[width=.32\textwidth]{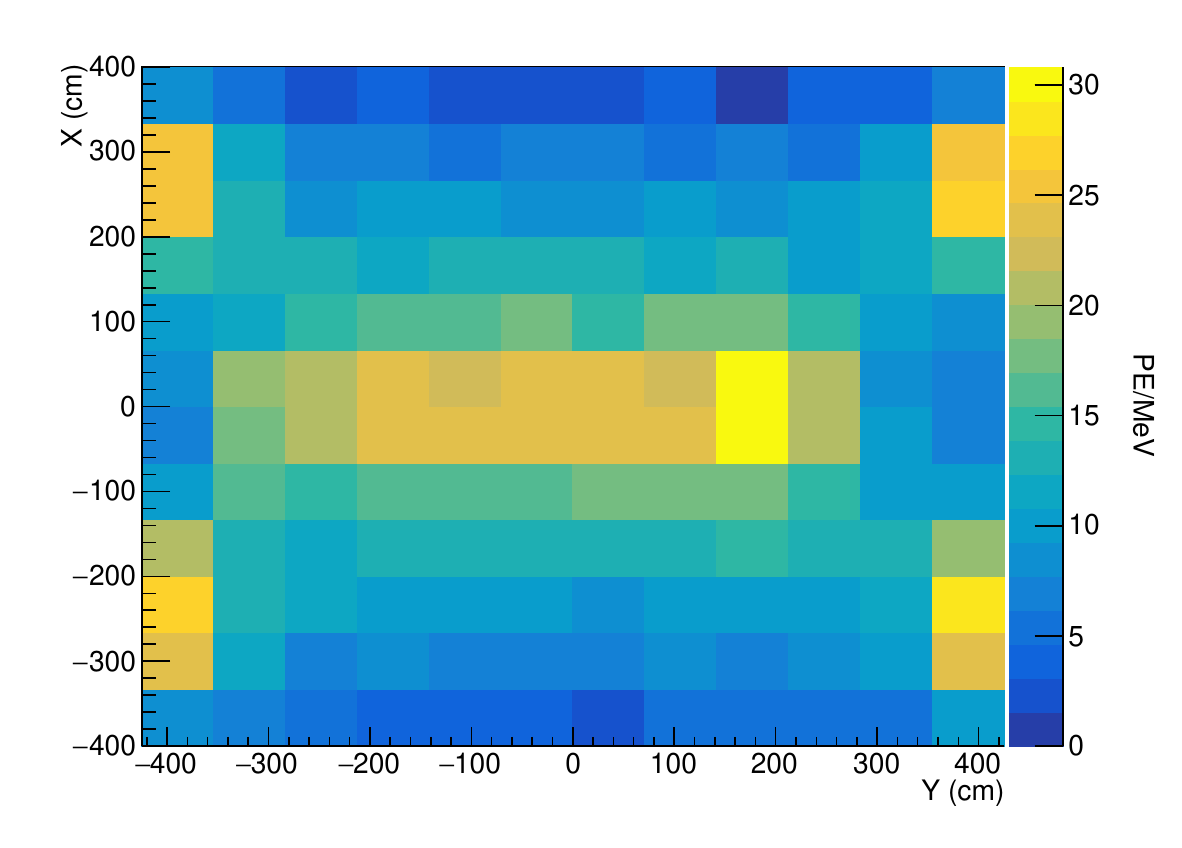}
\includegraphics[width=.32\textwidth]{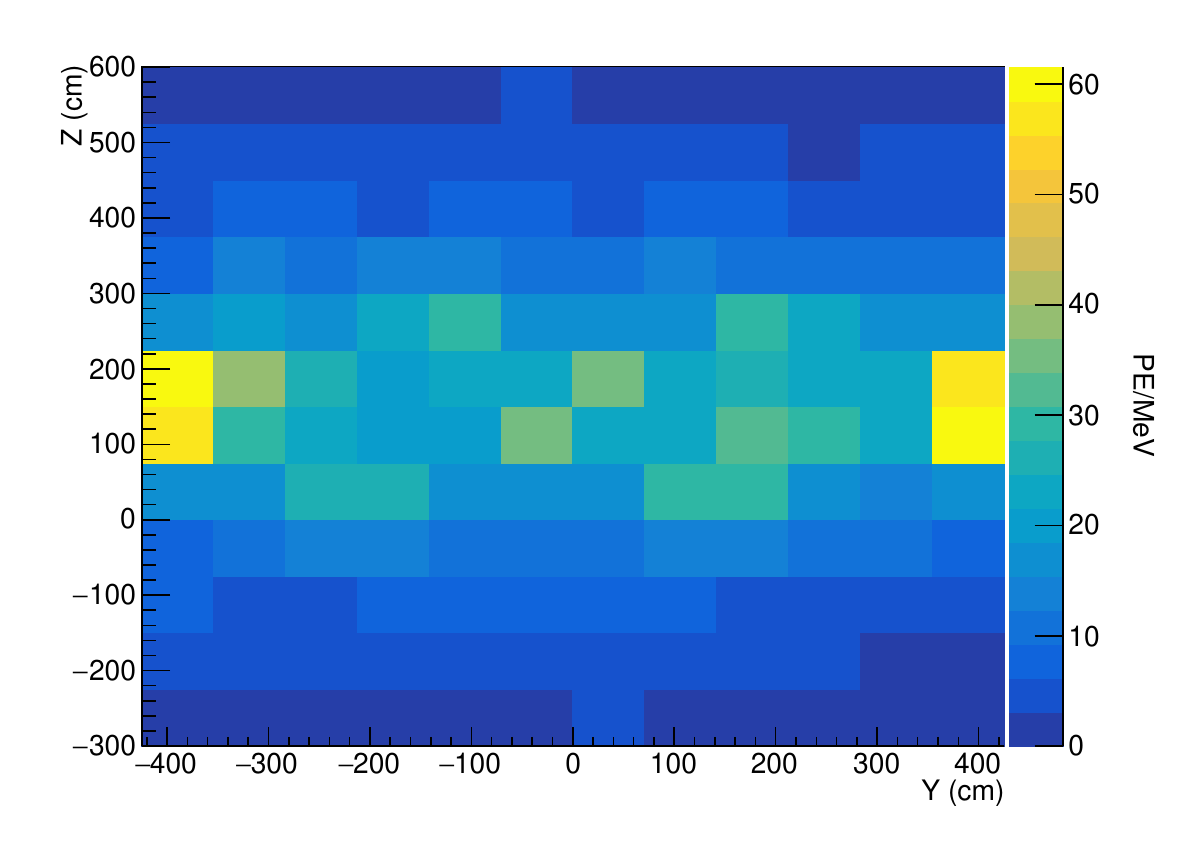}
\caption{Average projections of the regularized light yield map reconstructed using the combined X, Y, and Z plane samples, with 90,000 events total. From left to right, the panels show the displayed ZX, YX, and YZ projections, averaged over the remaining coordinate. Combining track directions provides complementary constraints on the spatial distribution of light yield, resulting in a better reconstruction compared to any single plane of crossing data.\label{fig:xyz_compare}}
\end{figure}

Figure~\ref{fig:plane_compare} shows that the crossing direction affects which features of the light yield distribution can be resolved. The ability to resolve the map depends on how these track directions intersect the spatial structure associated with the detector geometry. Tracks that repeatedly cross regions of enhanced and suppressed light yield together constrain their combined response more strongly than their individual contributions. This limits the reconstruction's ability to locate the enhancements and recover their amplitudes. In general, resolving a feature requires tracks that intersect it with sufficiently different paths and relative path lengths to separate its contribution from neighboring regions. Regularization stabilizes the remaining ambiguity by favoring smooth solutions, but cannot replace those independent constraints. Figure~\ref{fig:xyz_compare} shows the reconstruction obtained by combining all three crossing-plane samples. Tracks from different directions cross the same voxels in different combinations, helping to separate contributions that remain ambiguous within an individual sample. The best resolved projection from each individual crossing plane dataset (YZ for X-plane, XZ for Y-plane, and XY for Z-plane) closely resembles the corresponding projection in the combined reconstruction. This agreement illustrates how combining track directions allows the reconstruction to recover spatial features constrained by each sample.

The configurations represented in this study do not reflect an accurate muon flux for a surface detector. The flux and accepted track distribution depend on the detector geometry, the method used to reconstruct track starting and ending points, and any data quality cuts that could limit track variation. Although the cosmic muon sample in ProtoDUNE-VD will be dominated by tracks traveling in the vertical X direction, using TPC-reconstructed trajectories can also include inclined tracks and tracks entering or exiting through different detector faces. Including these tracks would provide complementary paths through the detector beyond those represented by the opposing X-plane configuration. With sufficient statistics and broader accepted angular coverage, such a sample should better constrain the spatial structure than the restricted X-plane sample studied here. The size of this improvement will depend on the track selection and coverage of the detector volume.

\section{Conclusions}
\label{sec:conclusions}

This study demonstrates a method for reconstructing voxelized 3D light yield maps in surface LArTPCs using crossing cosmic muons. The path length of each selected muon through the detector is converted to an estimated deposited energy in each crossed voxel, producing a linear system that relates the unknown voxel light yields to the total detected photon signal. Imposing a nonnegative constraint ensures that the reconstructed light yields remain physical, while the sparse structure of the track coverage matrix allows the method to be extended to larger event samples and finer voxelizations. An example implementation of the sparse bounded least squares reconstruction used in this study, together with the ROOT macro used to visualize the reconstructed light yield maps, is publicly available on GitHub\footnote{The code is available at the LY Mapping GitHub repository at: \url{https://github.com/AlexHeindel/ly-mapping}}. The input for this code is configured for the ROOT tree and branch structure used in this study but can be adapted to other datasets and file types by modifying the relevant input and detector geometry definitions.

Although this study uses straight line crossing muon trajectories, as would be obtained from two CRT crossing points, the reconstruction framework is not limited to this track model. Charge-reconstructed tracks can be incorporated by either taking start and end points to form straight line tracks, or representing each trajectory as a sequence of short piecewise-linear segments and applying the Liang-Barsky clipping algorithm to each segment independently. The total path length through each voxel is then obtained by summing the segment contributions. This makes the method adaptable to different track reconstruction inputs and detector geometries. Together with the demonstrated sensitivity to changes in photon detector configuration, this flexibility indicates that crossing muon light yield reconstruction could be applied broadly to surface LArTPCs with sufficient cosmic muon statistics.

\acknowledgments

We thank Jose Soto, Peter Filip, and Viktor Pěč for insightful discussions. We thank the members of the DUNE Collaboration whose work in developing and maintaining the ProtoDUNE simulation framework made this study possible.


\bibliographystyle{JHEP}
\bibliography{biblio.bib}

\end{document}